\documentclass[prd,aps,superscriptaddress,showpacs,preprintnumbers,amsmath,amssymb,floatfix,12pt]{revtex4}
\usepackage[dvips]{graphicx}
\unitlength=1.2mm \preprint{CU-TP-1154,BNL-HET-06/7,KEK-CP-179}
\begin{document}

\newcommand{\tr}{\mbox{tr}\,}
\newcommand{\DDs}{$D$/$D_{ s}$\ }
\newcommand{\mres}{m_{\rm res}}
\newcommand{\simg}{\rlap{\raise -4pt \hbox{$\sim$}}
                   \raise 3pt \hbox{$>$}}
\newcommand{\siml}{\rlap{\raise -4pt \hbox{$\sim$}}
                   \raise 3pt \hbox{$<$}}
\title
{Charm as a domain wall fermion in quenched lattice QCD}
\author{Huey-Wen Lin}
\email{hwlin@theory1.phys.columbia.edu}
\affiliation{Physics Department, Columbia University, New York, NY 10027, USA}
\author{Shigemi Ohta}
\email{shigemi.ohta@kek.jp}%
\affiliation{Institute of Particle and Nuclear Studies, KEK, Ibaraki 305-0801, Japan}
\affiliation{The Graduate University for Advanced Studies (Sokendai), Tsukuba, Ibaraki 305-0801, Japan}
\affiliation{RIKEN BNL Research Center, Brookhaven National Laboratory, Upton, NY 11973, USA}
\author{Amarjit Soni}
\email{soni@bnl.gov}%
\affiliation{Department of Physics, Brookhaven National Laboratory, Upton, NY 11973, USA}
\author{Norikazu Yamada}
\email{norikazu.yamada@kek.jp}%
\affiliation{Institute of Particle and Nuclear Studies, KEK, Ibaraki 305-0801, Japan}
\affiliation{The Graduate University for Advanced Studies (Sokendai), Tsukuba, Ibaraki 305-0801, Japan}
\date{July 20, 2006}
\begin{abstract}
We report a study describing the charm quark by a domain-wall
fermion (DWF) in lattice quantum chromodynamics (QCD).  Our study
uses a quenched gauge ensemble with the DBW2 rectangle-improved
gauge action at a lattice cutoff of $a^{-1} \sim 3$~GeV. We
calculate masses of heavy-light (charmed) and heavy-heavy
(charmonium) mesons with spin-parity \(J^P = 0^\mp\) and
\(1^\mp\), leptonic decay constants of the charmed pseudoscalar
mesons (\(D\) and \(D_s\)), and the \(D^0\)-\(\overline{D^0}\)
mixing parameter.  The charm quark mass is found to be
$m^{\overline{\rm MS}}_{ c}(m_{ c})=1.24(1)(18)$~GeV. The mass
splittings in charmed-meson parity partners \(\Delta_{q,J=0}\) and
\(\Delta_{q, J=1}\) are degenerate within statistical errors, in
accord with experiment, and they satisfy a relation $\Delta_{q=ud,
J} > \Delta_{q=s, J}$, also consistent with experiment. A C-odd
axial vector charmonium state, \(h_c\), lies 22(11) MeV above the
\(\chi_{c1}\) meson, or \(m_{h_{c}} =  3533(11)_{\rm stat.}\) MeV
using the experimental \(\chi_{c1}\) mass. However, in this
regard, we emphasize significant discrepancies in the calculation
of hyperfine splittings on the lattice. The leptonic decay
constants of $D$ and $D_s$ mesons are found to
be $f_D=232(7)_{\rm stat.}(^{+6}_{-0})_{\rm chiral}(11)_{\rm syst.}$~MeV 
and $f_{D_s}/f_{D} = 1.05(2)_{\rm stat.}(^{+0}_{-2})_{\rm
chiral}(2)_{\rm syst.}$, where the first error is statistical, the
second a systematic due to chiral extrapolation and the third
error combination of other known systematics. The
$D^0$-$\overline{D^0}$ mixing bag parameter, which enters the
$\Delta C = 2$ transition amplitude, is found to be %
$B_D(2\mbox{ GeV})=0.845(24)_{\rm stat.}(^{+24}_{-6})_{\rm
chiral}(105)_{\rm syst.}$.
\end{abstract}
\maketitle

\section{Introduction}

Properties of hadrons containing one or more charm quarks have
been intensively investigated in recent experiments
\cite{Bonvicini:2004gv,Arms:2003ra,Eisenstein:2004bg,Besson:2003jp,Flood:2005da,Aubert:2003fg,Benussi:2005db,Danilov:2004td,Abe:2005hd,Drutskoy:2005zr}.
While the main purpose of these experiments is to acquire
information useful or necessary to CKM phenomenology, it is also
exciting that several new hadron states have been discovered and
confirmed in these experiments. For the latest reviews on these
new states, see, for example,
Refs.~\cite{Swanson:2005tq,Quigg:2005tv,Barnes:2005zy}. Lattice
QCD should be able to describe various features of these new
states. However, in lattice calculations of heavy quark systems,
the discretization error associated with masses as large as the
lattice cutoff makes the interpretation of numerical results
ambiguous. One possible way to avoid this systematic error is to
rely on an effective theory such as
HQET~\cite{Eichten:1989zv,Eichten:1989kb,Eichten:1990vp} or
NRQCD~\cite{Thacker:1990bm,Lepage:1992tx}. With such an approach,
however, one must discard either the ability to study quarkonia or
to take the continuum limit. An alternative is to rely on a
relativistic method with brute numerical force. Thanks to rapid
growth of computational resources, discretization errors in this
approach for a relatively light heavy quark like charm are
beginning to be brought under control. Several studies have
already been made in this direction
\cite{deDivitiis:2003iy,deDivitiis:2003wy,Juttner:2003ns,Lin:2005ze,Kuramashi:2005ww}.

In this work, we explore the feasibility of applying the
domain-wall fermion
(DWF)~\cite{Kaplan:1992bt,Shamir:1993zy,Furman:1994ky} as a heavy
quark within the quenched approximation. Domain-wall fermions have
been successfully applied to calculations of the light hadron
spectrum and weak matrix
elements~\cite{Blum:1996jf,Blum:1997mz,Blum:2000kn,AliKhan:2000iv,Aoki:2002vt}.
The primary advantage of working with this fermion action is its
ability to retain continuum-like chiral symmetry even at a finite
lattice spacing, by adding a fifth dimension to the lattice. The
size of the symmetry breaking is represented by the residual mass,
$am_{\rm res}$, which is adjusted by changing the
fifth-dimensional extent of the lattice and is of the order of
$10^{-3}$ in lattice units in state-of-the-art calculations. It is
known that the presence of exact chiral symmetry guarantees the
absence of $O(a)$ discretization error, and in the domain-wall
case, the violation of the order of $am_{\rm res}$ implies that
the leading error is significantly suppressed. Thus it seems to be
a natural extension to apply the DWF formalism to heavier, \(c\)
and \(b\), quarks. To make the next-to-leading-order
discretization error ($\sim O((a m_q)^2)$) as small as possible,
our calculations are carried out on relatively fine lattices with
$a^{-1}\sim 3$~GeV. Although this study is done in the quenched
approximation, the experience should be useful in the future
studies with dynamical fermions.

One way to test a given heavy quark formalism is to see how
consistently the formalism describes both heavy-heavy and
heavy-light systems with a single value of the mass parameter.
While a single mass parameter does not have to describe both
systems completely consistently in the quenched approximation and
at a finite lattice spacing, making a quantitative test at some
point will offer a reference for future work. Although we do not
take a continuum limit here, such a study should become easier in
the near future as available computational resources grow.

Once the charm quark mass is determined, it is interesting to see
how well the lattice calculation reproduces the mass spectrum of
recently discovered hadrons: the excited $D$ meson states with
spin-parity $J^{P}=0^{+}$ and $1^{+}$. The spectrum of these
mesons had been studied before their discoveries by two groups
based on chiral quark models~\cite{Nowak:1992um,Bardeen:1993ae}.
These turned out to describe some qualitative features well. After
the experimental discoveries, these groups refined their analyses
quantitatively~\cite{Bardeen:2003kt,Nowak:2003ra}. The studies
suggest that chiral symmetry in the light quark sector plays an
important role in the spectrum of these mesons. Since we apply the
domain-wall formalism to both heavy and light quarks, we expect to
get better descriptions of these mesons than earlier works with
Wilson-type light quarks.

We would also like to make a lattice study of flavor physics,
especially weak matrix elements of $D$ mesons. As an example, we
present the $D$ meson leptonic decay constants and the bag
parameter, $B_D$, relevant to $D^0$-$\overline{D}^0$ mixing.
Experiments are currently searching for evidence of mixing in the
neutral $D$-meson system~\cite{Flood:2005da}. Since the expected
amplitude is very small in the Standard Model, once it has been
discovered, it would provide an important probe of physics beyond
the Standard Model(See, e.g.,\ Ref.\ \cite{Petrov:2004rf}.) Unlike
the neutral $B$ meson system, in $D^0$-$\overline{D}^0$ mixing it
is not clear whether the short-distance contribution mediated by a
local four-quark operator dominates over long-distance
contributions. Even so, it is still sensible to have an idea about
the size of the short-distance contribution. Knowing the size of
$B_D$ will be also useful when evaluating $B_B$ by extrapolation
in quark mass. In this paper we present the first DWF charm
lattice study of this quantity. Note there were Wilson fermions
studies earlier such as Refs.~\cite{Bernard:1988dy,Gupta:1996yt}.

The first exploratory study of massive DWF was done in
Refs.~\cite{Christ:2004gc,Liu:thesis}, which looked into the
low-lying eigenmodes of the five-dimensional Hermitian domain-wall
Dirac operator in detail and their dependence on the bare quark
mass, $am_q$. All eigenmodes are classified into one of two kinds
of states. The first are the physical, or ``decaying'', states
which are bound to the four-dimensional domain walls located at
either end of the fifth dimension. Their wavefunctions fall
exponentially away from the wall. When $am_q \ll 1$, these states
dominate the low-lying eigenmodes of the whole system and describe
four-dimensional physics. The other class contains unphysical, or
``propagating'', states, which have non-zero momentum in the fifth
dimension.  Their eigenvalues are large, $\sim O(1/a)$. From the
viewpoint of the four-dimensional effective theory, these states
are unphysical, a source of non-locality that can invalidate the
effective theory. The gap between these two types of states is
controlled by the domain wall height ``$M_5$'' parameter. In the
study of Refs.~\cite{Christ:2004gc,Liu:thesis}, it turned out that
as $am_q$ increases the absolute values of the eigenvalues of the
decaying states rapidly increase, and the binding to the domain
walls becomes less tight or even unbound. On the other hand, the
eigenvalues of the propagating states increase only slowly. Thus,
if $am_q$ further increases, at some point the lowest eigenmode in
the system will be one of the propagating states. Then one might
worry that something wrong happens to the
four-dimensional effective theory. %
In the past quenched studies~\cite{Christ:2004gc,Liu:thesis}, this
mass threshold can be as low as $0.2$ (in lattice units) for both
$\beta=6.0$ Wilson gauge action and $\beta=0.87$ DBW2 gauge
action, and 0.4 for $\beta=1.04$ DBW2 gauge action with $M_5=1.8$.
In the present work, as will be described, a higher lattice
coupling  is used for the DBW2 gauge action. This results in a
threshold of \(a m_q \sim 0.5\) or higher in lattice units, while
the bare charm quark mass is below 0.4.

The rest of the paper is organized as follows. In
Sec.~\ref{SecSimulation} we summarize our numerical methods and
their parameters. Then meson mass spectra and charm mass analyses
are presented in Sec.~\ref{SecSpectroscopy}.
Section~\ref{SecDecayConst} presents results for decay constants.
We will also present the first DWF calculation of the mixing
parameter, $B_D$, in Section~\ref{SecBD}. The $SU(3)$-breaking
ratio is discussed in Section~\ref{Sec:SU(3)breaking}.
Extrapolations in \(1/m_{\rm heavy}\) to the static limit are
reported in Section ~\ref{Bmeson}. Some systematic errors are
discussed in Sec.~\ref{SecSysErr}. Section~\ref{SecFuture}
concludes with a summary and future outlook for this program.
Preliminary results on spectrum and decay constants in this work
are reported in Refs.~\cite{Yamada:2003bk,Ohta:2005cn,Lin:2006kg}.

\section{Numerical Method and Parameters}
\label{SecSimulation}

The numerical lattice QCD calculations reported in this paper were
performed in the quenched approximation using the QCDSP computers
at RIKEN-BNL Research Center and Columbia University. The gauge
configurations were generated in a previous RBC work determining
the neutral kaon mixing bag parameter, $B_K$~\cite{Aoki:2005ga},
with the DBW2 rectangle-improved gauge
action~\cite{Takaishi:1996xj,deForcrand:1999bi} with gauge
coupling \(\beta=1.22\). With the 106 gauge configuration in total
used in that study, the lattice cutoff measured from the $\rho$
meson mass is $a^{-1}_{\rho}=2.914(54)$~GeV. This implies a
physical spatial volume for the $24^{3}\times 48$ lattices of
about $({\rm 1.6~fm})^{3}$. We will use this cutoff estimate as a
standard in this report. If we use the static quark potential
instead, we obtain
$a^{-1}_{r_0}=3.07$~GeV~\cite{Aoki:2005ga}. Notice that the
difference between these two determination is about 5\%, which is
smaller than found on coarser lattices.

Of the 106 gauge configurations reported in the earlier study, in
the present work we use only 103, due to the shutdown of the QCDSP
computers in early 2006. This results in slight differences in the
estimations of observables. We confirmed they are not significant.

\begin{figure}
\includegraphics[width=0.8\columnwidth,angle = 0]{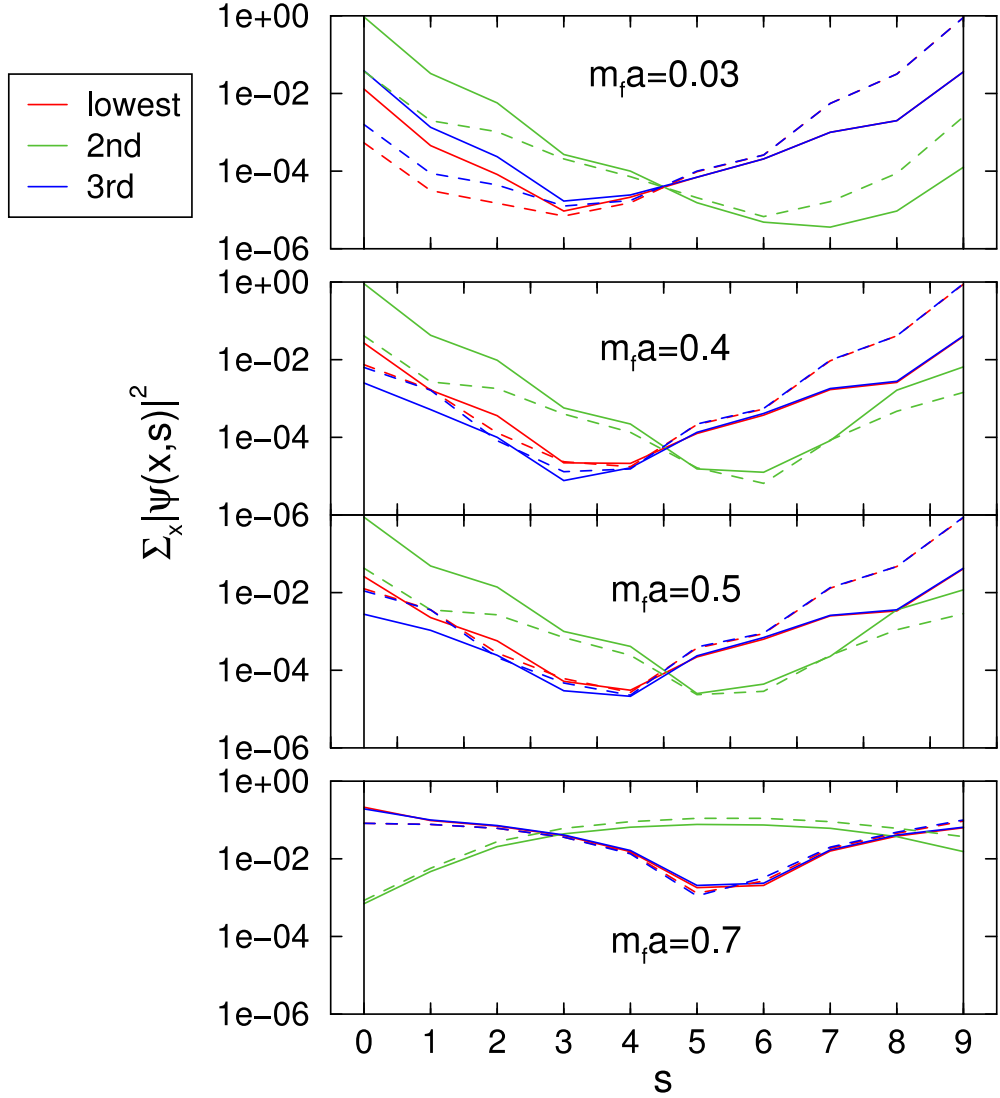}
 \caption{\small\small
Fifth-dimensional $s$-dependence of the three eigenvectors of
$D_H$ with smallest eigenvalues (displayed in right- and
left-projection pairs): $| \Psi_s^2| $ (with $y$-axis in log
scale) with $m_f \in \{0.03, 0.4, 0.5, 0.7\}$. The propagating
states only arise for $am_f > 0.5$. }\label{fig:dwfEigenLog}
\vskip -3mm
\end{figure}

We use domain-wall fermions
(DWF)~\cite{Kaplan:1992bt,Shamir:1993zy,Furman:1994ky} to describe
both heavy (charm) and light (up, down, strange) quarks.  The
number of sites in the fifth dimension, $L_s=10$ and the
domain-wall height, $M_5=1.65$, are taken to be the same as in the
previous work~\cite{Aoki:2005ga}. The simulation parameters and
some numerical results obtained in Ref.~\cite{Aoki:2005ga} which
are relevant to this work are summarized in
Table~\ref{tab:params}. As seen from the Table, our choice of
parameters results in a residual mass of $O(10^{-4})$ in lattice
units, about $0.3$~MeV. To study the quark mass dependence in a
comprehensive way, we use five values each for heavy and light
quark masses as listed in Table~\ref{tab:params}.  They
respectively cover approximate ranges of $[\frac{1}{4} m_{\rm
charm}, \frac{5}{4}m_{\rm charm}]$ and $[\frac{1}{4} m_{\rm
strange}, \frac{5}{4}m_{\rm strange}]$.
Note that our choice of the heavy mass appears reasonable:
Figure~\ref{fig:dwfEigenLog} shows the eigenfunctions $|\Psi_s^2|$
in the fifth dimension for the lowest three eigenvalues with
various masses: 0.03, 0.4, 0.5, 0.7. The largest mass in the
present work for the heavy quark is 0.5, and as shown in the plot,
its lowest three eigenstates drop exponentially away from the
walls and so do not appear to be unphysical.

Table~\ref{tab:meson_op} lists the meson operators, \(\bar{\psi}
\Gamma\psi\), used in this work.  We used source/sink combinations
of wall-wall (denoted as ${\cal C}^{\Gamma_1\Gamma_2}_{\rm ww}$)
and wall-point (denoted as ${\cal C}^{\Gamma_1\Gamma_2}_{\rm wp}$)
types for two- and three-point functions.  The source position is
set at $t_{\rm src}=7$. The gauge fields are fixed to Coulomb
gauge. The quark propagators are obtained by solving under
periodic and anti-periodic boundary conditions in time and
averaging over them.  This procedure gives us a twice larger
period in time.  In the calculations of two-point functions, we
double the statistics by using additional set of correlators with
$t_{\rm src}=41$.

As was mentioned in the introduction, we calculate the masses of
heavy-light (\(D\) and \(D_s\)) and heavy-heavy (charmonium)
mesons with spin-parity \(J^P = 0^\mp\) and \(1^\mp\), decay
constants of the pseudo-scalar mesons, and the
\(D^0\)-\(\overline{D^0}\) mixing parameter. We use the standard
single-elimination jackknife method for statistical error
analysis.  Further details of numerical methods used in
calculating respective quantities are summarized at the beginning
of the sections reporting the results.

\section{Spectroscopy}
\label{SecSpectroscopy}

The meson masses are extracted by fitting the wall-point
two-point correlators to the hyperbolic form
\begin{equation}
{\cal A} \cosh(m(t-L_t)),
 \label{eq:fit form}
\end{equation}
after shifting the wall source time slice of the correlators to
$t=0$.

Using the 103 configurations, we first repeated the measurement of
light meson spectroscopy as in Ref.~\cite{Aoki:2005ga}, and
confirmed that the differences from that reported in
Ref.~\cite{Aoki:2005ga} using full 106 configurations are
negligible. For example and for later use, we present the bare
strange quark mass in lattice units here. We choose to use $m_K$
and $m_\rho$ as inputs to determine the strange quark mass.
Following the same fit ranges and the same functional forms in
chiral extrapolation as in Ref.~\cite{Aoki:2005ga}, we obtain
$am_{\rm strange}=0.0298(13)$, which is consistent with \(am_{\rm
strange}=0.0295(14)\) quoted in Ref.~\cite{Aoki:2005ga}.

\subsection{Charmed Mesons}
\label{SubsecDs}

\begin{figure}
\includegraphics[width=0.95\columnwidth,angle = 0]{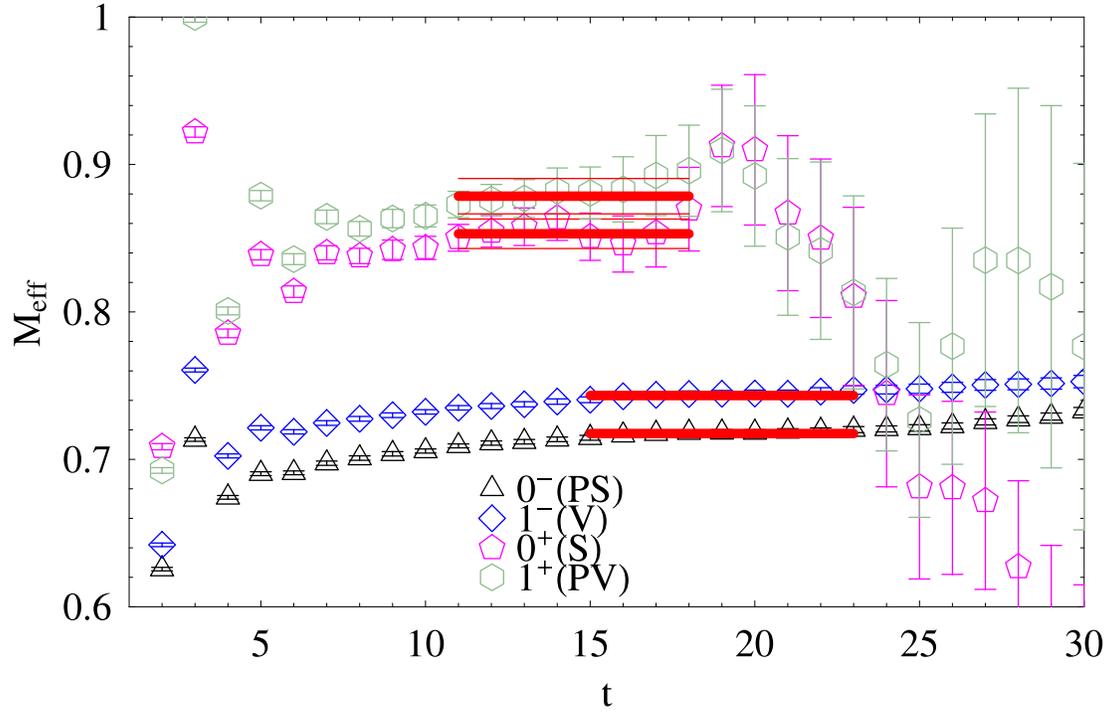}
 \caption{\small\small
Effective mass plots with ``close-to-quark-mass'' simulation
points: $am_{\rm Heavy}=0.4$ and $am_{\rm light}=0.032$.%
}\label{fig:effMass}  \vskip -3mm
\end{figure}

The effective mass plots for the heavy-light mesons with the four
different spin-parity states we are discussing are shown in
Figure~\ref{fig:effMass} for $am_{\rm heavy}=0.4$ and $am_{\rm
light}=0.032$, which are the closest combination to the bare charm
(obtained below) and strange (discussed above) mass, respectively.
As seen from the figure, the plots for all four mesons show
reasonable plateaux with this mass combination. Similar quality of
plateaux are obtained for the other combinations of $am_{\rm
heavy}$= 0.2, 0.3, 0.4 and 0.5 and $am_{\rm light}$=0.024, 0.032
and 0.040 as well. In addition, for the pseudoscalar ($0^-$) and
vector ($1^-$) channels, we could also extract the masses for the
mass combinations of $am_{\rm heavy}=0.1$ and $am_{\rm
light}=0.008$ and 0.016. We take the fit ranges as \(15\le t\le
23\) for \(0^{-}\) and \(1^{-}\) states, and \(10\le t\le 16\) for
\(0^{+}\) and \(1^{+}\). The numerical results for meson masses
are summarized in
Tables~\ref{tab:HLPSmass},~\ref{tab:HLVmass},~\ref{tab:HLSmass}
and~\ref{tab:HLAmass}.

\begin{figure}
\includegraphics[width=0.95\columnwidth,angle = 0]{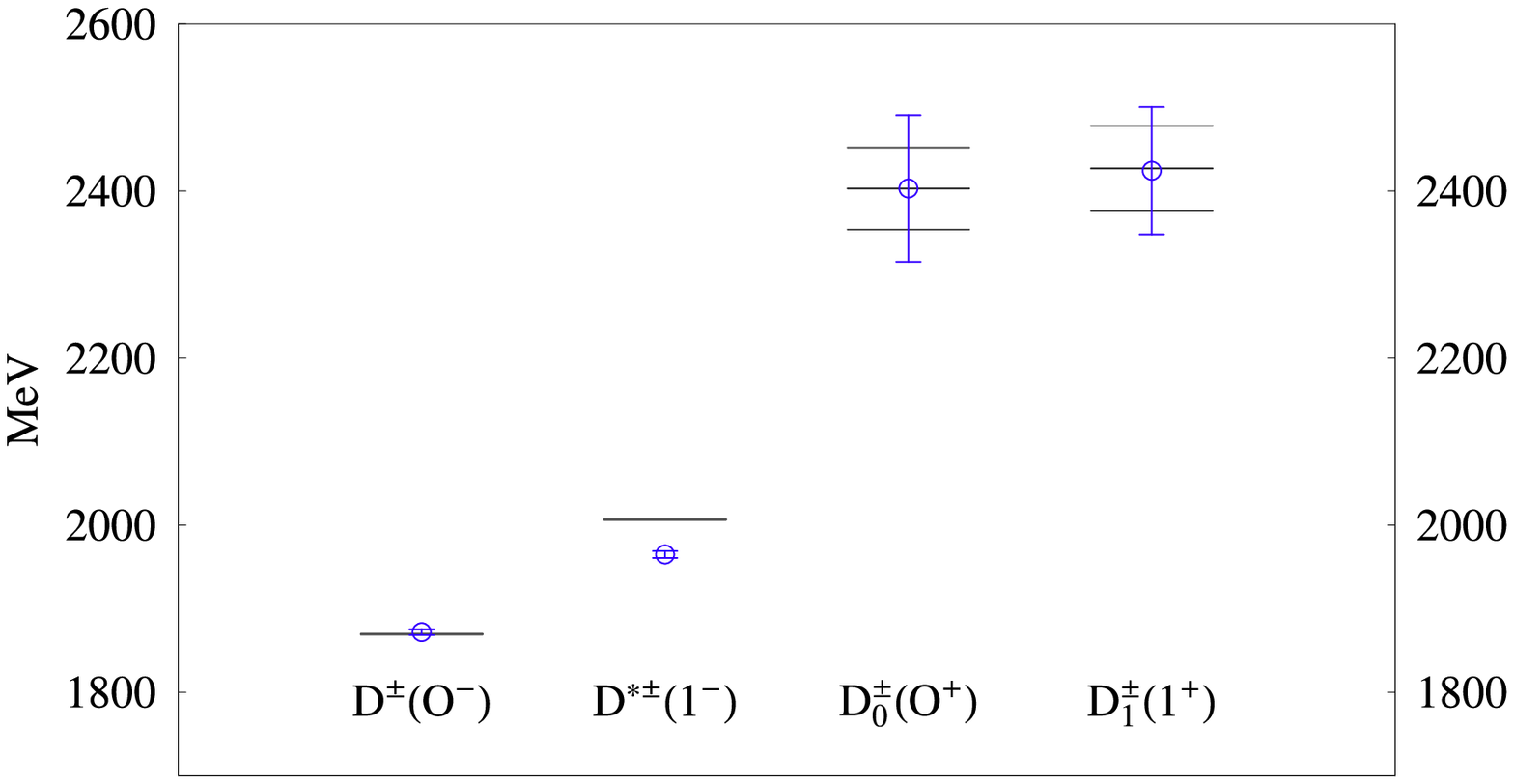}
\includegraphics[width=0.95\columnwidth,angle = 0]{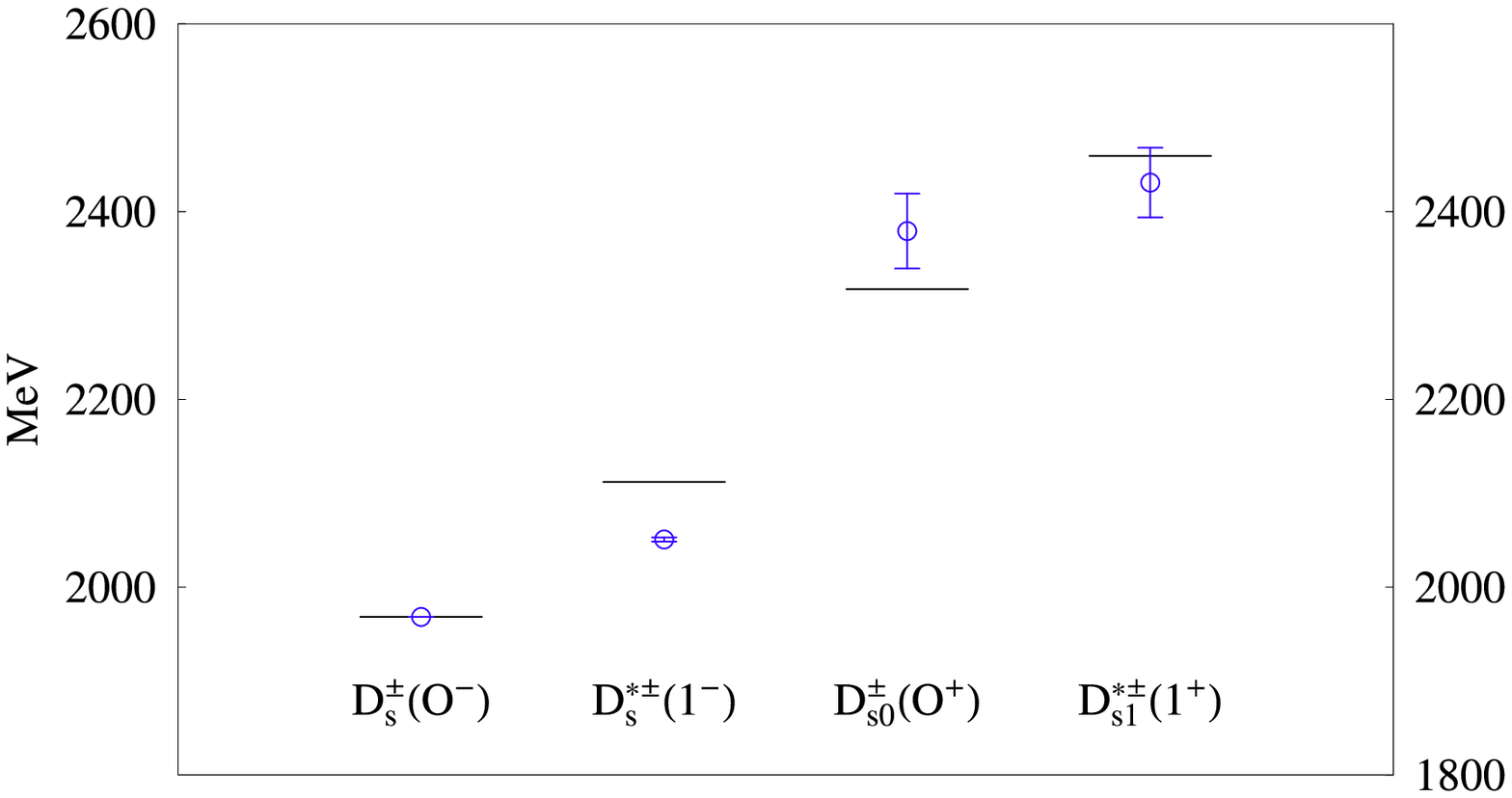}
 \caption{\small\small
The spectrum of the $D_{\rm s}$ (above) and $D$ (below) systems.
The circles are our results with statistical errorbars and the
horizontal lines correspond to experimental data with one $\sigma$
error.} \label{fig:stectrum} \vskip -3mm
\end{figure}

We use the pseudoscalar $D_s$ meson mass (1968.3(5)~MeV), the bare
strange quark mass estimate of $am_{\rm strange}=0.0298(13)$
\cite{Aoki:2005ga}, and \(\rho\)-meson mass $m_\rho=770$~MeV as
inputs to determine the bare charm quark mass. The determination
using quarkonium system will be discussed later. For this purpose,
we first interpolate the heavy-light pseudoscalar masses in light
quark mass to $am_{\rm strange}$, and then interpolate between the
resulting data for $am_{\rm heavy}=0.3$ and 0.4. Through this
procedure we obtain an estimate of $am_{\rm
charm}^{D_s/m_\rho}=0.3583(22)$, where the superscript denotes the
inputs used. Using $am_{\rm charm}^{D_s/m_\rho}$, we estimate the
masses of the four different spin-parity states, resulting in
Figure~\ref{fig:stectrum}, which shows the comparison of our
estimates (circles) with the experimental values (vertical lines),
and in Table~\ref{tab:spectrum}. Not surprisingly, the
calculations are in reasonable agreement with the experiments, to
within a few \%.

A more stringent test is provided by the parity splittings,
$\Delta_{q,J}=m_{D_q(J^+)}-m_{D_q(J^-)}$, rather than the masses
themselves. The numerical results in physical unit via
$a^{-1}_{\rho}$ are shown in Table~\ref{tab:splitting} together
with their experimental values. First we extract some features
from the experiments:
\begin{enumerate}
\item $\Delta_{q,0}\approx\Delta_{q,1}$ for both $q=ud$ and $s$.
\item $\Delta_{ud,J}>\Delta_{s,J}$ for both $J=0$ and 1.
\item The difference, $\Delta_{ud,J}-\Delta_{s,J}$, is close to $m_{D_s(J^-)}-m_{D_{ud}(J^-)}$.
This means that the positive parity state masses depend on the
mass of the light spectator only weakly while the negative parity
states change by $m_{D_s(J^-)}-m_{D_{ud}(J^-)}\approx
m_s-m_{ud}\approx m_s$ between $D_s(J^-)$ and $D_{ud}(J^-)$.
\end{enumerate}

There have been attempts to understand some of these features
using model analyses. As we noted in the Introduction, Bardeen
et~al.~\cite{Bardeen:2003kt} described these charmed heavy-light
mesons using a chiral quark model in a way that respected heavy
quark symmetry, and predicted that
$\Delta_{q,0}\approx\Delta_{q,1}$ with an assumption that
$\Lambda_{\rm QCD}/m_{\rm charm}$ corrections are small. Nowak
et~al.~\cite{Nowak:2003ra} made similar predictions with a
slightly different model. Becirevic et~al.~\cite{Becirevic:2004uv}
noted difficulties in understanding the experimental observation
of $\Delta_{ud,J}>\Delta_{s,J}$ in terms of a version of chiral
perturbation theory extended to include the four spin-parity
states of the heavy-light mesons.

Importantly, many previous lattice calculations have
systematically overestimated the parity splitting by about 50 to
200~MeV for
$\Delta_{s,0}$~\cite{Boyle:1997rk,Hein:2000qu,diPierro:2003iw,Bali:2003jv,Dougall:2003hv,Green:2003zz}.
These works are summarized in Figure~\ref{summary of lattice
calc}, which is the same plot given in Ref.~\cite{Green:2003zz}
but the results from Dougall et~al.~\cite{Dougall:2003hv} is
modified by applying $r_0=0.50$ fm instead of their 0.55 fm.

Now let us turn to our results. Both $\Delta_{s,0}$ and
$\Delta_{s,1}$ are overestimated compared with their experimental
values by 1 -- 1.5 standard deviations, which corresponds to a
difference of 35 -- 60 MeV. Although our results for
$\Delta_{ud,J}$ are not as precise as $\Delta_{s,J}$ since the
extrapolations of the positive parity states to the chiral limit
of the light quark suffer from large statistical uncertainty (as
described below), they show a consistency with the experimental
values. The degeneracy between $\Delta_{q,0}$ and $\Delta_{q,1}$
are well reproduced for both $q=ud$ and $s$ within the statistical
uncertainties. A comparison with the previous lattice calculations
is made in Figure~\ref{summary of lattice calc}. Looking at the
data around $m_c/m_Q=1$, the present result obtained at our
relatively fine lattice cutoff turns out to be consistent with
that of Dougall et~al.\ \cite{Dougall:2003hv}, which is obtained
in the continuum limit.

\begin{figure}
\includegraphics[width=0.95\columnwidth,angle = 0]{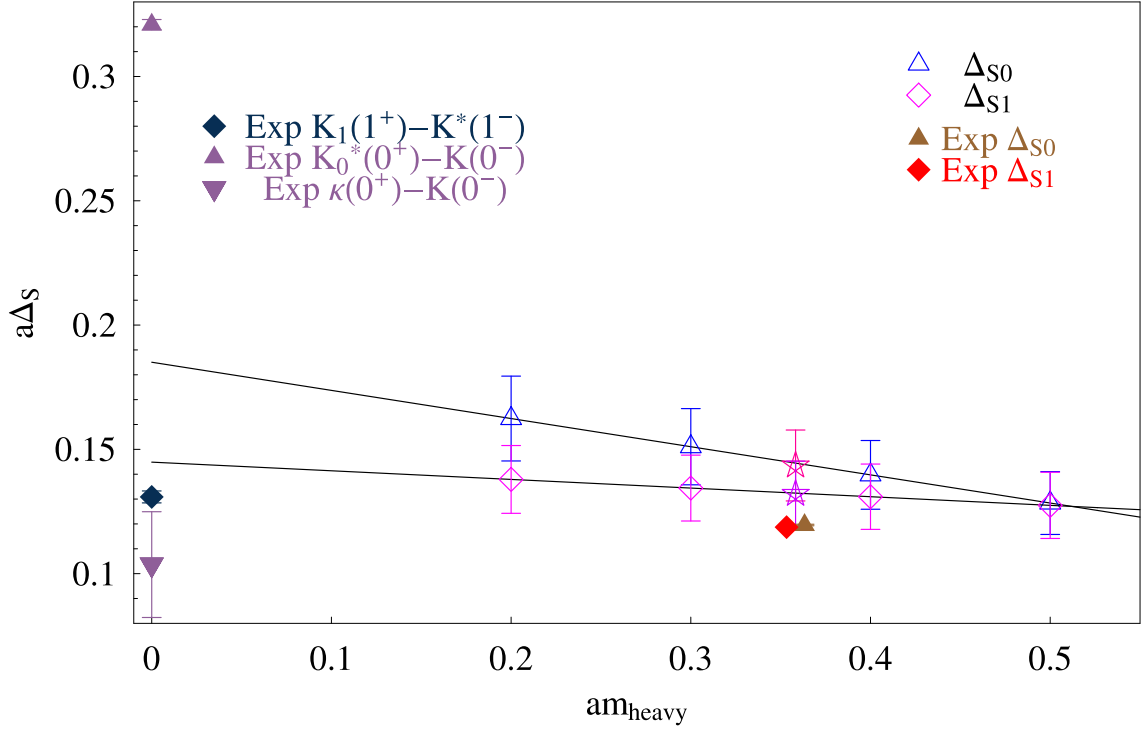}
 \caption{\small\small
The heavy quark mass dependence of parity splitting in lattice
units, where $am_{\rm light}=am_{\rm strange}$.
The stars denote $am_{\rm heavr}=am_{\rm charm}$. %
} \label{deltaH}\vskip -3mm
\end{figure}

The heavy quark mass dependence of $\Delta_{s,1}$ and
$\Delta_{s,0}$ are shown in Figure~\ref{deltaH} together with the
available experimental data. The light quark mass is fixed to the
strange quark mass. It is seen that \(\Delta_{s,0}\) and
\(\Delta_{s,1}\) are statistically indistinguishable for $am_{\rm
heavy}>$ 0.3, but that \(\Delta_{s,0} > \Delta_{s,1}\) becomes
clearer for smaller \(am_{\rm heavy}\) as \(\Delta_{s,0}\)
increases while \(\Delta_{s,1}\) stays more or less constant. The
latter behavior may be supported by the experimental values of the
\(K_{1}(1270)\) and \(K^{*}(892)\) masses.

\begin{figure}
\includegraphics[width=0.95\columnwidth,angle = 0]{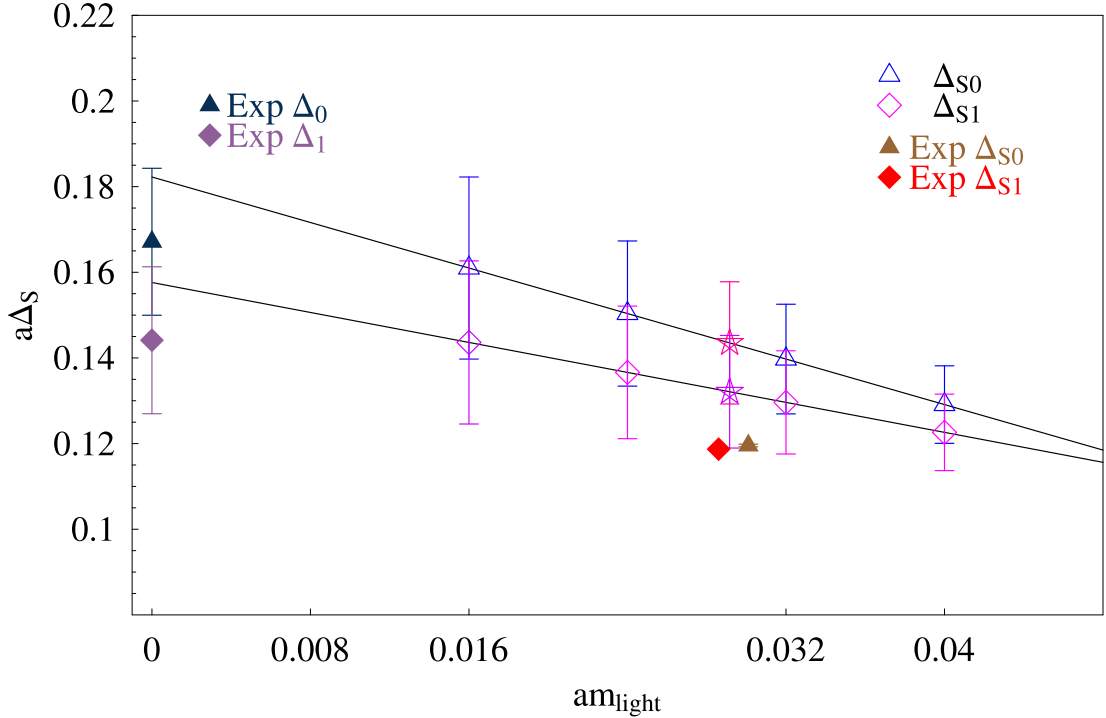}
 \caption{\small\small
The light quark mass dependence of parity splitting in lattice
units where $am_{\rm heavy}=am_{\rm charm}^{D_s/m_\rho}$.
The stars denote $am_{\rm light}=am_{\rm strange}$. %
} \label{deltaL}\vskip -3mm
\end{figure}

The light quark mass dependence of the splittings is shown in
Figure~\ref{deltaL}, where $am_{\rm heavy}$ is fixed to $am_{\rm
charm}^{D_s/m_\rho}$. Both \(\Delta_{J=0}\) and \(\Delta_{J=1}\)
moderately increase toward lighter quark mass, which is consistent
with the experimental observation.

The splitting between the pseudoscalar \(D_s\) and \(D\) mesons,
$m_{D_{ s}}-m_D$, is calculated to be $101(5)$~MeV. The central
value is slightly above but completely consistent with the
experimental value of 99~MeV.

As indicated in Figure~\ref{fig:stectrum}, the estimates of the
hyperfine splittings are significantly smaller than the
experimental value. We find the $1S$ hyperfine splitting
   $m_{D^*}      - m_D        = 93(4)$~MeV and
   $m_{D^*_{ s}} - m_{D_{ s}} = 82(2)$~MeV, while the experimental
results are 142 and 144~MeV respectively. This has been a
long-standing problem in lattice calculations of heavy quark
systems. Take the $D_{s}$
system, for example: UKQCD~\cite{Bowler:2000xw} 
studied the splitting using a nonperturbatively $O(a)$-improved
Wilson fermion action on quenched Wilson gauge action lattices at
$\beta=6.2$, giving $m_{D^*_{ s}} - m_{D_{ s}}=95(6)$~MeV using
$m_{\rho}$ to set the scale. In later studies, A.~Dougall
et~al.\cite{Dougall:2003hv}, using the Sommer scale $r_0 =
0.55$~fm from the $K^*/K$ mass ratio, obtain $m_{D^*_{\rm s}} -
m_{D_{\rm s}} = 97(6)$~MeV after taking the continuum limit. They
also report a result of $96(2)$~MeV from
two-flavor lattices at cutoff $\approx 1.7$~GeV. %
Di~Pierro et~al.\cite{diPierro:2003iw} from 2+1 staggered
dynamical fermion
$20^3\times48$, $a \approx 0.13$~fm (from $r_0=0.5$~fm) 
lattices report the result $m_{D^*_{\rm s}} - m_{D_{\rm s}}
\approx 112(20)$~MeV.

There are many possible reasons for this underestimate in our
calculation.
Besides the quenched approximation, ambiguity in the lattice
spacing, and the light quark mass being extrapolated from rather
heavy values to the physical point, one possible reason is the
absence of the ``clover term''. If one requires accuracy through
$O(a^2m_{\rm charm}\Lambda_{\rm QCD})$ for on-shell quantities, it
turns out that one needs to incorporate the clover term into the
DWF with proper coefficients~\cite{Yamada:2004ri}. Although
naively it is expected that the $O(a^2m_{\rm charm}\Lambda_{\rm
QCD})$ error is fairly small in the present calculation, this
needs to be confirmed in future work.

From the above observations, it is interesting to look at the
ratio,
\begin{equation}
\frac{\Delta_{q,J}}{\Delta_{\rm hyp}},
\end{equation}
to see how well a given heavy quark formalism describes the whole
heavy-light system. Lattice calculations have overestimated the
numerator while underestimating the denominator, thus obtaining
the ratio closer to the experimental value seems to be extremely
difficult for any formalism of lattice heavy quark. In our case,
we obtain about 5 for \(am_{\rm heavy}=0.4\), 4 for 0.3, and 3 for
0.2, and presumably any value of the heavy quark mass would not
reproduce the experimental value of 2.4 for the charm-strange
system. Although we need to take into account the mixing with the
other $0^+$ or $1^+$ states before a reliable conclusion can be
drawn, as $am_{\rm heavy}$ becomes smaller, it is worthwhile to
keep watching this ratio in future calculations.

\subsection{Charmonium}

\begin{figure}
\includegraphics[width=0.95\columnwidth,angle = 0]{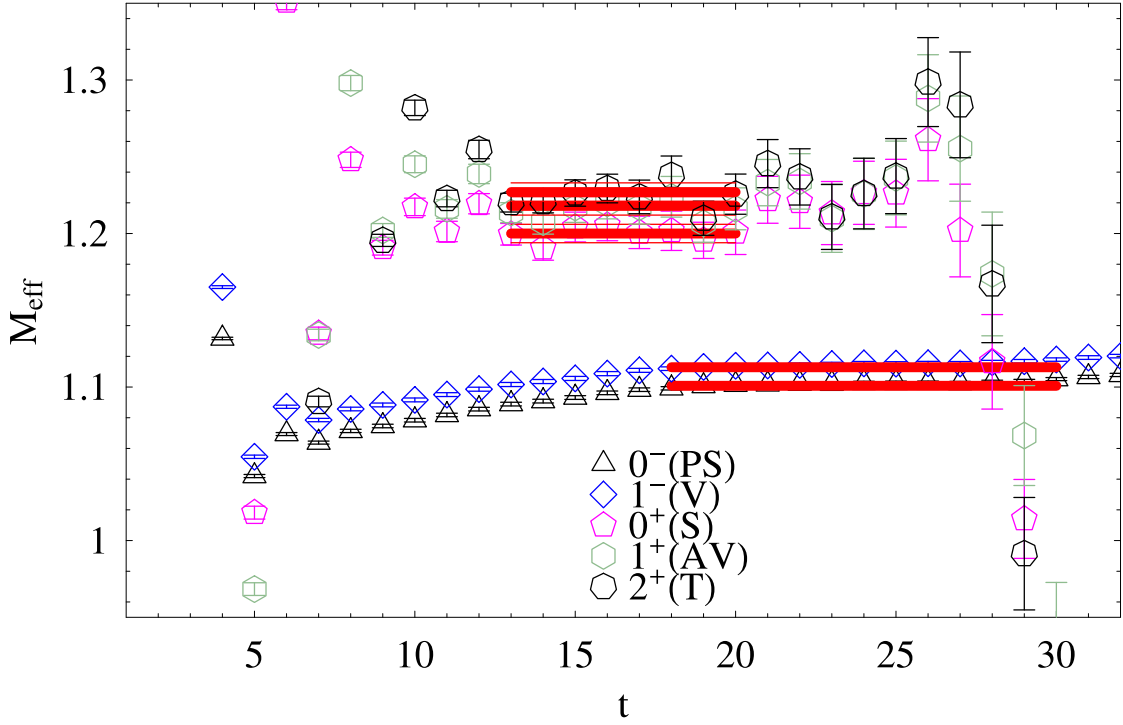}
 \caption{\small\small
Effective mass plots in the heavy-heavy sector at the simulation
point $am_{\rm Heavy}=0.4$%
}\label{fig:HHeffMass}  \vskip -3mm
\end{figure}

The effective mass plots for the four spin-parity channels of the
heavy-heavy meson system with $am_{\rm heavy}=0.4$ are shown in
Figure~\ref{fig:HHeffMass}. Similarly good plateaux are observed
for other heavy quark masses of 0.1, 0.2 and 0.3. We summarize the
meson mass estimates obtained from them in Table~\ref{tab:Row
HHspectrum}.

\begin{figure}
\includegraphics[width=0.95\columnwidth,angle = 0]{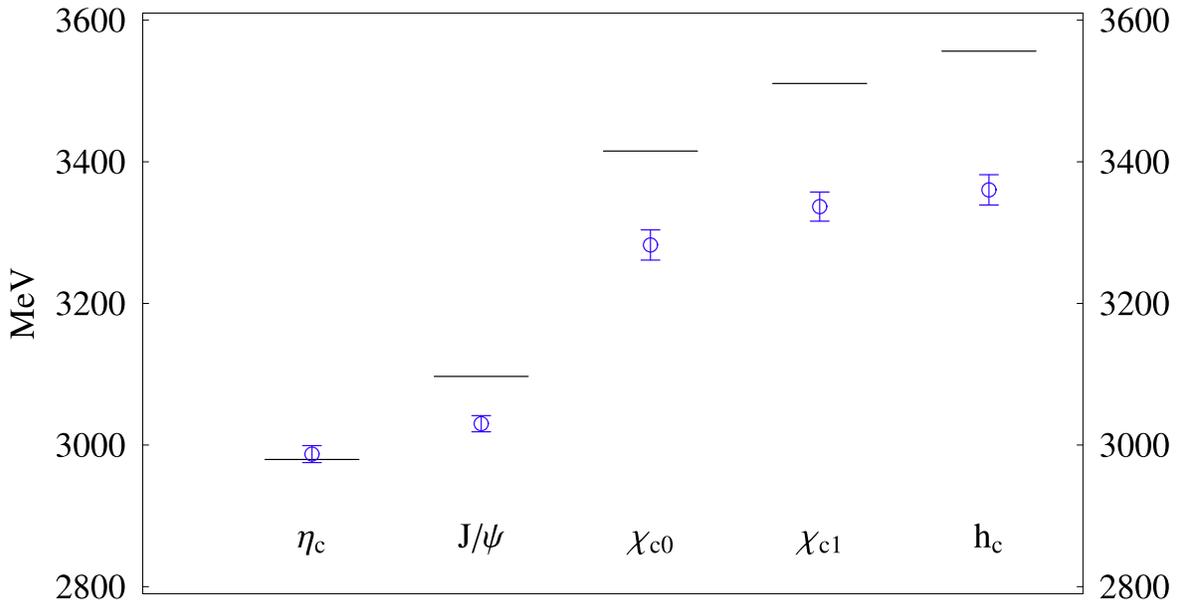}
 \caption{\small\small
The spectrum of the charmonium system. The circles are our results
with statistical errorbars and the horizontal lines correspond to
experimental values.} \label{fig:HHspectrum} \vskip -3mm
\end{figure}

If we interpolate to $am_{\rm charm}^{D_s/m_\rho}=0.3583(22)$ from
subsection \ref{SubsecDs}, we obtain the charmonium mass estimates
shown in Table~\ref{tab:HHspectrum} and
Figure~\ref{fig:HHspectrum}, where the experimental values are
shown together. Notice that with these inputs the mass of  the
$\eta_c$ (\(J^P=0^-\)) state is consistent with the experimental
value. Alternatively we can get an estimate of \(m_{\rm charm} =
0.3561(11)\) for the bare charm quark mass from the \(\eta_c\)
(\(J^P=0^-\)) and \(a_\rho\). This of course is consistent with
the \(am_{\rm charm}^{D_s/m_\rho}\) estimate.

On the other hand the calculated hyperfine splitting of 43(1)~MeV
is significantly smaller than the experimental value of 116~MeV,
just like in the charmed meson cases. One of the reasons is what
we described in the heavy-light case that the lack of  ``clover''
term might cause the smallness of the hyperfine splitting.
Secondly, heavy quarkonia hyperfine splittings are notoriously
difficult to reproduce by lattice calculations. In the quenched
approximation, a detailed study of the relation of hyperfine
splitting and the lattice scale was carried out by QCD-TARO
collaboration~\cite{Choe:2003wx}. Their result is 77(2)(6)~MeV
after continuum extrapolation. In the dynamical three-flavor
staggered case, S.~Gottlieb et.~al.~\cite{Gottlieb:2005me}
reported their hyperfine splitting to be 107(3)~MeV from the
``fine'' MILC lattice configurations ($a\approx0.086$ fm), with
lattice scale obtained from bottomonium system. In the charmonium
system, the problem appears independent of the heavy quark action
adopted and is not solved even using dynamical configurations. For
more details, see the review article in
Ref.~\cite{Brambilla:2004wf}.

The masses of the $P$-wave states, in contrast to the charmed
meson cases where they are overestimated, are several standard
deviations underestimated from the experimental values.

Also we note a result of the mass of a yet-to-be-established
\(C\)-odd axial vector state, \(h_c (J^{PC}=1^{+-})\):
experimentally it has been a long-standing puzzle
\cite{Brambilla:2004wf}. Recently two positive results with a mass
of \(m_{h_{c}} = 3524.4(6)(4)\) MeV
\cite{Rosner:2005ry,Rubin:2005px} and 3526.2(0.15)(0.2) MeV
\cite{Andreotti:2005vu} were reported. A third experiment
\cite{Fang:2006bz}, though, did not confirm this. In the present
lattice calculation the mass difference between \(h_c\) and
\(\chi_{c1}\) is estimated as 22(11) MeV (see Table
\ref{tab:HHspectrum}). This is obtained from the ratio of the
\(h_c\) and \(\chi_{c1}\) correlators in the same fitting range
used for extracting the \(\chi_{c0}\) and \(\chi_{c1}\) masses, so
is likely more reliable than the absolute values of the
excited-state meson mass themselves. Note these masses are
significantly underestimated in the present work. If we add this
difference to the experimentally known \(\chi_{c1}\) mass of
3510.59(10) MeV, we get a value of \(3533(11) _{\rm stat.}\) MeV.
We quote only the purely statistical error here. The \(h_{c}\)
correlator by itself yields \(3361(21)_{\rm stat.}\) MeV, also
with only purely statistical error.

Finally we discuss the charm mass and lattice cutoff estimations
solely by charmonium.  First we can determine the bare charm mass
by looking at some charmonium mass ratio.  For example, if we take
a traditional choice  spin-weighted  mass ratios,
\([(m_{\chi_{c0}}+3m_{\chi_{c1}}) -
(m_{\eta_{c}}+3m_{J/\psi})]/(m_{\eta_{c}}+3m_{J/\psi})=0.13664(6)\),
we obtain an estimation of \(m_{\rm charm}a = 0.235(2)\).   Then
by matching the experimental mass of
\((m_{\eta_{c}}+3m_{J/\psi})/4=3.0676(12)\)~GeV and the
corresponding interpolation of the calculated mass,
0.804(8)~\(a^{-1}\), we obtain a cutoff estimate of \(a^{-1} =
3.82(4)\) GeV.  If we use spin-0 mesons alone we obtain \(m_{\rm
charm}a = 0.225(2)\) and \(a^{-1} = 3.90(3)\) GeV and from spin-1
mesons \(m_{\rm charm}a = 0.239(2)\) and \(a^{-1} = 3.79(4)\) GeV.
This estimate from charmonium is about 30\% larger than the one
from \(\rho\) meson mass \cite{Aoki:2005ga}. The most likely cause
of this is of course the quenched approximation. The rectangular
improvement of the DBW2 action may also be playing some role here.

\subsection{Charm Quark Mass}
Presently available estimates of the charm quark mass are given in
the PDG review~\cite{Eidelman:2004wy}; the charm quark mass is
estimated to be in the range
\begin{equation}
1.15~{\rm GeV}\leq m^{\overline{\rm MS}}_{ c} (m_{ c}) \leq
1.35 ~{\rm GeV}.
\end{equation}
The renormalized mass can be obtained from
\begin{equation}
m_{\rm ren} = Z_m^{\overline{\rm MS}} (m_f + m_{\rm res})
\end{equation}
where $Z_m^{\overline{\rm MS}}= Z^{\rm match} Z_m^{\rm lat}$ is
the mass renormalization factor between lattice and continuum.
$Z_m^{\rm lat}$ can be obtained with lattice perturbation theory
or RI/MOM-scheme non-perturbative renormalization either from
matching the lattice quark propagator to the continuum one or from
the scalar bilinear operator renormalization factor as $1/Z_S$
\cite{Martinelli:1994ty,Dawson:1997ic,Blum:2001sr}. $Z^{\rm
match}$ matches between $\overline{\rm MS}$ in the continuum and
whatever scheme was used on the lattice; the matching between
$\overline{\rm MS}$ and RI/MOM scheme can be obtained from
Ref.~\cite{Chetyrkin:1999pq}. Aoki~et~al.\cite{Aoki:2002iq}
calculate one-loop $Z_m^{\overline{\rm MS}}$ with DWF fermions in
the DBW2 gauge action, giving $Z_m^{\overline{\rm MS}} =
0.989441$; at the scale $\mu = 2$~GeV it gives $m_{
c}=1.0330(64)$~GeV from the \(D_s\) estimate of $am_{\rm
charm}^{D_s/m_\rho}$=0.3583(22). Chetyrkin calculates the
anomalous dimension of the running of the quark mass to
$O(\alpha_s^4)$~\cite{Chetyrkin:1997dh,Chetyrkin:2000yt} in
$\overline{\rm MS}$ scheme:
\begin{equation}
\mu^2\frac{d}{d\mu^2}m^{(n_f)}(\mu)=m^{(n_f)}(\mu)\gamma_m^{(n_f)}
\left(\alpha_s^{(n_f)}\right).
\end{equation}
The renormalization group invariant mass~\cite{Capitani:1998mq}
\begin{equation}
m_{\rm RI}=\lim_{\mu \rightarrow \infty} m (\mu)
\left[\frac{33-2N_f}{6\pi}\alpha_s\right]^{\frac{-12}{33-2N_f}},
\end{equation}
The running charm mass is then given
by\cite{Chetyrkin:1999pq,Vermaseren:1997fq}
\begin{equation}
m_{ c}(\mu^2)= m_{c, {\rm RI}}\left(\alpha_s\right)^{4/11}
\left[1+0.68733\left(\alpha_s\right)+ 1.51211
\left(\alpha_s\right)^2 + 4.05787\left(\alpha_s\right)^3\right];
\end{equation}
using $\alpha_s$ from Eq.~\ref{eq:alphaS}, we find at the scale of
$m_{ c}$, our $m^{\overline{\rm MS}}_{ c}(m_{ c})=1.239(8)$~GeV.
The perturbative renormalization contains a systematic error of
$O(\alpha_s^2) \approx 4\%$, giving us $m^{\overline{\rm MS}}_{
c}(m_{ c})=1.24(1)(4)$~GeV. If we start from the charmonium
estimate of \(m_{c}a = 0.235(2)\), we obtain $m^{\overline{\rm
MS}}_{ c}(m_{ c})=1.07(1)(4)$~GeV. We find that the difference
between these two charm quark mass estimates dominates our
systematic error; after combining this with the other known
systematics, we find the charm quark mass to be 1.24(1)(18).

For the convenience of the reader, some past estimations from
quenched lattice QCD with continuum extrapolation are listed in
Table \ref{tab:charmmass}. Note, UKQCD recently
\cite{Dougall:2005ev} used a two-flavor clover sea to obtain
$m^{\overline{\rm MS}}_{ c}(m_{ c}) = 1.29(7)(13)$~GeV after
continuum extrapolation.

\section{Leptonic Decay Constants }
\label{SecDecayConst}

The leptonic decay constants are obtained in a standard procedure
from two-point correlation functions with heavy-light current
$A_4=\bar{\Psi}_l \gamma_\mu \gamma_5 \Psi_c$:
\begin{equation}
\langle 0\left| A_4 \right|\mbox{\rm PS at rest}\,\rangle
              = if_{\rm PS}\cdot m_{\rm PS}\, .
\label{EqDecCont}
\end{equation}
We calculated both ${\cal C}^{A_4P}_{\rm pw}$ and ${\cal
C}^{PP}_{\rm ww}$ correlators for pseudoscalar mesons. Since these
two correlators give the same pseudoscalar meson mass ($m_{\rm
PS}$), we can extract $m_{\rm PS}$ along with amplitudes from
fitting these two correlators simultaneously. That is, we minimize
the $\chi^2$ given by
\begin{eqnarray}
 \chi^2 &=& \sum_{t=t_{\rm min}}^{t_{\rm max}}
 \left\{
 \left[\frac{{\cal C}^{A_4P}_{\rm pw}(t,0) -
   {\cal A}^{A_4P}_{\rm pw}\sinh (m_{\rm PS}(t-L_t))}
   {\sigma^{A_4P}_{\rm pw}(t)}\right]^2 \right. \nonumber \\
& &\hspace{-0.5 cm} + \left. \left[\frac{{\cal C}^{PP}_{\rm
ww}(t,0) - {\cal A}^{PP}_{\rm ww}\cosh (m_{\rm PS}(t-L_t))}
   {\sigma^{PP}_{\rm ww}(t)}\right]^2 \right\},
\label{EqSimulFit}
\end{eqnarray}
where $\sigma(t)$ is the jackknife error of the correlator at $t$.
The simultaneous fit gives us more stable amplitudes for two-point
correlators than individual fits for the single correlators. The
decay constants can be obtained from
\begin{eqnarray}
f^{\rm lat}_{\rm PS} &=&  \frac{{\cal A}^{A_4P}_{\rm pw}}
{\sqrt{\frac{m_{\rm PS}}{2}V {\cal A}^{PP}_{\rm ww}}},
\label{EqDecayConst}
\end{eqnarray}
where $V$ denotes the spatial volume of the lattice.

\begin{figure}
\includegraphics[width=0.95\columnwidth,angle = 0]{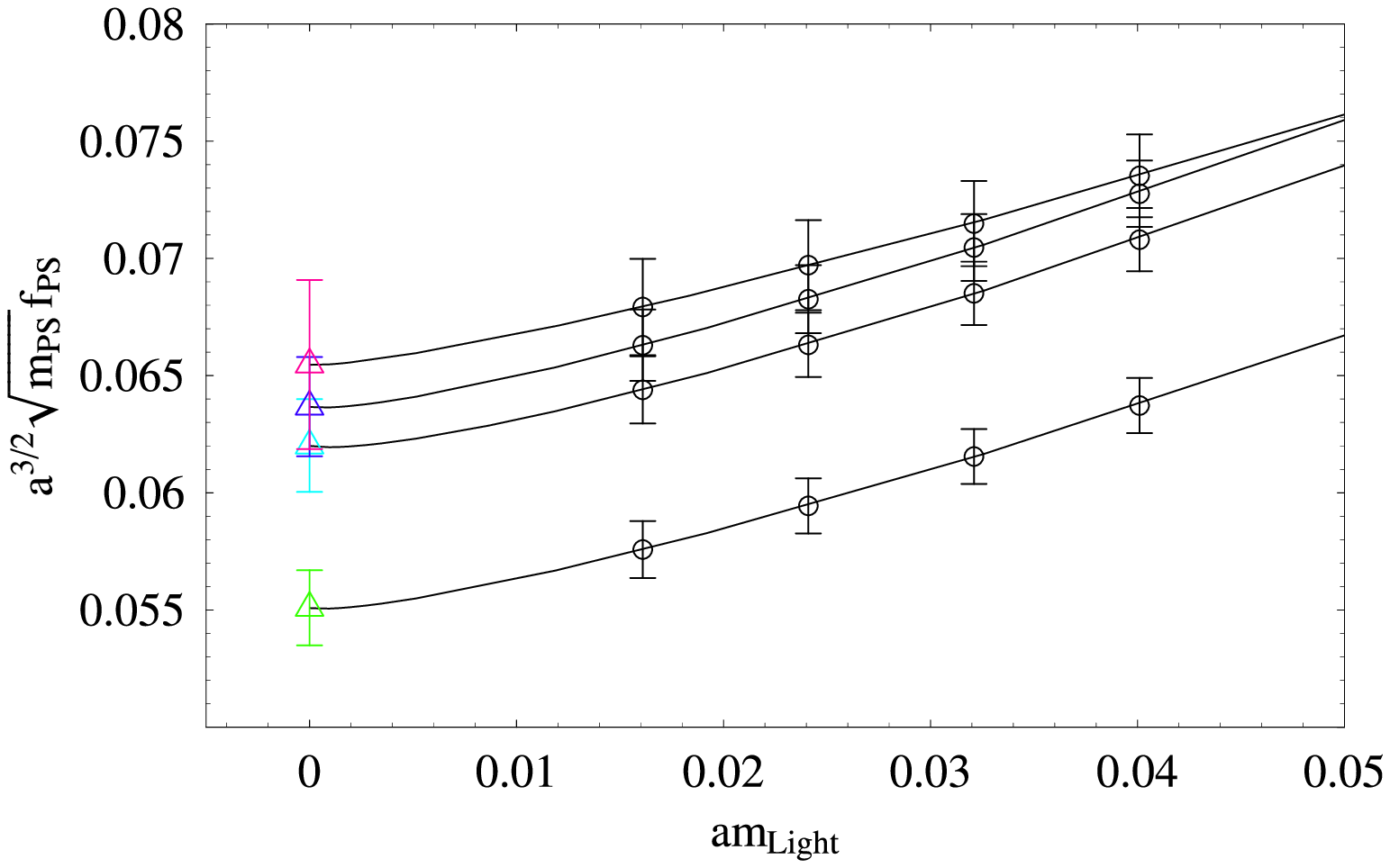}
\includegraphics[width=0.95\columnwidth,angle = 0]{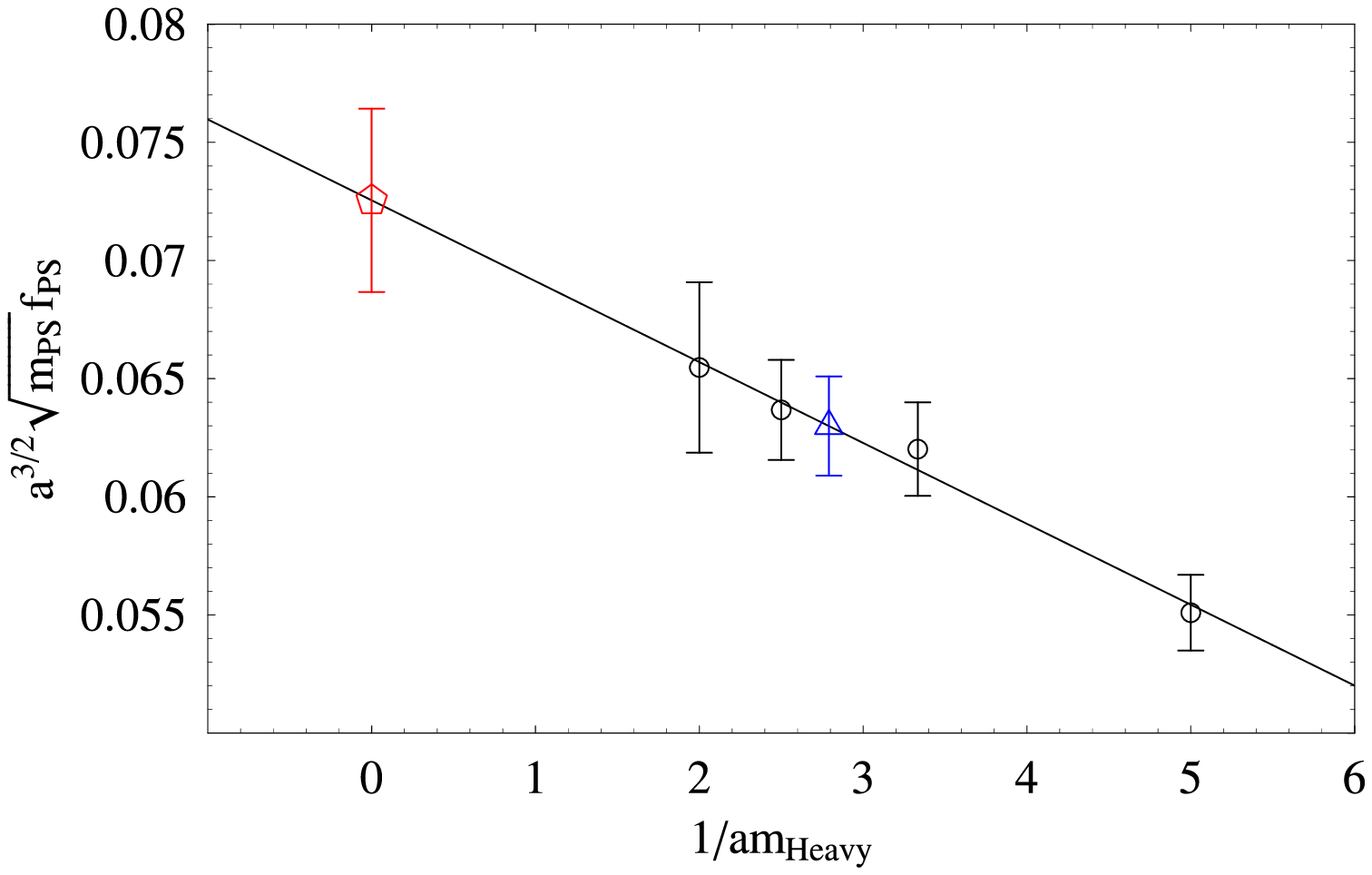}
 \caption{\small\small
The light (above) and heavy (below) quark mass dependence of
$a^{3/2}\sqrt{m_D}f_D$ with the chiral expression: $\sqrt{m_{Qq}}
f_{Qq} =
     F_1 + F_2 m_{q} + \left(F_3 m_{q} \right) \ln {m_{q}}$ and
linear (in $1/m_{\rm Heavy}$) fit respectively. The top figure
displays fixed $am_{\rm Heavy} \in \{0.2, 0.3, 0.4, 0.5\}$ from
top to bottom. In the bottom figure, the triangle (blue) point
represents the physical $D$ meson point, 0.0630(21),
and the pentagon (red) point represents the static quark limit point of 0.0725(39). %
 }\label{fig:sMDfDX} \vskip -3mm
\end{figure}

\begin{figure}[hbt]
\includegraphics[width=0.95\columnwidth,angle = 0]{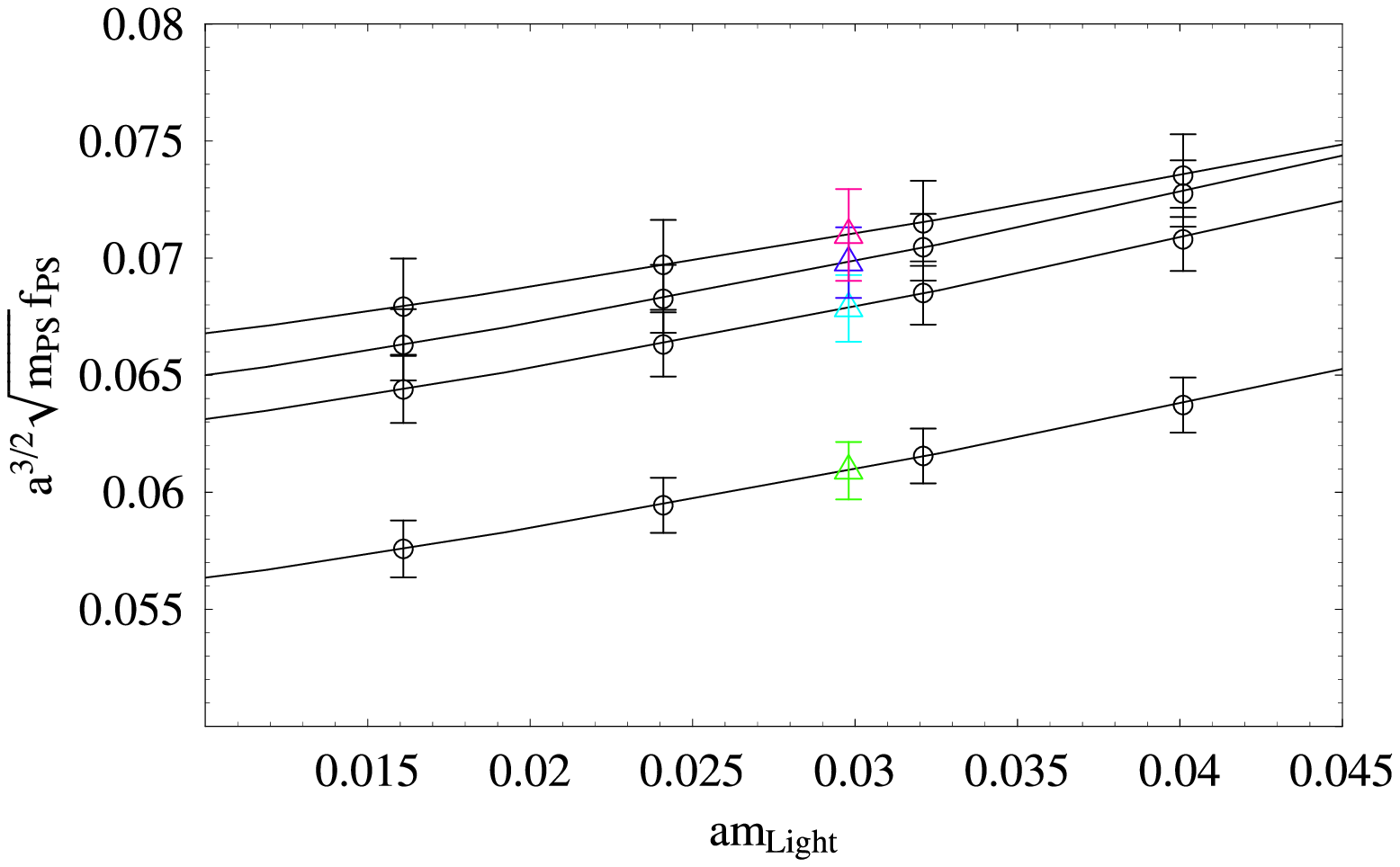}
\includegraphics[width=0.95\columnwidth,angle = 0]{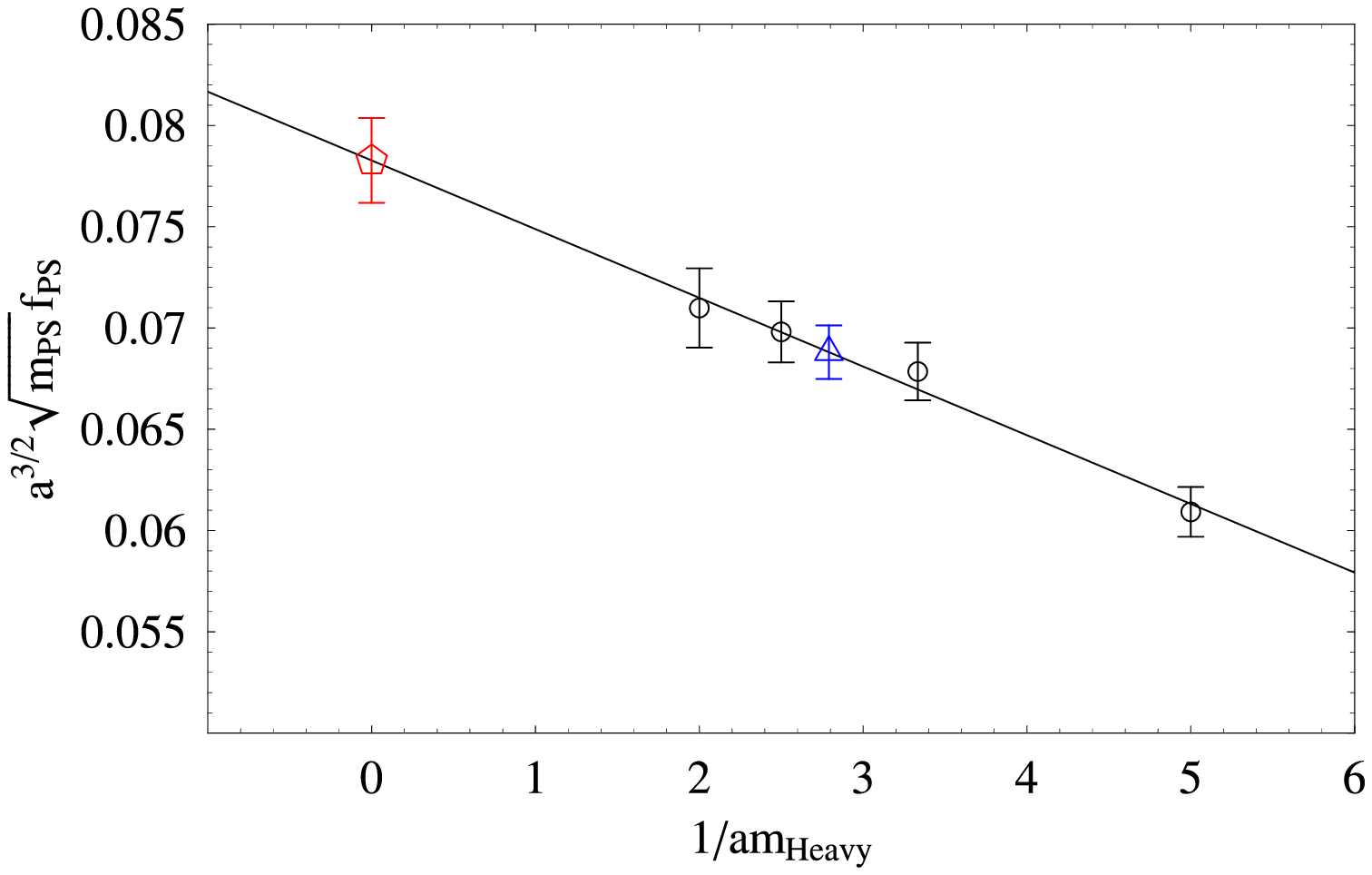}
 \caption{\small\small
The light (above) and heavy (below) quark mass dependence of
$a^{3/2}\sqrt{m_{D_{\rm s}}}f_{D_{\rm s}}$ with the chiral
expression: $\sqrt{m_{Qq}} f_{Qq} =
     F_1 + F_2 m_{q} + \left(F_3 m_{q} \right) \ln {m_{q}}$
     and linear (in $1/m_{\rm Heavy}$) fit respectively.
In the bottom figure, the triangle (blue) point represents the
physical ${D_{\rm s}}$ meson point, 0.0688(13), and the pentagon
(red) point represents the static quark limit point of 0.0783(21).
}\label{fig:sMDsfDsX} \vskip -3mm
\end{figure}

Booth~\cite{Booth:1994hx} and Sharpe and Zhang~\cite{Sharpe:1995qp}
calculate the chiral behavior of the heavy-light decay constants
in the quenched approximation to be:
\begin{equation}\label{eq:Mpsfps}
\sqrt{m_{Qq}} f_{Qq} =
     F_1 + F_2 m_{q} +  (F_3 m_{q} +F_4)\ln {m_{q}},
\end{equation}
after replacement of $m_{qq}^2 \propto m_q$. McNeile and
Michael~\cite{McNeile:2004wn} reported that the effect of the
chiral log on $f_B$ with static-light is small. This indicates
that we may be able to drop the most divergent term $\ln {m_{q}}$.
To be concrete, we enumerate below various fitting functions that
we will use for the purpose of extrapolating/interpolating the
light quark mass dependence:
\begin{equation}\label{eq:MpsfpsChi}
\sqrt{m_{Qq}} f_{Qq} =
     F_1 + F_2 m_{q} +  F_3 m_{q} \ln {m_{q}},
\end{equation}
\begin{equation}\label{eq:MpsfpsP2}
\sqrt{m_{Qq}} f_{Qq} =
     F_1 + F_2 m_{q} + F_3 m_{q}^2,
\end{equation}
\begin{equation}\label{eq:MpsfpsLin}
\sqrt{m_{Qq}} f_{Qq} =
     F_1 + F_2 m_{q},
\end{equation}
at fixed $am_{\rm heavy}.$ Then we examine the $1/am_{\rm heavy}$
behavior and interpolate to $am_{\rm charm}$. The upper half of
Figure~\ref{fig:sMDfDX} shows the light quark mass dependence at
fixed $am_{\rm heavy}$ of the quantity $\sqrt{m_{Qq}}f_{Qq}$
extrapolated to $-m_{\rm res}$ according to Eq.~\ref{eq:MpsfpsChi}
and Figure~\ref{fig:sMDsfDsX} shows the light quark mass
dependence of the quantity $\sqrt{m_{Qq}}f_{Qq}$ interpolated to
$m_{\rm strange}$. Then we further linearly interpolate (in $1/M$)
the heavy quark mass dependence to $am_{\rm heavy}=am_{\rm
charm}$, and summarize the results in Table~\ref{tab:decayFit}.
The results from various expressions used for light quark
extrapolation/interpolation are consistent among the fits. Here we
take the fit from the quadratic expression, since this falls
roughly in the middle of the range from the other expressions, and
incorporate the difference from the other fits into the systematic
error. Thus, we obtain
 \begin{equation}
   f_{D}^{\rm lat}     = 225(7)(^{+6}_{-0}) {\rm MeV},
 \end{equation}
 \begin{equation}
   f_{D_{{ s}}}^{\rm lat} = 243(4)(^{+0}_{-3}) {\rm MeV},
 \end{equation}
before renormalization.

In order to compare our lattice calculations with continuum
results, we need to properly renormalize our lattice decay
constants by a factor of:
\begin{eqnarray}
Z_A^{hl} & = & Z_A^{ll}\times \sqrt{\frac{Z_{q, {\rm DWF}}(am_{\rm
heavy})}{Z_{q, {\rm DWF}}(am_{\rm light})}}.
\end{eqnarray}
The nonperturbative light-light current renormalization in the
chiral limit, $m_f=-m_{\rm res}$, is calculated in
Ref.~\cite{Aoki:2005ga}, as $0.88813(19)$. Here we employ a quark
mass-dependent renormalization factor
~\cite{Yamada:2004ri}:
\begin{equation}\label{eq:wavefunc}
      {Z_{q,{\rm DWF}}^{\rm lat}}^{(0)}(am_f,\omega) =  \frac{am_f \left(1+(am_f)^2\right)
\cosh(m^{(0)}_p) -2(am_f)^2 \omega {\rm sinh}(m^{(0)}_p)}
           {\left(1-(am_f)^2 \right) {\rm sinh}(m^{(0)}_p)},
\end{equation}
in terms of the tree-level on-shell pole mass
\begin{equation}\label{eq:Polemass}
m^{(0)}_p = \ln\left|\frac{-am_f \omega^2
 +\sqrt{(1+am_f^2)^2+am_f^2 \omega^2(\omega^2-4)}}{1+am_f^2-2am_f
 \omega}\right|,
\end{equation}
with $\omega = 2-M_5$, and ${Z_{q,{\rm DWF}}^{\rm lat}}$ defined
as
\begin{equation}
q_R = \sqrt{Z_{q,{\rm DWF}}^{\rm lat}} q_{\rm lat}.
\end{equation}
Therefore, the heavy-light current renormalization is $1.0492(22)$
with $m_{\rm charm} =0.3583(22)$; thus $f_D=232(7)(^{+6}_{-0}) $
and $f_{D_{\rm s}}=254(4)(^{+0}_{-3}) $~MeV. Recent experimental
measurements for these values are $f_{D^+}=
222.6\pm16.7^{+2.8}_{-3.4}$~MeV\cite{Artuso:2005ym} and $f_{D_{
s}} = 267 \pm 33$~MeV~\cite{Eidelman:2004wy} and the recent
lattice three-flavor dynamical results of MILC and
Fermilab~\cite{Aubin:2005ar} are: $f_{D^+} = 201 \pm 3 \pm
17$~MeV, and $f_{D_{ s}} = 249 \pm 3 \pm 16$~MeV.

\begin{figure}[hbt]
\includegraphics[width=0.95\columnwidth,angle = 0]{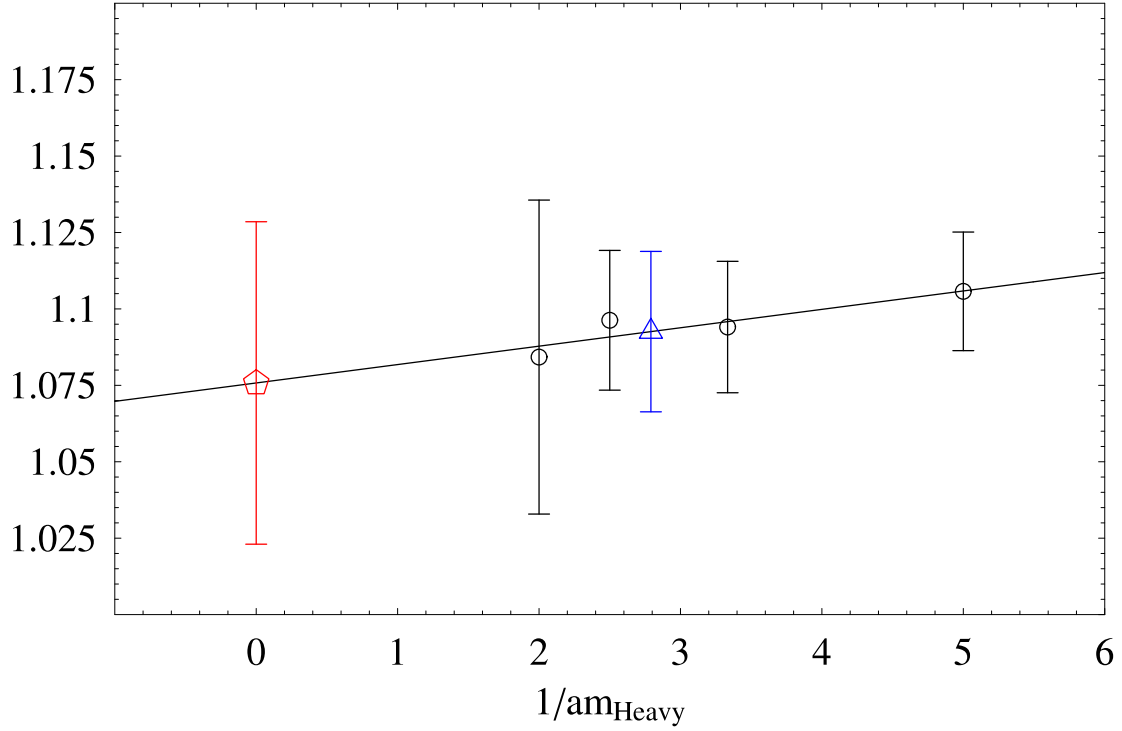}
 \caption{\small\small
The heavy quark mass dependence of the $SU(3)$ breaking ratio
$\frac{\sqrt{m_{D_{\rm s}}}f_{D_{\rm s}}}{\sqrt{m_{D}}f_{D}}$ with
chiral interpolation/extrapolation (Eq.~\ref{eq:MpsfpsChi}) in
light quark mass (black points). The triangle (blue) point
represents the physical charm quark point, 1.092(26),
and the pentagon (red) point represents the static quark limit point of 1.076(53). 
} \label{fig:sMDsfDsRatioX} \vskip -3mm
\end{figure}

To calculate the ratio of the decay constants, we first use the
chiral form to obtain
\begin{equation}
f_{D_{{ s}}}\sqrt{m_{D_{ s}}}/f_{D}\sqrt{m_{D}} =
1.11(2)(^{+0}_{-2});
\end{equation}
Figure~\ref{fig:sMDsfDsRatioX} shows an example of our heavy quark
mass interpolations of this ratio, with chiral
interpolation/extrapolation on the light quark mass according to
Eq.~\ref{eq:MpsfpsChi}. We may then retrieve the ratio of decay
constants for the \DDs systems by multiplying in appropriate
factors of mass:
\begin{equation}
f_{D_{{ s}}}/f_{D} = 1.05(2)(^{+0}_{-2}).
\end{equation}
The experimental value suggesting the above ratio is $1.19(25)$
and the MILC number is 1.24(22). Our result has a smaller central
value but agrees with the experimental and dynamical result within
one $\sigma$.

\section{$D$-$\overline{D}$ mixing}
\label{SecBD}%
In the Standard Model, $D^0$-$\overline{D^0}$ mixing is strongly
suppressed by CKM and GIM factors; if such a mixing is observed,
it might be evidence for physics beyond the Standard
Model~\cite{Burdman:2003rs}. The $D$-mixing is the only probe for
the dynamics for $c$-quark. The effective Hamiltonian of the QCD
contribution to $\Delta C =2 $ in the Standard Model comes from
the box diagram
\begin{equation}\label{eq:EffH}
{\mathcal H}_{\rm eff}^{\Delta C=2} =\frac{G_F^2}{4 \pi^2} |
V_{cs}^*V_{cd}|^2\frac{(m_s^2-m_d^2)^2}{m_c^2}({\cal O} + 2{\cal
O^{\prime}})
\end{equation}
where ${\cal O}=\bar{c}\gamma^{\mu}(1-\gamma_{5})u
\bar{c}\gamma_{\mu}(1-\gamma_{5})u$ and ${\cal
O^{\prime}}=\bar{c}(1+\gamma_{5})u \bar{c}(1+\gamma_{5})u$ %
results from non-negligible external momentum. The $b$-quark
contribution is proportional to $V_{ ub} m_{ b}$ highly suppressed
by the very small CKM matrix factor $V_{ ub}$, which is ignored
here. The BaBar $B$-factory studies $D-\overline{D^0}$ mixing from
the semileptonic decay modes of $D^{*+}\rightarrow \pi^{+}D^{0} $
and $D^{0}\rightarrow Ke\nu$~\cite{Burdman:2003rs}. In this work,
we focus on the bag parameter of $D$ meson, often defined for
historical reasons as
\begin{equation}\label{eq:Bag}
B_D = \frac{\langle \overline{D^0}|{\cal
O}|D^0\rangle}{\frac{8}{3}m_D^2f_D^2},
\end{equation}
where ${\cal O}=\bar{c}\gamma^{\mu}(1-\gamma_{5})u
\bar{c}\gamma_{\mu}(1-\gamma_{5})u$ and  $B_D$ is 1 in the
vacuum-insertion approximation.

The parity-even operator for $B_D$ in the continuum limit is
\begin{equation}
{\cal O}_{VV+AA} = (\bar{c}\gamma_\mu u )(\bar{c}\gamma_\mu u) +
(\bar{c}\gamma_5\gamma_\mu u)(\bar{c}\gamma_5\gamma_\mu u).
\label{EqOvv+aa}
\end{equation}
However, since DWF does not have exact chiral symmetry, ${\cal
O}_{VV+ AA}$ mixes with four other operators
\begin{eqnarray}
{\cal O}_{VV-AA}&=& (\bar{c}\gamma_\mu u)(\bar{c}\gamma_\mu u)
- (\bar{c}\gamma_5\gamma_\mu u)(\bar{c}\gamma_5\gamma_\mu u),\label{EqOvv-aa}\\
{\cal O}_{SS\pm PP}&=& (\bar{c} u)(\bar{c}u)
\pm (\bar{c}\gamma_5 u)(\bar{c}\gamma_5 u),\label{EqOsspp}\\
{\cal O}_{TT}&=& (\bar{c}\sigma_{\mu\nu} u)(\bar{c}\sigma_{\mu\nu}
u). \label{EqOtt}
\end{eqnarray}
The mixing coefficients for these operators do not go to zero as
$m_f$ approaches the chiral limit, but are highly suppressed by
$O((a\mres)^2)$, at least in the case for $B_K$ with DWF. This has
been theoretically and numerically demonstrated in
Ref.~\cite{Aoki:2005ga}, and one can simply ignore the
contribution from the above four operators. However, in the finite
quark mass region, the correct way of solving this problem is to
measure the matrix elements of all the operators in Eqns.\
\ref{EqOvv-aa}, \ref{EqOsspp} and \ref{EqOtt} directly on the
lattice. These operators are relevant to beyond the Standard Model
sources of mixing in \(\Delta F=2\) processes \cite{Aoki:2005ga}.
Since we did not do this measurement while the data were taken, we
quote the maximal uncertainly from the mixing coefficients, which
are expected to be $O((am_{f})^2)$, due to the soft chiral
symmetry breaking from the non-small $m_f$ term.

\begin{figure}[hbt]
\includegraphics[width=0.55\columnwidth,angle = 0]{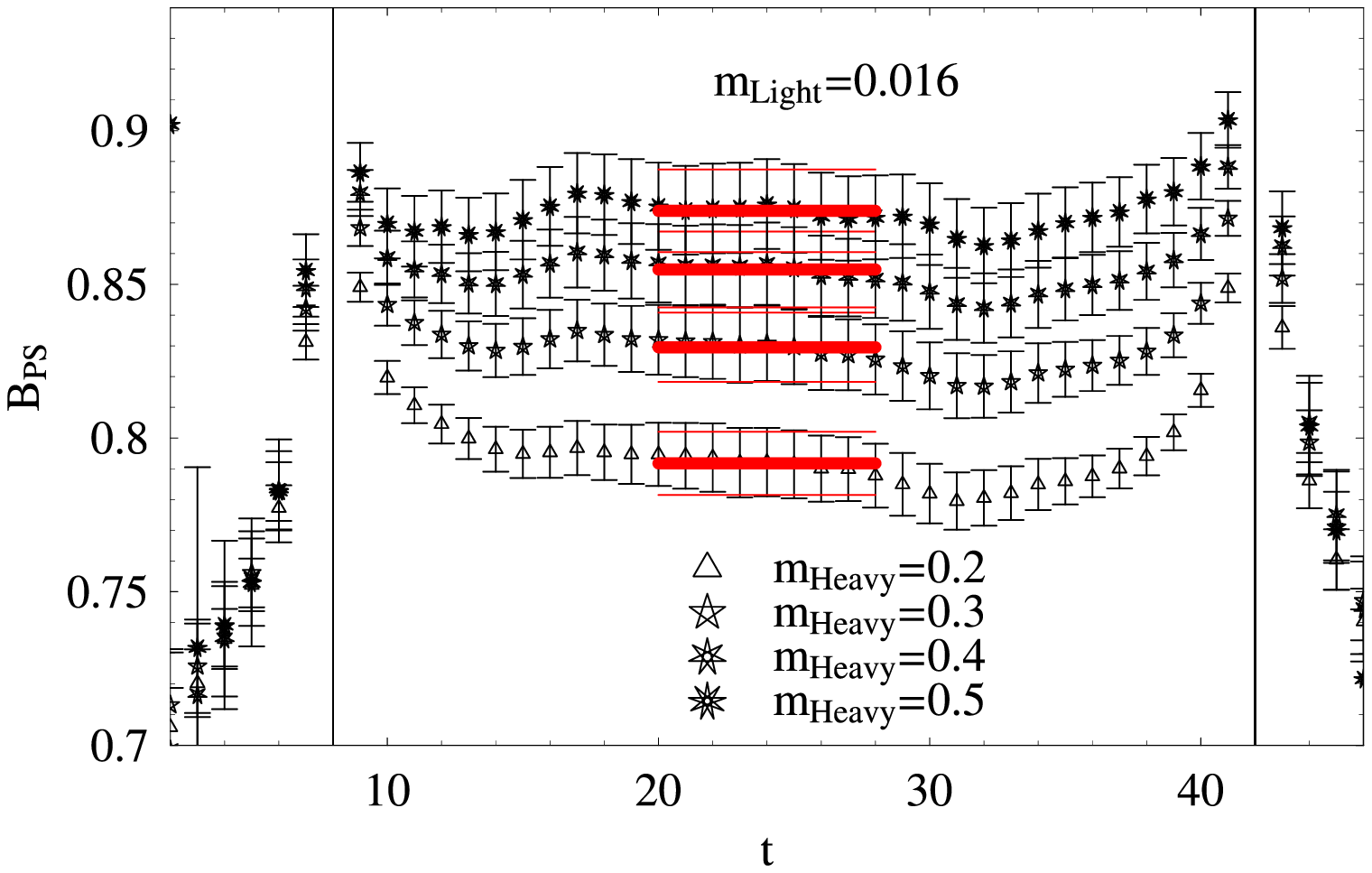}
\includegraphics[width=0.55\columnwidth,angle = 0]{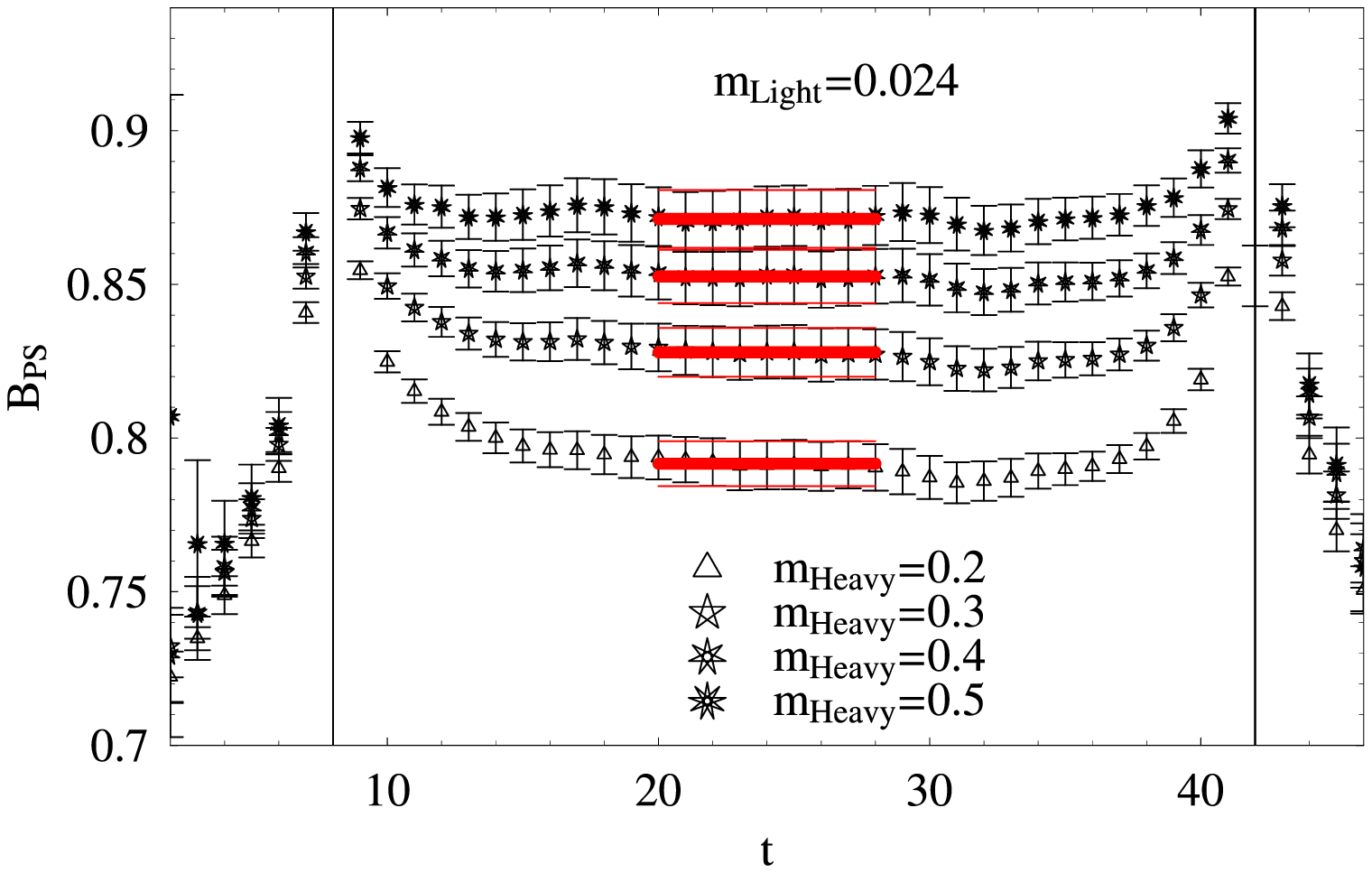}
\includegraphics[width=0.55\columnwidth,angle = 0]{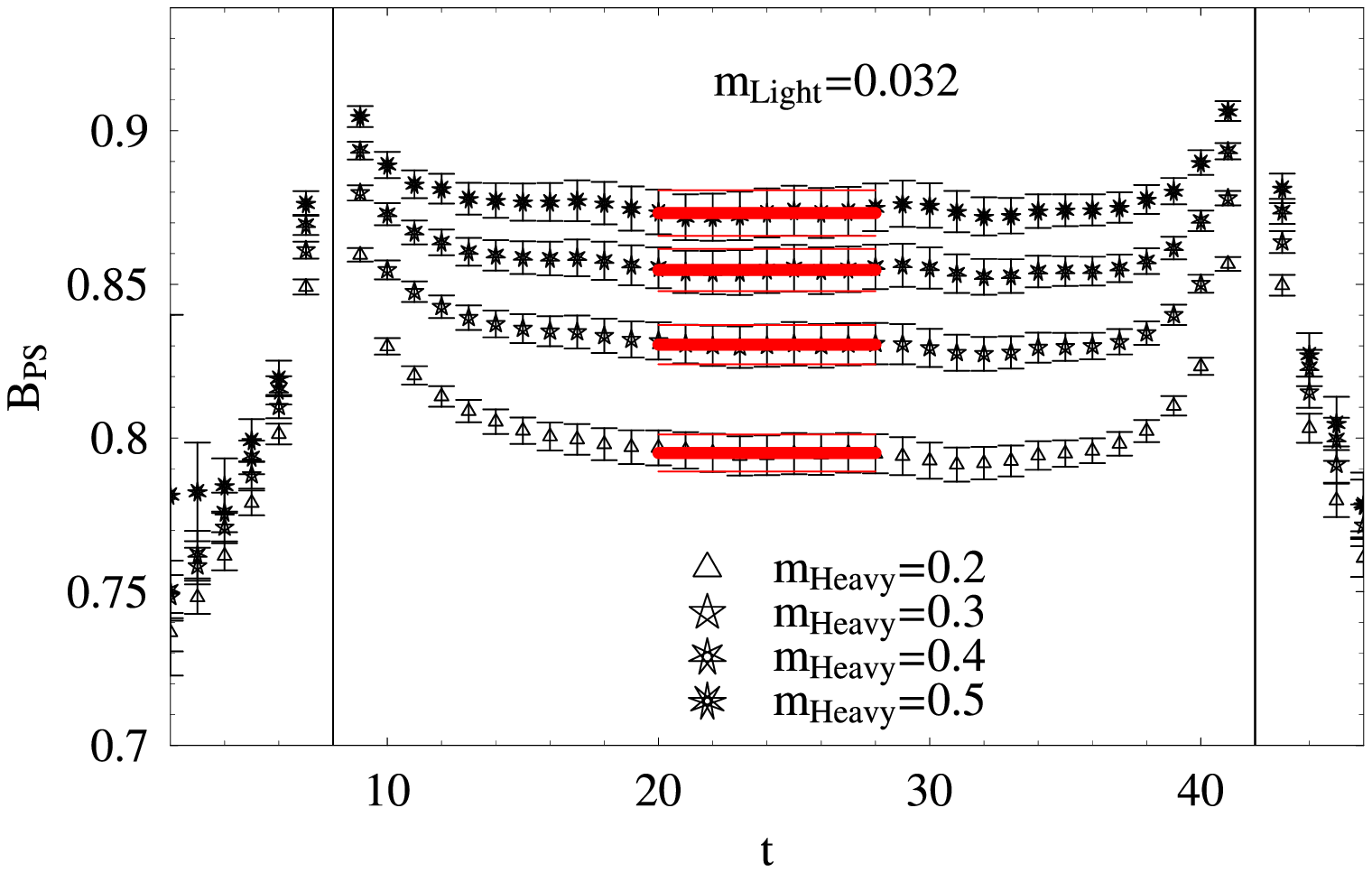}
\includegraphics[width=0.55\columnwidth,angle = 0]{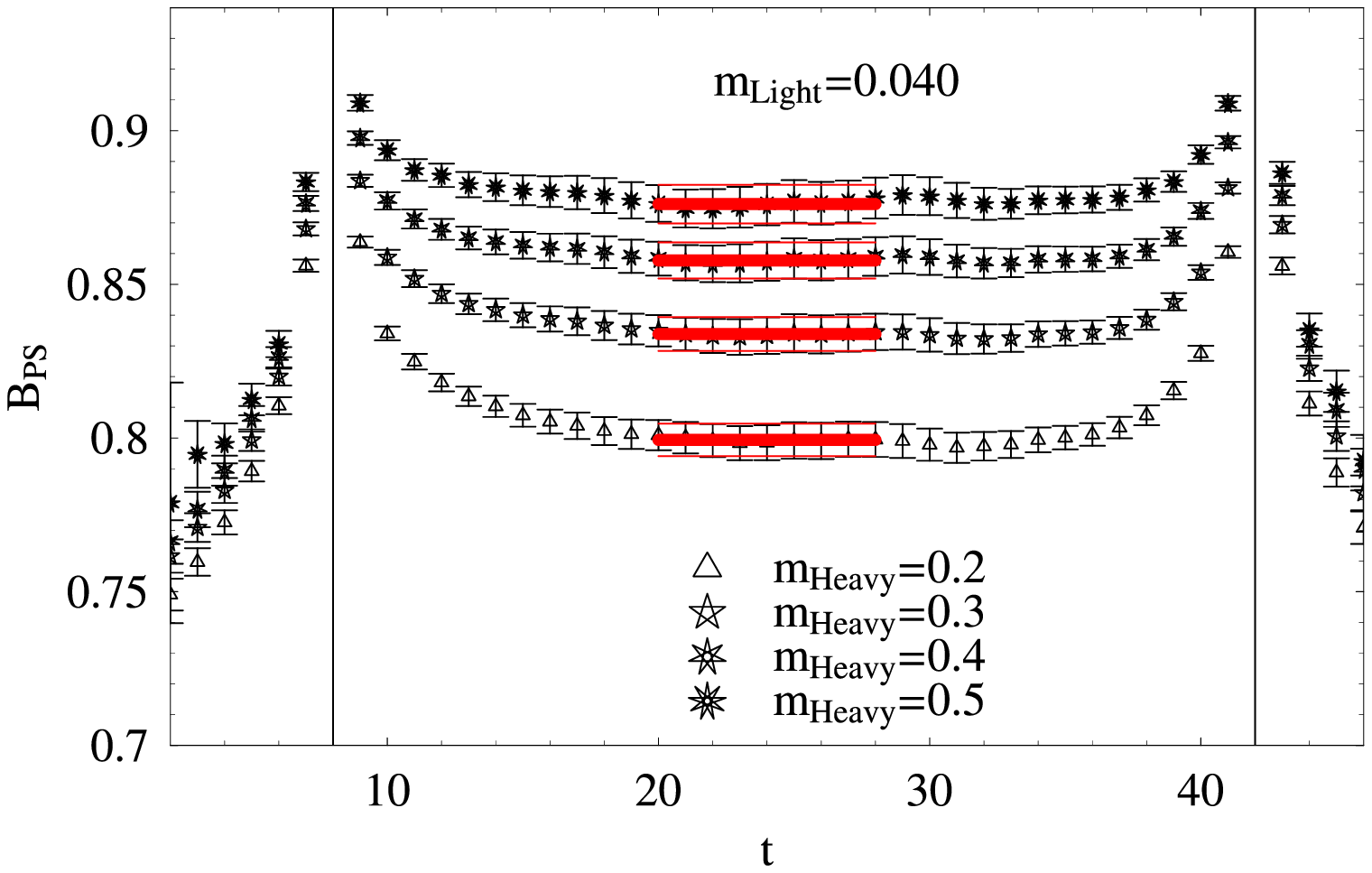}
\vskip -10mm \caption{\small\small The time dependence of
pseudoscalar meson bag parameters at fixed light quark mass:
0.016, 0.024, 0.032, 0.040 from top to bottom. In each subgraph,
we display the
heavy quark mass dependence of $B_{\rm PS}$. %
} \label{fig:Bps}
\end{figure}

On the lattice we can rewrite Eq.~\ref{eq:Bag} directly in terms
of the correlation functions obtained on the lattice:
\begin{eqnarray}\label{EqBps}
B_{\rm PS}^{(\rm lat)} &= &\frac{\left\langle 0\left|{\chi^\dagger
(t_{\rm snk}) {\cal O}(t)\chi^\dagger (t_{\rm
src})}\right|0\right\rangle }{ {\frac{8}{3}{\cal C}^{A_4P}_{\rm
pw}(t,t_{\rm snk}){\cal C}^{A_4P}_{\rm pw}(t,t_{\rm src})}}
    \Biggr|_{t_{\rm src}\ll t \ll t_{\rm snk}},
\end{eqnarray}
where $t_{\rm src}$ and $t_{\rm snk}$ are the source and the sink
location of the quarks, and the $\chi(t)$ are quark bilinear
interpolating meson fields. The results can be found in
Figure~\ref{fig:Bps} with various $m_{\rm heavy}$ and $m_{\rm
light}$. The plateau looks pretty good for extracting the value of
$B_{\rm PS}^{\rm lat}$, and details are listed in
Table~\ref{tab:Bps}.

\begin{figure}[hbt]
\includegraphics[width=0.95\columnwidth,angle = 0]{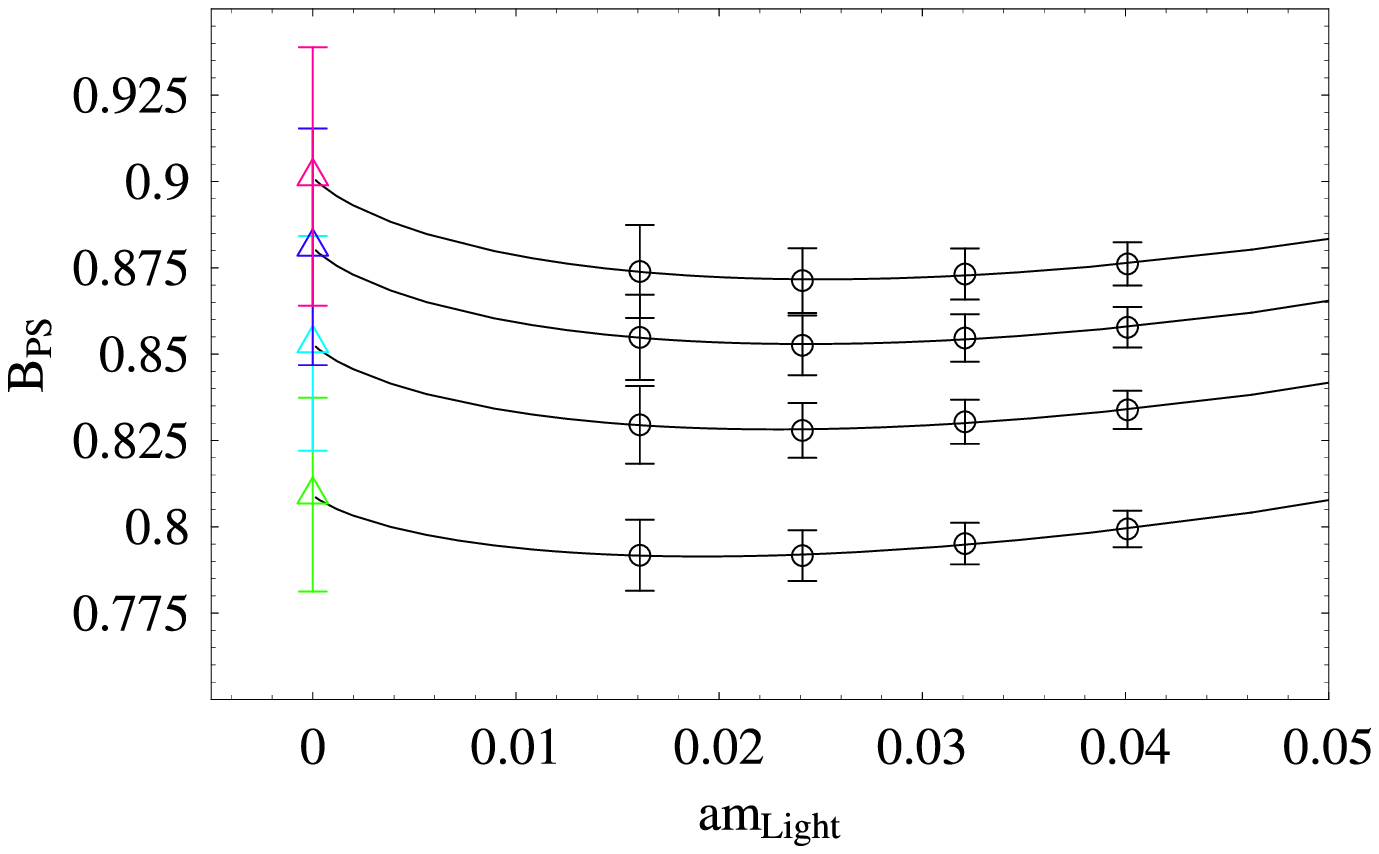}
\includegraphics[width=0.95\columnwidth,angle = 0]{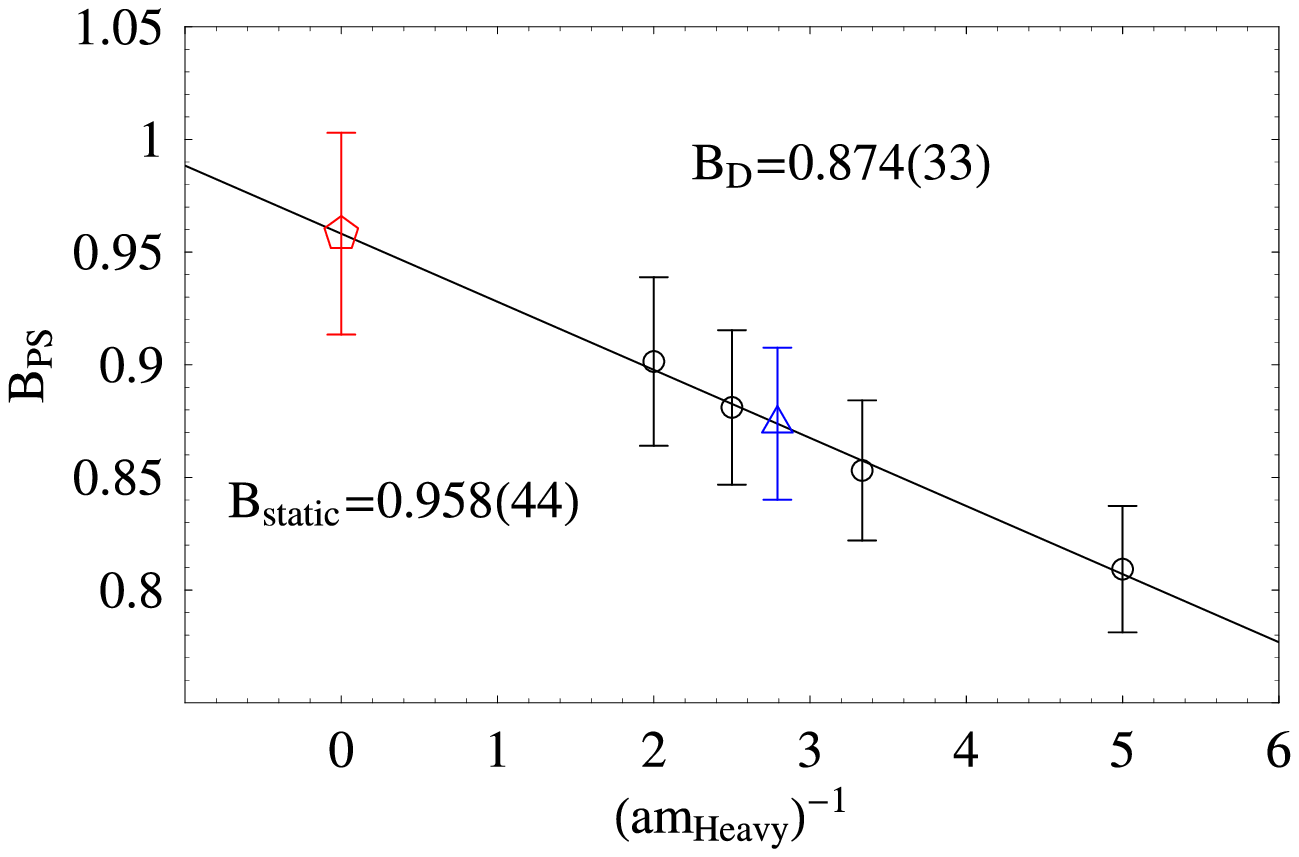}
 \caption{\small\small
The light (top) and heavy (bottom) quark mass dependence of
$B_{\rm PS}$ with the chiral: $B_{Qq} = B_1 + B_2 m_{q} + B_3
m_{q} \ln m_{q} $ and linear (in $1/M$) fit respectively. The top
figure displays light quark mass interpolation to $-m_{\rm res}$
at fixed $am_{\rm heavy}$: 0.2, 0.3, 0.4, 0.5 from top to bottom.
In the bottom figure, the triangle (blue) point represents the
physical $D$ meson point and the pentagon (red) point represents
the static quark limit point. } \label{fig:BDX} \vskip -3mm
\end{figure}

\begin{figure}[hbt]
\includegraphics[width=0.95\columnwidth,angle = 0]{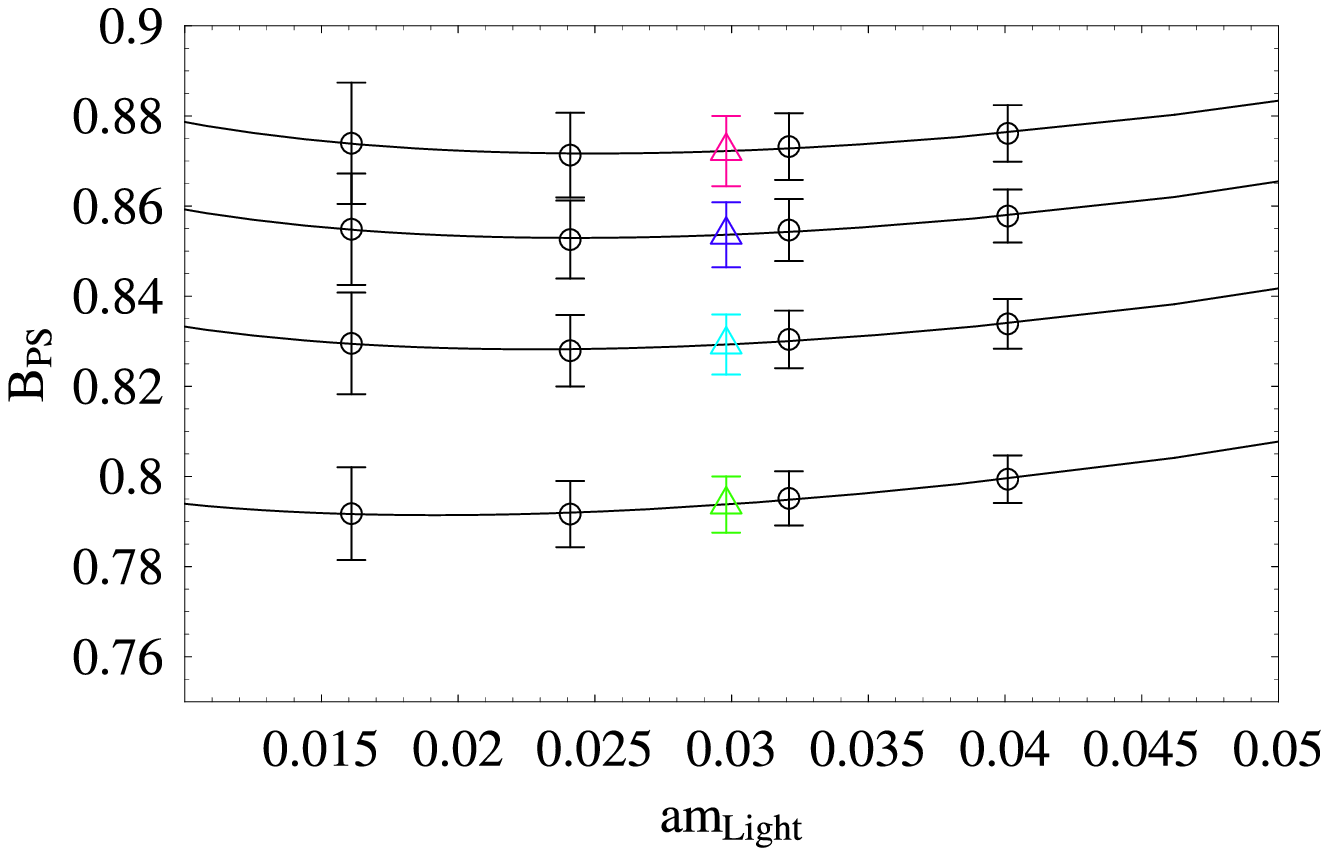}
\includegraphics[width=0.95\columnwidth,angle = 0]{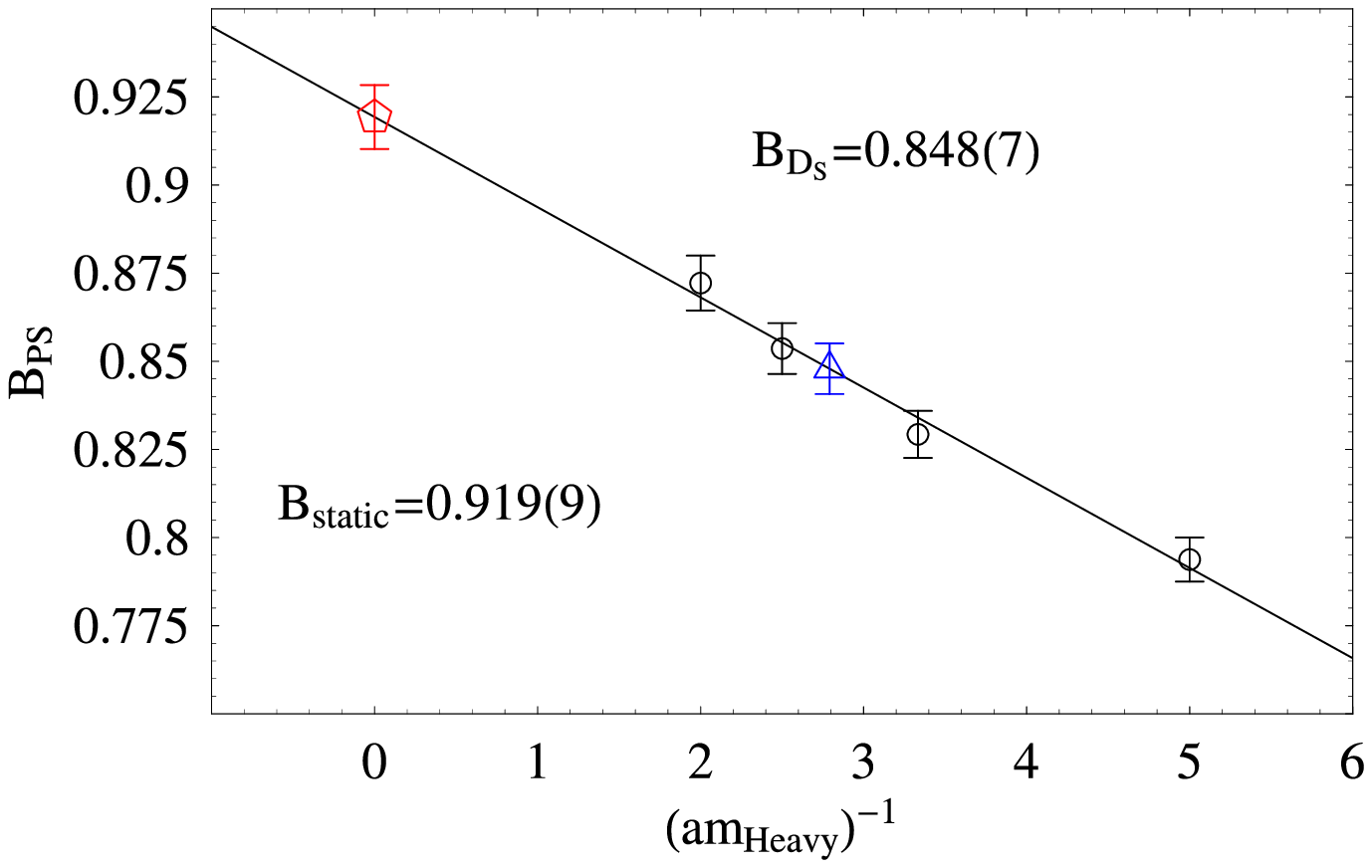}
 \caption{\small\small
The light (top) and heavy (bottom) quark mass dependence of
$B_{\rm PS}$ with the chiral: $B_{Qq} = B_1 + B_2 m_{q} + B_3
m_{q} \ln m_{q} $ and linear (in $1/M$) fit respectively. The top
figure displays light quark mass interpolation to $m_{\rm strage}$
at fixed $am_{\rm heavy}$: 0.2, 0.3, 0.4, 0.5 from top to bottom.
In the bottom figure, the triangle (blue) point represents the
physical $D_{\rm s}$ meson point and the
pentagon (red) point represents the static quark limit point.%
} \label{fig:BDsX} \vskip -3mm
\end{figure}

Booth~\cite{Booth:1994hx} and Sharpe \& Zhang~\cite{Sharpe:1995qp}
calculate the chiral behavior of the heavy-light meson bag
parameter in the quenched approximation to be:
\begin{equation}\label{eq:qBps}
B_{Qq} = B_1 + B_2 m_{q} +  B_3 m_{q} \ln {m_{q}} + B_4 \ln
{m_{q}}.
\end{equation}
The coefficient of the most divergent term, $\ln (m_q)$, is
accompanied by a factor of $1-3g^2$, where $g$ is the coupling of
$D-D^*-\pi$ that can be obtained from $D^*\rightarrow D\pi$
decay~\cite{Stewart:1998ke}, which yields a value of $0.2(7)$.
Note that this $g$ for $B-B^*-\pi$ is about the same due to
heavy-quark symmetry.  However, it is still hard to judge the
effect of this logarithmic dependence in our range of light quark
mass. We will ignore it for the rest of paper and incorporate it
into the systematic error by comparing with the unquenched
calculation, where the $\ln (m_q)$ term is absent. In order to
estimate the systematic error due to our choice of fitting form,
we adopt different expressions to extrapolate/interpolate the
light quark dependence:
\begin{equation}\label{eq:qBpsfitX}
B_{Qq} = B_1 + B_2 m_{q} +  B_3 m_{q} \ln {m_{q}};
\end{equation}
\begin{equation}\label{eq:qBpsfitQua}
B_{Qq} = B_1 + B_2 m_{q} +  B_3 m_{q}^2;
\end{equation}
\begin{equation}\label{eq:qBpsfit}
B_{Qq} = B_1 + B_2 m_{q}.
\end{equation}
Here we show the extrapolations using the Eq.~\ref{eq:qBpsfitX}
 in the upper graphs in
Figure~\ref{fig:BDX} to the chiral limit, $m_f=-\mres$ and in
Figure~\ref{fig:BDsX} to the strange quark mass. The results from
various fits are summarized in Table~\ref{tab:BpsFit} and
different analyses agree with each other within statistical error
bars. Therefore, we take the quadratic fit as the central value
and absorb the other fits into the systematic error; thus we have
\begin{equation}
B_{D}^{\rm lat}  = 0.859(24)(^{+24}_{-6})
 \end{equation}
 \begin{equation}
B_{D_{s}}^{\rm lat} = 0.848(7)(^{+2}_{-0})
\end{equation}
before proper renormalization. Note that since $D_{ s}$ is a
charged meson, there cannot be oscillations between $D_{ s}$ and
$\overline{D_{ s}}$. Here we calculate its bag parameter merely
for the interest of studying the $SU(3)$ breaking effect in
charmed systems. Also, the extrapolation to the bag parameters
from $B$ and $B_{ s}$ mesons are simply side products of this
study. It is interesting to check the goodness of $1/M$
extrapolation with the static quark lattice studies.

We take advantage of the bag parameter nonperturbative
renormalization (NPR) done in Ref~\cite{Aoki:2005ga} with RI/MOM
scheme and further convert into $\overline{\rm MS}$.
Ref.~\cite{Ciuchini:1995cd,Ciuchini:1997bw} calculated the
conversion factors in NLO perturbation theory. First, we convert
the renormalization factor into an RI value
\begin{equation}
Z_B^{\rm RI}=[\alpha_s(\mu^{\rm lat})]^{-2/l1}
\left[1+\frac{\alpha_s(\mu^{\rm lat})}{4\pi} J_{\rm RI/MOM}
\right] Z_B^{\rm lat, RI/MOM}(\mu^{\rm lat})
\end{equation}
where
\begin{equation}\label{eq:alphaS}
\alpha_s(\mu)=\frac{12 \pi }{(33-3N_f) \ln(\mu^2/\Lambda_{\rm
QCD}^2)}\left(1-\frac{918-90N_f-24N_f^2}{(33-2N_f)^2}\frac{\ln
\ln(\mu^2/\Lambda_{\rm QCD}^2)}{\ln(\mu^2/\Lambda_{\rm
QCD}^2)}\right),
\end{equation}
with $N_f=0$, and $\Lambda_{\rm QCD}^{0}=238$~$MeV$ and $J_{\rm
RI/MOM}(N_f=0)=2.883$; that gives $Z_B^{\rm RI}=1.409(5)$.%
Then we calculate the factor for $N_f=4$, as
\begin{equation}
Z_B^{\rm \overline{MS}}= \alpha_s(\mu)^{25/3}
\left[1+\frac{\alpha_s(\mu^{\rm lat})}{4\pi} J_{\rm \overline{MS}}
\right]^{-1} \hat{Z_B}.
\end{equation}
The $\Lambda_{\rm QCD}$ with $N_f=5$ is
217~MeV~\cite{Eidelman:2004wy}, suggesting that the $\Lambda_{\rm
QCD}$ is 276~MeV for $N_f=4$. Then $\alpha_s(\mu=2~{\rm GeV})$ is
0.283. Therefore, $Z_{B}^{\rm \overline{MS}}$, is 1.000(4) and our
bag parameters are:
\begin{equation}
B_{D} = 0.845(24)(^{+24}_{-6}),
\end{equation}
\begin{equation}
B_{D_{{ s}}}= 0.835(7)(^{+2}_{-0}),
\end{equation}
\begin{equation}
B_{D_{{ s}}}/B_{D} = 0.987(22)(^{+0}_{-27}).
\end{equation}

\section{$SU(3)$ breaking ratios }
\label{Sec:SU(3)breaking}%

In the long run, lattice QCD should provide a high-precision
determination of $SU(3)$ flavor-breaking of the $\Delta F =2$
heavy-light matrix element
\begin{equation}
\frac{{\cal M}_{ Qs}}{{\cal M}_{{ Q}l}} =\frac{\langle
\overline{P}_{ Qs}| \overline{Q} \gamma_{\mu}(1-\gamma_5)s
\overline{Q} \gamma_{\mu}(1-\gamma_5)s |P_{ Qs}
\rangle}{\langle \overline{P}_{{ Q}l}| \overline{Q}
\gamma_{\mu}(1-\gamma_5)l \overline{Q} \gamma_{\mu}(1-\gamma_5)l
|P_{{ Q}l} \rangle},
\end{equation}
where $Q$ stands for heavy quark, $l$ stands for light quark
($u$,$d$), and ${P}_{{ Q}l}$ represents a pseudoscalar meson
composed of $Q$ and $l$. There are two ways of obtaining this
ratio: first, by calculating the matrix element
directly\cite{Bernard:1998dg}; secondly, by calculating the decay
constants
and bag parameters separately and combining them with Eq.~\ref{eq:Bag}. %
In this work, we will focus on the second method (often referred
to as the ``indirect'' approach) to get the ratios by the
combination of ratios of bag parameters and decay constants:
\begin{equation}
r_Q=\frac{{\cal M}_{ Qs}}{{\cal M}_{{ Q}l}} = \frac{ B_{{
Qs}}(m_{{ Qs}}f_{{ Qs}})^2 }{ B_{{ Q}l}(m_{{
Q}l}f_{{ Q}l})^2 }.
\end{equation}
Therefore, it is important to obtain the ratio of
\begin{equation}
\frac{f_{Q{ s}}}{ f_{Ql}} \sqrt{\frac{B_{Ql}}{B_{Ql}}} = \xi_Q,
\end{equation}
Table~\ref{tab:SU3ratio} summarizes $\xi_{ Q}$ and the $SU(3)$
breaking for different chiral extrapolation formulae. From the
charm quark sector, we obtain
\begin{equation}
\xi_{ c} =
1.071(23)(^{+0}_{-31}),
\end{equation}
and
\begin{equation}
r_{ c}=1.273(65)(^{+0}_{-67}).
\end{equation}
\section{Extrapolation to Static $B$ Mesons}
\label{Bmeson}%
Although our main goal in this work is to apply the DWF to the
charm quark physics directly, it is of some interest to
extrapolate the heavy quark mass to the static limit. In the past,
as is pointed out in Ref.~\cite{Yamada:2002wh,Hashimoto:1999ck},
some static limit $B$ parameters obtained from extrapolation from
charm region with simple linear function in \(1/m_{\rm heavy}\)
were found disagreeing with direct static calculations: after all
the charm mass may not be sufficiently heavy to justify such a
simple extrapolation to the static limit. However, as a mere
by-product of our work on charm, the extrapolation may still be
instructive.

\subsection{Decay constants}
Since we do not know how to renormalize the heavy-light decay
constant for a static quark with DWF as the light quark action, we
will only focus on ratios of decay constants. Since the bottom
quark is so heavy compared to the scale of our physics, we may
extrapolate to the static quark limit to get the result pertaining
to $B$ mesons. The ratio is
\begin{equation}
(f_{B_{{ s}}}\sqrt{m_{B_{ s}}}/f_{B}\sqrt{m_{B}})^{\rm static} = 1.08(4)(^{+1}_{-2})
\end{equation}
Taking $m_{B_{ s}}/m_{B}$ as
1.017~\cite{Eidelman:2004wy}, we have
\begin{equation}
(f_{B_{{ s}}}/f_{B})^{\rm static} = 1.06(4)(^{+1}_{-2}).
\end{equation}
A previous quenched study using the static approximation for the
heavy fermion action and the step-scaling
technique~\cite{Heitger:2003xg} gives 1.11(16) when extrapolated
to the continuum limit, which is consistent with our extrapolation
result here. A recent study using partially quenched two-flavor
DWF lattices with the static approximation \cite{Gadiyak:2005ea}
(with lattice cutoff $\approx 1.7$~GeV) yields 1.29(4)(4)(2) for
the same quantity, giving us an idea of the systematic error due
to the quenched approximation. HPQCD use a 2+1 staggered fermion
sea with NRQCD heavy quark action and lattice cutoffs $\approx
1.6$~GeV and $\approx 2.6$~GeV to obtain the ratio
1.20(2)(1) \cite{Gray:2005ad}.

\subsection{Bag parameters}
Following the analyses in Sec.~\ref{SecBD}, we extrapolate lattice
bag parameter to the static limit
\begin{equation}
(B_{B}^{\rm lat})^{\rm static}  = 0.940(33)(^{+30}_{-6})
\end{equation}
\begin{equation}
(B_{B_{ s}}^{\rm lat})^{\rm static} = 0.919(9)(^{+3}_{-0}).
\end{equation}
We set the scale $\mu$ at the mass of $b$-quark, 4.5~GeV, and the
renormalization factor $Z_{B}^{\rm \overline{MS}}(m_{ b})$ (with
$N_f=5$) is 0.920(3), which suggests
\begin{equation}
(B_{B})^{\rm static} = 0.865(33)(^{+30}_{-6})
\end{equation}
\begin{equation}
(B_{B_{ s}})^{\rm static} = 0.845(9)(^{+3}_{-0})
\end{equation}
Ref.~\cite{Christensen:1996sj} uses static heavy and Wilson light
quark actions to get $B_B(m_{ b})=0.98(4)(^{+3}_{-18})$ at
$\approx 1.8$~GeV lattice cutoff. Another quenched study with
static heavy and overlap light quarks
gives~\cite{Becirevic:2005sx} (again, with the finest lattice
cutoff $\approx 1.8$~GeV)
\begin{equation}
(B_{B_{ s}}(m_{ b}))^{\rm static} =0.940(16)(22).
\end{equation}
Our number does not agree too well with previous static results.
This might be due to the coarse lattices used in their
simulations. However, our number agrees better with JLQCD's
quenched calculation with NRQCD~\cite{Aoki:2002bh}, where the
finest lattice spacing is 2.3~GeV:
\begin{equation}
B_{B_{ d}}(m_{ b})=0.84(3)(5),
\end{equation}
\begin{equation}
B_{B_{ s}}/B_{B_{ d}}=1.020(21)(^{+15}_{-16})(^{+5}_{-0}),
\end{equation}
\begin{equation}
B_{S_{ s}}(m_{ b})=0.85(1)(5)(^{+1}_{-0}),
\end{equation}
after taking the continuum limit.
Our results for the $B$ case are consistent with the two-flavor
DWF static-light calculation in Ref.~\cite{Gadiyak:2005ea}:
\begin{equation}
(B_{B})^{\rm static}  = 0.812(48)(67)(^{+0}_{-300}),
\end{equation}
\begin{equation}
(B_{B_{ s}})^{\rm static} = 0.864(28)(71)(^{+0}_{-320}),
\end{equation}
\begin{equation}
(B_{B_{ s}}/B_{B})^{\rm static} = 1.06(6)(3)(1),
\end{equation}
and the two-flavored $O(a)$-improved Wilson fermion sea with NRQCD
heavy quark study~\cite{Aoki:2003xb} done by JLQCD
\begin{equation}
(B_{B}) = 0.836(27)(^{+56}_{-62})\\
(B_{B_{ s}}/B_{B}) = 1.017(16)(^{+56}_{-17}).
\end{equation}

\subsection{$SU(3)$ breaking ratio}

\begin{figure}[hbt]
\includegraphics[width=0.95\columnwidth,angle = 0]{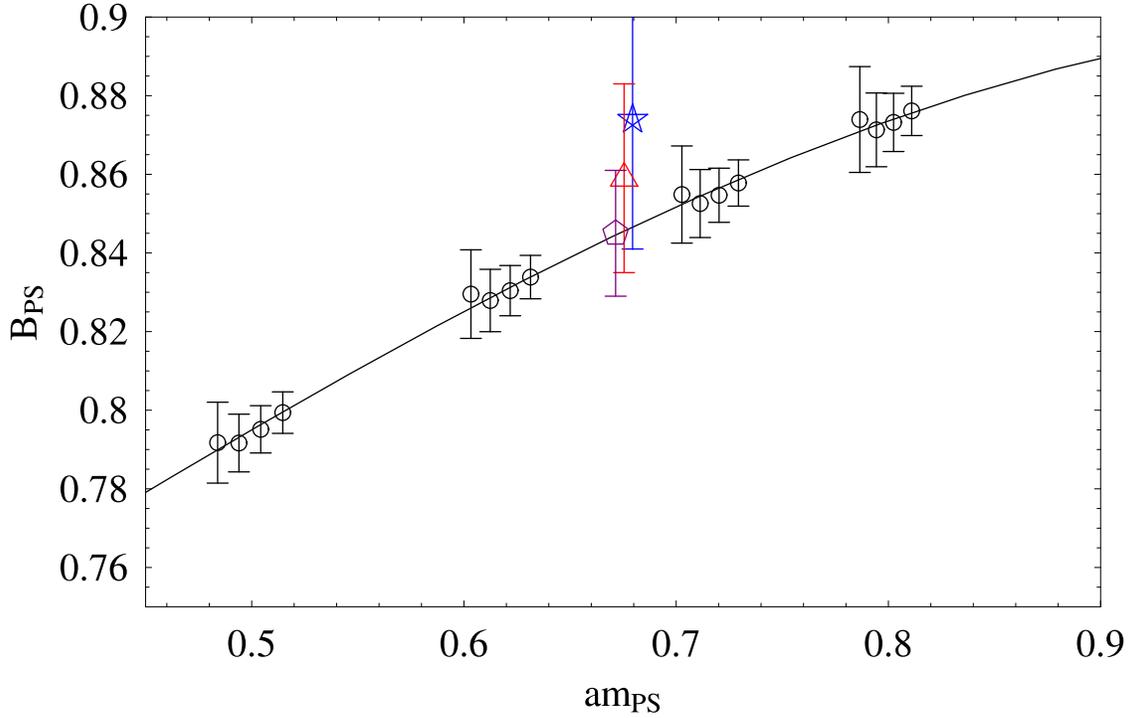}
\caption{\small\small The $B_{PS}^{\rm lat}$ as a function of the
pseudoscalar meson. The circle points are the data points and the
black line is the fit of the form: $B_{\rm PS}( m_{\rm PS})= B_1 +
B_2 m_{\rm PS}^2 + B_3 m_{\rm PS}^2 \ln m_{\rm PS}^2 $. The star
(blue) point is the $B_D^{lat}$ point obtained from
Fig.\ref{fig:BDX}, the triangle (red) point is the $B_D^{lat}$
from quadratic fit, and the pentagon (purple) point is the
$B_D^{lat}$ from linear fit. } \label{fig:BpsMps} \vskip -3mm
\end{figure}

Extending the discussions in Sec.~\ref{Sec:SU(3)breaking}, we have
the results for the bottom sector: $\xi_{  b} =
1.019(37)(^{+16}_{-34}) $ and the $SU(3)$ breaking ratio
$r_{b}=1.274(94)(^{+32}_{-180})$. It is useful to recall that in
the first calculation 
in Ref.\ \cite{Bernard:1998dg}, using a Wilson fermion action,
found $\xi_{  b}=1.30(4)(^{+21}_{-15})$  
and $1.17(2)(^{+12}_{-6})$ after linear continuum extrapolations
from direct and indirect methods, respectively.
A previous static quark study on a two-flavor DWF sea gives:
$\xi_{ b}= 1.33(8)(8)$~\cite{Gadiyak:2005ea}. Using JLQCD ratio of
bag parameters \cite{Aoki:2003xb} and HPQCD 2+1 result
\cite{Aubin:2005ar} on decay constants, the Lattice '05 review
\cite{Okamoto:2005zg} gives $\xi_{ b}=1.210(^{+47}_{-35})$ and the
previous world average \cite{Hashimoto:2004hn} gives $\xi_{
b}=1.23(6)$.
Our central value is smaller than these previous results but quite
consistent with other studies given our relatively large errors.
We first take the light quark mass limit according to the
heavy-light meson bag parameter chiral formula and then take the
heavy quark mass to $m_{\rm charm}^{\rm lat}$;
Figures~\ref{fig:BDX} and \ref{fig:BDsX} show that the chiral
behavior of the light quark part makes the bag parameters for
mesons with up and down quarks somewhat larger than those with a
strange quark. However, the other studies get their bag parameters
by fitting the bag parameter as a function of heavy-light
pseudoscalar mass, instead of the heavy-light chiral forms; see
the dependence for our data in Figure~\ref{fig:BpsMps}. The bag
parameter increases with the pseudoscalar mass. Therefore, if one
fits the bag parameter function as input of heavy-light
pseudoscalar mass, from a relation $m_{\{B,D\}_{ s}}/m_{\{B,D\}} >
1$, which always holds, it follows that $B_{\{B,D\}_{
s}}/B_{\{B,D\}} > 1$. Therefore, $\xi_{ Q}$ in these studies must
come out larger than our values.

\section{Systematic errors}
\label{SecSysErr}%
In the previous sections, we present our calculation with
statistical errors which are computed using the jackknife
procedure, along with systematic errors mostly caused by chiral
extrapolation. There are other potential sources of systematic
errors caused by the quenched approximation, finite-volume
effects, operator mixing and matching calculations. Here is a
detailed estimate of the various
systematic errors.
\begin{description}
\item[Finite volume:]
The proper way to estimate this error is to perform the
calculation on at least two different volumes and extrapolate to
the infinite-volume limit. However, in this work, we only perform
our calculation in one lattice volume, which is about $(1.6~{\rm
fm})^3$. Since the scale of $D$ system is in between $K$ and $B$,
we will estimate our finite-volume error by quoting whichever
system ($K$ or $B$) has larger finite-volume error in previously
published studies. Ref.~\cite{Aoki:2005ga}, which uses the same
lattice box we do, quoted a 2\% error in $B_K$ by comparing with a
2.4~fm box. We also compare our $B_B$ found by extrapolation to
the static limit with a quenched, larger-volume ($\approx
2.4$~fm), calculation \cite{Aoki:2002bh}, which gives $B_{B_d}
(m_b) = 0.84(3)(5)$ with NRQCD fermions; this gives us an
estimation of 0.4\% error. Taking the larger of these two
estimates, we conclude that the finite-volume effect on the $B_D$
should be smaller than 2\%.

\item[Continuum extrapolation:]
In this paper, we only perform our calculation at one lattice
spacing, and therefore we cannot extrapolate our result to the
continuum limit. This should be checked carefully in future works.
The next-best thing we can do is to estimate the error from a
continuum extrapolation study from the same gauge configuration.
Such a study has been performed on the decay constants and matrix
elements of the $K$ system in Ref.~\cite{Aoki:2005ga} with an
additional lattice cutoff at 1.982(30)~GeV. The result shows mild
dependence from $a^{-1} \approx 3$~GeV to the continuum.
Therefore, we add 0.2\% to our systematic errors due to continuum
extrapolation.

\item[RI formulation:]
We need the $\alpha_s$ to first convert the scale-dependence
RI/MOM NPR factor, $Z_B^{\rm RI/MOM}$ to an RI value and further
convert it to an $\overline{\rm MS}$ one. In the expression, we
use the NLO order formulation, which leaves the remaining leading
error up to $O(\alpha_s^2) \approx 4 \%$.

\item[NPR renormalization factor in bag parameters:]
We adopt the RI/MOM NPR renormalization factor from
Ref.~\cite{Aoki:2005ga}, and we quote the estimation within that
paper as 1\%.

\item[Operators mixing:]
We have discussed mixing with wrong chirality operators, which
contributes around 10\% uncertainties to our final $B_D$
calculation.

\item[Quenched approximation:]
The quenched approximation ignores sea quark loop contributions
which in general are considered to be a major contribution to
systematic errors. Ref.~\cite{Aoki:2005ga}, which uses the same
quenched lattice ensemble as in the present work, quotes a 6\%
error on the $B_K$ factor due to this approximation, after
comparing with the number calculated on two-flavor DWF lattices.
Similarly, we can compare our static quark limit value with the
one calculated on two-flavor DWF lattices \cite{Gadiyak:2005ea},
which gives us 6\%. Therefore, it seems reasonable to use a 6\%
systematic quenching error for $B_D$ as well.

\end{description}

\section{Conclusion }
\label{SecFuture}

In this work, we use the domain-wall fermions (DWF) formulation
for charm quark as well as the three lighter flavors, up, down and
strange, on the lattice with a relatively high cutoff (around
3~GeV). We use the mass ratio $m_K/m_\rho$ to set the bare strange
quark mass $m_{\rm strange}a=0.0298(13)$ in lattice units. Then
combining this and another hadronic mass ratio $m_{D_{s}}/m_\rho$
we obtain the bare charm mass,  \(m_{\rm charm}a=0.3583(22)\).

Using these bare quark mass values we found, we conclude the following
for charmed and charmed-strange meson states:
\begin{itemize}
\item the masses of the  \(J^{P}=0^{\pm}\) and \(1^{\pm}\) \(D\), \(D^{*}\),
\(D_{s}\) and \(D_{sJ}\) states are well reproduced to within a
few percent.
\item Their parity splitting, \(\Delta_{J}\), are better reproduced than
previous works, with only 10-20\% over estimations.
\item The experimental observation of \(\Delta_{ud}>\Delta_{s}\) is reproduced.
\item The hyperfine splittings are only 60-65\% reproduced.
\end{itemize}
Regarding the dependence on heavy quark mass,
\begin{itemize}
\item \(\Delta_{J=0}\) and \(\Delta_{J=1}\) are degenerate for $m_{\rm heavy}a > 0.2-0.3$.
\item \(\Delta_{J=0}\) increases as \(m_{\rm heavy}\) decreases further, while
\item \(\Delta_{J=1}\) does not.
\end{itemize}
Also with the bare charm quark mass we worked out the
renormalization factor, $Z_m$, from the existing one-loop
domain-wall perturbation calculation. After RG scaling, the
$m^{\overline{\rm MS}}_{ c}(m_{ c})=1.239(8)$~GeV. The
perturbative renormalization contains a systematic error of
$O(\alpha_s^2) \approx 4\%$, giving us $m^{\overline{\rm MS}}_{
c}(m_{ c})=1.24(1)(18)$~GeV.

In the charmonium system we find the $\eta_c$ mass agrees well
with the experimental value.  The $J/\psi$ mass (hence the
hyperfine splitting) is smaller than the experimental one,
confirming the long puzzle in lattice QCD.  We also note a
prediction for the mass of yet-to-be-discovered C-odd \(h_c\)
state: the mass difference between \(h_c\) and \(\chi_{c1}\) is
estimated as 22(11)~MeV.  This would translate to the mass of
\(3533(11) _{\rm stat.}\)
 MeV for this meson.

The leptonic decay constants are also calculated.
We tried to estimate the systematic uncertainty from the associated
quenched logarithm by adopting
three different fitting formulations. Our
result after considering other systematic uncertainties due to the
quenched approximation and continuum extrapolation give us
\begin{equation}
f_D=232(7)(^{+6}_{-0})(11)~{\rm MeV}
 \end{equation}
 \begin{equation}
f_{D_{{ s}}}/f_{D} = 1.05(2)(^{+0}_{-2})(2).
\end{equation}
We also discussed extrapolation to the static limit in terms of
\(1/m_{\rm heavy}\) which gives results consistent with previous
static calculations.

The bag parameters in the $D$ and $D_{ s}$ (purely theoretical but
interesting in regard of $SU(3)$ breaking effect) are studied. The
use of domain-wall fermions on the current fine lattice with only
softly broken chiral symmetry gives us some advantages such as
absence of complicated mixing and the availability of the RI/MOM
nonperturbative renormalization techniques. We include a detailed
estimation of systematic uncertainties. The biggest systematic
errors come from the quenched approximation and the mixing of
wrong-chirality operators. Our result is:
\begin{equation}
B_D(2{\rm~GeV})=0.845(24)(^{+24}_{-6})(105),
 \end{equation}
 \begin{equation}
B_{D_{{ s}}}/B_{D} = 0.987(22)(^{+0}_{-27})(23),
\end{equation}
where the first error is statistical, the second systematic from
fitting, and the third combining all other known systematics. Thus
a minor $SU(3)$-breaking is seen in our calculation. We also
discussed extrapolation to the static limit in terms of \(1/m_{\rm
heavy}\) which suggests the \(D\) and \(D_{s}\) meson results are
reliable.

In conclusion, using DWF to simulate the charm quark on a quenched
ensemble at a moderately high cutoff of about 3 GeV obtained with
the DBW2 action results in reasonable descriptions for most of the
calculated meson observables: the masses and their splittings,
leptonic decay constant, and \(\Delta C=2\) mixing.  It seems the
charm quark propagation is successfully described with the current
set up.  This obviously is an attractive direction to proceed,
especially with dynamical QCD ensembles that are becoming
available.

\section*{ACKNOWLEDGMENTS}
We thank Junichi Noaki for generating the gauge configurations and
useful discussions, various RBC members for physics discussions
and  RIKEN, Brookhaven National Laboratory and the U.S. Department
of Energy for providing the facilities essential for the
completion of this work. N.~Yamada is supported in part by the
Grant-in-Aid of the Ministry of Education (Nos.18034011, 18340075,
18740167). H.-W.~Lin is supported by DOE grant DE-FG02-92ER40699.
A.~Soni is supported in part by DOE grant DE-AC02-98CH10886.
 \vspace{-0.1in}


\bibliography{paper}

\begin{thebibliography}{100}
\expandafter\ifx\csname natexlab\endcsname\relax\def\natexlab#1{#1}\fi
\expandafter\ifx\csname bibnamefont\endcsname\relax
  \def\bibnamefont#1{#1}\fi
\expandafter\ifx\csname bibfnamefont\endcsname\relax
  \def\bibfnamefont#1{#1}\fi
\expandafter\ifx\csname citenamefont\endcsname\relax
  \def\citenamefont#1{#1}\fi
\expandafter\ifx\csname url\endcsname\relax
  \def\url#1{\texttt{#1}}\fi
\expandafter\ifx\csname urlprefix\endcsname\relax\def\urlprefix{URL }\fi
\providecommand{\bibinfo}[2]{#2}
\providecommand{\eprint}[2][]{\url{#2}}

\bibitem[{\citenamefont{Bonvicini et~al.}(2004)}]{Bonvicini:2004gv}
\bibinfo{author}{\bibfnamefont{G.}~\bibnamefont{Bonvicini}}
  \bibnamefont{et~al.} (\bibinfo{collaboration}{CLEO}), \bibinfo{journal}{Phys.
  Rev.} \textbf{\bibinfo{volume}{D70}}, \bibinfo{pages}{112004}
  (\bibinfo{year}{2004}), \eprint{hep-ex/0411050}.

\bibitem[{\citenamefont{Arms et~al.}(2004)}]{Arms:2003ra}
\bibinfo{author}{\bibfnamefont{K.}~\bibnamefont{Arms}} \bibnamefont{et~al.}
  (\bibinfo{collaboration}{CLEO}), \bibinfo{journal}{Phys. Rev.}
  \textbf{\bibinfo{volume}{D69}}, \bibinfo{pages}{071102}
  (\bibinfo{year}{2004}), \eprint{hep-ex/0309065}.

\bibitem[{\citenamefont{Eisenstein et~al.}(2004)}]{Eisenstein:2004bg}
\bibinfo{author}{\bibfnamefont{B.~I.} \bibnamefont{Eisenstein}}
  \bibnamefont{et~al.} (\bibinfo{collaboration}{CLEO}) (\bibinfo{year}{2004}),
  \eprint{hep-ex/0408055}.

\bibitem[{\citenamefont{Besson et~al.}(2004)}]{Besson:2003jp}
\bibinfo{author}{\bibfnamefont{D.}~\bibnamefont{Besson}} \bibnamefont{et~al.}
  (\bibinfo{collaboration}{CLEO}), \bibinfo{journal}{AIP Conf. Proc.}
  \textbf{\bibinfo{volume}{698}}, \bibinfo{pages}{497} (\bibinfo{year}{2004}),
  \eprint{hep-ex/0305017}.

\bibitem[{\citenamefont{Flood}(2005)}]{Flood:2005da}
\bibinfo{author}{\bibfnamefont{K.~T.} \bibnamefont{Flood}}
  (\bibinfo{collaboration}{BABAR}), \bibinfo{journal}{Int. J. Mod. Phys.}
  \textbf{\bibinfo{volume}{A20}}, \bibinfo{pages}{3686} (\bibinfo{year}{2005}).

\bibitem[{\citenamefont{Aubert et~al.}(2003)}]{Aubert:2003fg}
\bibinfo{author}{\bibfnamefont{B.}~\bibnamefont{Aubert}} \bibnamefont{et~al.}
  (\bibinfo{collaboration}{BABAR}), \bibinfo{journal}{Phys. Rev. Lett.}
  \textbf{\bibinfo{volume}{90}}, \bibinfo{pages}{242001}
  (\bibinfo{year}{2003}), \eprint{hep-ex/0304021}.

\bibitem[{\citenamefont{Benussi}(2005)}]{Benussi:2005db}
\bibinfo{author}{\bibfnamefont{L.}~\bibnamefont{Benussi}}
  (\bibinfo{collaboration}{FOCUS}), \bibinfo{journal}{Int. J. Mod. Phys.}
  \textbf{\bibinfo{volume}{A20}}, \bibinfo{pages}{549} (\bibinfo{year}{2005}).

\bibitem[{\citenamefont{Danilov}(2004)}]{Danilov:2004td}
\bibinfo{author}{\bibfnamefont{M.}~\bibnamefont{Danilov}}
  (\bibinfo{collaboration}{BELLE}) (\bibinfo{year}{2004}),
  \bibinfo{note}{prepared for 32nd International Conference on High-Energy
  Physics (ICHEP 04), Beijing, China, 16-22 Aug 2004}.

\bibitem[{\citenamefont{Abe et~al.}(2005)}]{Abe:2005hd}
\bibinfo{author}{\bibfnamefont{K.}~\bibnamefont{Abe}} \bibnamefont{et~al.}
  (\bibinfo{year}{2005}), \eprint{hep-ex/0507019}.

\bibitem[{\citenamefont{Drutskoy et~al.}(2005)}]{Drutskoy:2005zr}
\bibinfo{author}{\bibfnamefont{A.}~\bibnamefont{Drutskoy}} \bibnamefont{et~al.}
  (\bibinfo{collaboration}{Belle}), \bibinfo{journal}{Phys. Rev. Lett.}
  \textbf{\bibinfo{volume}{94}}, \bibinfo{pages}{061802}
  (\bibinfo{year}{2005}).

\bibitem[{\citenamefont{Swanson}(2006)}]{Swanson:2005tq}
\bibinfo{author}{\bibfnamefont{E.}~\bibnamefont{Swanson}},
  \bibinfo{journal}{AIP Conf. Proc.} \textbf{\bibinfo{volume}{814}},
  \bibinfo{pages}{203} (\bibinfo{year}{2006}), \eprint{hep-ph/0509327}.

\bibitem[{\citenamefont{Quigg}(2006)}]{Quigg:2005tv}
\bibinfo{author}{\bibfnamefont{C.}~\bibnamefont{Quigg}}, \bibinfo{journal}{PoS}
  \textbf{\bibinfo{volume}{HEP2005}}, \bibinfo{pages}{400}
  (\bibinfo{year}{2006}), \eprint{hep-ph/0509332}.

\bibitem[{\citenamefont{Barnes}(2006)}]{Barnes:2005zy}
\bibinfo{author}{\bibfnamefont{T.}~\bibnamefont{Barnes}}, \bibinfo{journal}{AIP
  Conf. Proc.} \textbf{\bibinfo{volume}{814}}, \bibinfo{pages}{735}
  (\bibinfo{year}{2006}), \eprint{hep-ph/0510365}.

\bibitem[{\citenamefont{Eichten and Hill}(1990{\natexlab{a}})}]{Eichten:1989zv}
\bibinfo{author}{\bibfnamefont{E.}~\bibnamefont{Eichten}} \bibnamefont{and}
  \bibinfo{author}{\bibfnamefont{B.}~\bibnamefont{Hill}},
  \bibinfo{journal}{Phys. Lett.} \textbf{\bibinfo{volume}{B234}},
  \bibinfo{pages}{511} (\bibinfo{year}{1990}{\natexlab{a}}).

\bibitem[{\citenamefont{Eichten and Hill}(1990{\natexlab{b}})}]{Eichten:1989kb}
\bibinfo{author}{\bibfnamefont{E.}~\bibnamefont{Eichten}} \bibnamefont{and}
  \bibinfo{author}{\bibfnamefont{B.}~\bibnamefont{Hill}},
  \bibinfo{journal}{Phys. Lett.} \textbf{\bibinfo{volume}{B240}},
  \bibinfo{pages}{193} (\bibinfo{year}{1990}{\natexlab{b}}).

\bibitem[{\citenamefont{Eichten and Hill}(1990{\natexlab{c}})}]{Eichten:1990vp}
\bibinfo{author}{\bibfnamefont{E.}~\bibnamefont{Eichten}} \bibnamefont{and}
  \bibinfo{author}{\bibfnamefont{B.}~\bibnamefont{Hill}},
  \bibinfo{journal}{Phys. Lett.} \textbf{\bibinfo{volume}{B243}},
  \bibinfo{pages}{427} (\bibinfo{year}{1990}{\natexlab{c}}).

\bibitem[{\citenamefont{Thacker and Lepage}(1991)}]{Thacker:1990bm}
\bibinfo{author}{\bibfnamefont{B.~A.} \bibnamefont{Thacker}} \bibnamefont{and}
  \bibinfo{author}{\bibfnamefont{G.~P.} \bibnamefont{Lepage}},
  \bibinfo{journal}{Phys. Rev.} \textbf{\bibinfo{volume}{D43}},
  \bibinfo{pages}{196} (\bibinfo{year}{1991}).

\bibitem[{\citenamefont{Lepage et~al.}(1992)\citenamefont{Lepage, Magnea,
  Nakhleh, Magnea, and Hornbostel}}]{Lepage:1992tx}
\bibinfo{author}{\bibfnamefont{G.~P.} \bibnamefont{Lepage}},
  \bibinfo{author}{\bibfnamefont{L.}~\bibnamefont{Magnea}},
  \bibinfo{author}{\bibfnamefont{C.}~\bibnamefont{Nakhleh}},
  \bibinfo{author}{\bibfnamefont{U.}~\bibnamefont{Magnea}}, \bibnamefont{and}
  \bibinfo{author}{\bibfnamefont{K.}~\bibnamefont{Hornbostel}},
  \bibinfo{journal}{Phys. Rev.} \textbf{\bibinfo{volume}{D46}},
  \bibinfo{pages}{4052} (\bibinfo{year}{1992}), \eprint{hep-lat/9205007}.

\bibitem[{\citenamefont{de~Divitiis
  et~al.}(2003{\natexlab{a}})\citenamefont{de~Divitiis, Guagnelli, Petronzio,
  Tantalo, and Palombi}}]{deDivitiis:2003iy}
\bibinfo{author}{\bibfnamefont{G.~M.} \bibnamefont{de~Divitiis}},
  \bibinfo{author}{\bibfnamefont{M.}~\bibnamefont{Guagnelli}},
  \bibinfo{author}{\bibfnamefont{R.}~\bibnamefont{Petronzio}},
  \bibinfo{author}{\bibfnamefont{N.}~\bibnamefont{Tantalo}}, \bibnamefont{and}
  \bibinfo{author}{\bibfnamefont{F.}~\bibnamefont{Palombi}},
  \bibinfo{journal}{Nucl. Phys.} \textbf{\bibinfo{volume}{B675}},
  \bibinfo{pages}{309} (\bibinfo{year}{2003}{\natexlab{a}}),
  \eprint{hep-lat/0305018}.

\bibitem[{\citenamefont{de~Divitiis
  et~al.}(2003{\natexlab{b}})\citenamefont{de~Divitiis, Guagnelli, Palombi,
  Petronzio, and Tantalo}}]{deDivitiis:2003wy}
\bibinfo{author}{\bibfnamefont{G.~M.} \bibnamefont{de~Divitiis}},
  \bibinfo{author}{\bibfnamefont{M.}~\bibnamefont{Guagnelli}},
  \bibinfo{author}{\bibfnamefont{F.}~\bibnamefont{Palombi}},
  \bibinfo{author}{\bibfnamefont{R.}~\bibnamefont{Petronzio}},
  \bibnamefont{and} \bibinfo{author}{\bibfnamefont{N.}~\bibnamefont{Tantalo}},
  \bibinfo{journal}{Nucl. Phys.} \textbf{\bibinfo{volume}{B672}},
  \bibinfo{pages}{372} (\bibinfo{year}{2003}{\natexlab{b}}),
  \eprint{hep-lat/0307005}.

\bibitem[{\citenamefont{Juttner and Rolf}(2003)}]{Juttner:2003ns}
\bibinfo{author}{\bibfnamefont{A.}~\bibnamefont{Juttner}} \bibnamefont{and}
  \bibinfo{author}{\bibfnamefont{J.}~\bibnamefont{Rolf}}
  (\bibinfo{collaboration}{ALPHA}), \bibinfo{journal}{Phys. Lett.}
  \textbf{\bibinfo{volume}{B560}}, \bibinfo{pages}{59} (\bibinfo{year}{2003}),
  \eprint{hep-lat/0302016}.

\bibitem[{\citenamefont{Lin and Christ}(2006)}]{Lin:2005ze}
\bibinfo{author}{\bibfnamefont{H.-W.} \bibnamefont{Lin}} \bibnamefont{and}
  \bibinfo{author}{\bibfnamefont{N.~H.} \bibnamefont{Christ}},
  \bibinfo{journal}{PoS} \textbf{\bibinfo{volume}{LAT2005}},
  \bibinfo{pages}{225} (\bibinfo{year}{2006}), \eprint{hep-lat/0510111}.

\bibitem[{\citenamefont{Kuramashi et~al.}(2006)}]{Kuramashi:2005ww}
\bibinfo{author}{\bibfnamefont{Y.}~\bibnamefont{Kuramashi}}
  \bibnamefont{et~al.} (\bibinfo{collaboration}{CP-PACS}),
  \bibinfo{journal}{PoS} \textbf{\bibinfo{volume}{LAT2005}},
  \bibinfo{pages}{226} (\bibinfo{year}{2006}).

\bibitem[{\citenamefont{Kaplan}(1992)}]{Kaplan:1992bt}
\bibinfo{author}{\bibfnamefont{D.~B.} \bibnamefont{Kaplan}},
  \bibinfo{journal}{Phys. Lett.} \textbf{\bibinfo{volume}{B288}},
  \bibinfo{pages}{342} (\bibinfo{year}{1992}), \eprint{hep-lat/9206013}.

\bibitem[{\citenamefont{Shamir}(1993)}]{Shamir:1993zy}
\bibinfo{author}{\bibfnamefont{Y.}~\bibnamefont{Shamir}},
  \bibinfo{journal}{Nucl. Phys.} \textbf{\bibinfo{volume}{B406}},
  \bibinfo{pages}{90} (\bibinfo{year}{1993}), \eprint{hep-lat/9303005}.

\bibitem[{\citenamefont{Furman and Shamir}(1995)}]{Furman:1994ky}
\bibinfo{author}{\bibfnamefont{V.}~\bibnamefont{Furman}} \bibnamefont{and}
  \bibinfo{author}{\bibfnamefont{Y.}~\bibnamefont{Shamir}},
  \bibinfo{journal}{Nucl. Phys.} \textbf{\bibinfo{volume}{B439}},
  \bibinfo{pages}{54} (\bibinfo{year}{1995}), \eprint{hep-lat/9405004}.

\bibitem[{\citenamefont{Blum and Soni}(1997{\natexlab{a}})}]{Blum:1997mz}
\bibinfo{author}{\bibfnamefont{T.}~\bibnamefont{Blum}} \bibnamefont{and}
  \bibinfo{author}{\bibfnamefont{A.}~\bibnamefont{Soni}},
  \bibinfo{journal}{Phys. Rev. Lett.} \textbf{\bibinfo{volume}{79}},
  \bibinfo{pages}{3595} (\bibinfo{year}{1997}{\natexlab{a}}),
  \eprint{hep-lat/9706023}.

\bibitem[{\citenamefont{Blum et~al.}(2004)}]{Blum:2000kn}
\bibinfo{author}{\bibfnamefont{T.}~\bibnamefont{Blum}} \bibnamefont{et~al.},
  \bibinfo{journal}{Phys. Rev.} \textbf{\bibinfo{volume}{D69}},
  \bibinfo{pages}{074502} (\bibinfo{year}{2004}), \eprint{hep-lat/0007038}.

\bibitem[{\citenamefont{Ali~Khan et~al.}(2001)}]{AliKhan:2000iv}
\bibinfo{author}{\bibfnamefont{A.}~\bibnamefont{Ali~Khan}} \bibnamefont{et~al.}
  (\bibinfo{collaboration}{CP-PACS}), \bibinfo{journal}{Phys. Rev.}
  \textbf{\bibinfo{volume}{D63}}, \bibinfo{pages}{114504}
  (\bibinfo{year}{2001}), \eprint{hep-lat/0007014}.

\bibitem[{\citenamefont{Aoki et~al.}(2004)}]{Aoki:2002vt}
\bibinfo{author}{\bibfnamefont{Y.}~\bibnamefont{Aoki}} \bibnamefont{et~al.},
  \bibinfo{journal}{Phys. Rev.} \textbf{\bibinfo{volume}{D69}},
  \bibinfo{pages}{074504} (\bibinfo{year}{2004}), \eprint{hep-lat/0211023}.

\bibitem[{\citenamefont{Blum and Soni}(1997{\natexlab{b}})}]{Blum:1996jf}
\bibinfo{author}{\bibfnamefont{T.}~\bibnamefont{Blum}} \bibnamefont{and}
  \bibinfo{author}{\bibfnamefont{A.}~\bibnamefont{Soni}},
  \bibinfo{journal}{Phys. Rev.} \textbf{\bibinfo{volume}{D56}},
  \bibinfo{pages}{174} (\bibinfo{year}{1997}{\natexlab{b}}),
  \eprint{hep-lat/9611030}.

\bibitem[{\citenamefont{Nowak et~al.}(1993)\citenamefont{Nowak, Rho, and
  Zahed}}]{Nowak:1992um}
\bibinfo{author}{\bibfnamefont{M.~A.} \bibnamefont{Nowak}},
  \bibinfo{author}{\bibfnamefont{M.}~\bibnamefont{Rho}}, \bibnamefont{and}
  \bibinfo{author}{\bibfnamefont{I.}~\bibnamefont{Zahed}},
  \bibinfo{journal}{Phys. Rev.} \textbf{\bibinfo{volume}{D48}},
  \bibinfo{pages}{4370} (\bibinfo{year}{1993}), \eprint{hep-ph/9209272}.

\bibitem[{\citenamefont{Bardeen and Hill}(1994)}]{Bardeen:1993ae}
\bibinfo{author}{\bibfnamefont{W.~A.} \bibnamefont{Bardeen}} \bibnamefont{and}
  \bibinfo{author}{\bibfnamefont{C.~T.} \bibnamefont{Hill}},
  \bibinfo{journal}{Phys. Rev.} \textbf{\bibinfo{volume}{D49}},
  \bibinfo{pages}{409} (\bibinfo{year}{1994}), \eprint{hep-ph/9304265}.

\bibitem[{\citenamefont{Bardeen et~al.}(2003)\citenamefont{Bardeen, Eichten,
  and Hill}}]{Bardeen:2003kt}
\bibinfo{author}{\bibfnamefont{W.~A.} \bibnamefont{Bardeen}},
  \bibinfo{author}{\bibfnamefont{E.~J.} \bibnamefont{Eichten}},
  \bibnamefont{and} \bibinfo{author}{\bibfnamefont{C.~T.} \bibnamefont{Hill}},
  \bibinfo{journal}{Phys. Rev.} \textbf{\bibinfo{volume}{D68}},
  \bibinfo{pages}{054024} (\bibinfo{year}{2003}), \eprint{hep-ph/0305049}.

\bibitem[{\citenamefont{Nowak et~al.}(2004)\citenamefont{Nowak, Rho, and
  Zahed}}]{Nowak:2003ra}
\bibinfo{author}{\bibfnamefont{M.~A.} \bibnamefont{Nowak}},
  \bibinfo{author}{\bibfnamefont{M.}~\bibnamefont{Rho}}, \bibnamefont{and}
  \bibinfo{author}{\bibfnamefont{I.}~\bibnamefont{Zahed}},
  \bibinfo{journal}{Acta Phys. Polon.} \textbf{\bibinfo{volume}{B35}},
  \bibinfo{pages}{2377} (\bibinfo{year}{2004}), \eprint{hep-ph/0307102}.

\bibitem[{\citenamefont{Petrov}(2005)}]{Petrov:2004rf}
\bibinfo{author}{\bibfnamefont{A.~A.} \bibnamefont{Petrov}},
  \bibinfo{journal}{Nucl. Phys. Proc. Suppl.} \textbf{\bibinfo{volume}{142}},
  \bibinfo{pages}{333} (\bibinfo{year}{2005}), \eprint{hep-ph/0409130}.

\bibitem[{\citenamefont{Bernard et~al.}(1988)\citenamefont{Bernard, Draper,
  Hockney, and Soni}}]{Bernard:1988dy}
\bibinfo{author}{\bibfnamefont{C.~W.} \bibnamefont{Bernard}},
  \bibinfo{author}{\bibfnamefont{T.}~\bibnamefont{Draper}},
  \bibinfo{author}{\bibfnamefont{G.}~\bibnamefont{Hockney}}, \bibnamefont{and}
  \bibinfo{author}{\bibfnamefont{A.}~\bibnamefont{Soni}},
  \bibinfo{journal}{Phys. Rev.} \textbf{\bibinfo{volume}{D38}},
  \bibinfo{pages}{3540} (\bibinfo{year}{1988}).

\bibitem[{\citenamefont{Gupta et~al.}(1997)\citenamefont{Gupta, Bhattacharya,
  and Sharpe}}]{Gupta:1996yt}
\bibinfo{author}{\bibfnamefont{R.}~\bibnamefont{Gupta}},
  \bibinfo{author}{\bibfnamefont{T.}~\bibnamefont{Bhattacharya}},
  \bibnamefont{and} \bibinfo{author}{\bibfnamefont{S.~R.}
  \bibnamefont{Sharpe}}, \bibinfo{journal}{Phys. Rev.}
  \textbf{\bibinfo{volume}{D55}}, \bibinfo{pages}{4036} (\bibinfo{year}{1997}),
  \eprint{hep-lat/9611023}.

\bibitem[{\citenamefont{Christ and Liu}(2004)}]{Christ:2004gc}
\bibinfo{author}{\bibfnamefont{N.~H.} \bibnamefont{Christ}} \bibnamefont{and}
  \bibinfo{author}{\bibfnamefont{G.}~\bibnamefont{Liu}},
  \bibinfo{journal}{Nucl. Phys. Proc. Suppl.} \textbf{\bibinfo{volume}{129}},
  \bibinfo{pages}{272} (\bibinfo{year}{2004}).

\bibitem[{\citenamefont{Liu}(2003)}]{Liu:thesis}
\bibinfo{author}{\bibfnamefont{G.}~\bibnamefont{Liu}}, Ph.D. thesis,
  \bibinfo{school}{Columbia University} (\bibinfo{year}{2003}),
  \bibinfo{note}{uMI-31-04827}.

\bibitem[{\citenamefont{Yamada}(2004)}]{Yamada:2003bk}
\bibinfo{author}{\bibfnamefont{N.}~\bibnamefont{Yamada}}
  (\bibinfo{collaboration}{RBC}), \bibinfo{journal}{Nucl. Phys. Proc. Suppl.}
  \textbf{\bibinfo{volume}{129}}, \bibinfo{pages}{376} (\bibinfo{year}{2004}),
  \eprint{hep-lat/0311013}.

\bibitem[{\citenamefont{Ohta et~al.}(2006)\citenamefont{Ohta, Lin, and
  Yamada}}]{Ohta:2005cn}
\bibinfo{author}{\bibfnamefont{S.}~\bibnamefont{Ohta}},
  \bibinfo{author}{\bibfnamefont{H.}~\bibnamefont{Lin}}, \bibnamefont{and}
  \bibinfo{author}{\bibfnamefont{N.}~\bibnamefont{Yamada}}
  (\bibinfo{collaboration}{RBC}), \bibinfo{journal}{PoS}
  \textbf{\bibinfo{volume}{LAT2005}}, \bibinfo{pages}{096}
  (\bibinfo{year}{2006}), \eprint{hep-lat/0510071}.

\bibitem[{\citenamefont{Lin et~al.}(2006)\citenamefont{Lin, Ohta, and
  Yamada}}]{Lin:2006kg}
\bibinfo{author}{\bibfnamefont{H.-W.} \bibnamefont{Lin}},
  \bibinfo{author}{\bibfnamefont{S.}~\bibnamefont{Ohta}}, \bibnamefont{and}
  \bibinfo{author}{\bibfnamefont{N.}~\bibnamefont{Yamada}}
  (\bibinfo{collaboration}{RBC}), \bibinfo{journal}{Nucl. Phys. Proc. Suppl.}
  \textbf{\bibinfo{volume}{153}}, \bibinfo{pages}{199} (\bibinfo{year}{2006}).

\bibitem[{\citenamefont{Aoki et~al.}(2006)}]{Aoki:2005ga}
\bibinfo{author}{\bibfnamefont{Y.}~\bibnamefont{Aoki}} \bibnamefont{et~al.},
  \bibinfo{journal}{Phys. Rev.} \textbf{\bibinfo{volume}{D73}},
  \bibinfo{pages}{094507} (\bibinfo{year}{2006}), \eprint{hep-lat/0508011}.

\bibitem[{\citenamefont{Takaishi}(1996)}]{Takaishi:1996xj}
\bibinfo{author}{\bibfnamefont{T.}~\bibnamefont{Takaishi}},
  \bibinfo{journal}{Phys. Rev.} \textbf{\bibinfo{volume}{D54}},
  \bibinfo{pages}{1050} (\bibinfo{year}{1996}).

\bibitem[{\citenamefont{de~Forcrand et~al.}(2000)}]{deForcrand:1999bi}
\bibinfo{author}{\bibfnamefont{P.}~\bibnamefont{de~Forcrand}}
  \bibnamefont{et~al.} (\bibinfo{collaboration}{QCD-TARO}),
  \bibinfo{journal}{Nucl. Phys.} \textbf{\bibinfo{volume}{B577}},
  \bibinfo{pages}{263} (\bibinfo{year}{2000}), \eprint{hep-lat/9911033}.

\bibitem[{\citenamefont{Becirevic et~al.}(2004)\citenamefont{Becirevic, Fajfer,
  and Prelovsek}}]{Becirevic:2004uv}
\bibinfo{author}{\bibfnamefont{D.}~\bibnamefont{Becirevic}},
  \bibinfo{author}{\bibfnamefont{S.}~\bibnamefont{Fajfer}}, \bibnamefont{and}
  \bibinfo{author}{\bibfnamefont{S.}~\bibnamefont{Prelovsek}},
  \bibinfo{journal}{Phys. Lett.} \textbf{\bibinfo{volume}{B599}},
  \bibinfo{pages}{55} (\bibinfo{year}{2004}), \eprint{hep-ph/0406296}.

\bibitem[{\citenamefont{Boyle}(1998)}]{Boyle:1997rk}
\bibinfo{author}{\bibfnamefont{P.}~\bibnamefont{Boyle}}
  (\bibinfo{collaboration}{UKQCD}), \bibinfo{journal}{Nucl. Phys. Proc. Suppl.}
  \textbf{\bibinfo{volume}{63}}, \bibinfo{pages}{314} (\bibinfo{year}{1998}),
  \eprint{hep-lat/9710036}.

\bibitem[{\citenamefont{Hein et~al.}(2000)}]{Hein:2000qu}
\bibinfo{author}{\bibfnamefont{J.}~\bibnamefont{Hein}} \bibnamefont{et~al.},
  \bibinfo{journal}{Phys. Rev.} \textbf{\bibinfo{volume}{D62}},
  \bibinfo{pages}{074503} (\bibinfo{year}{2000}), \eprint{hep-ph/0003130}.

\bibitem[{\citenamefont{di~Pierro et~al.}(2004)}]{diPierro:2003iw}
\bibinfo{author}{\bibfnamefont{M.}~\bibnamefont{di~Pierro}}
  \bibnamefont{et~al.}, \bibinfo{journal}{Nucl. Phys. Proc. Suppl.}
  \textbf{\bibinfo{volume}{129}}, \bibinfo{pages}{328} (\bibinfo{year}{2004}),
  \eprint{hep-lat/0310045}.

\bibitem[{\citenamefont{Bali}(2003)}]{Bali:2003jv}
\bibinfo{author}{\bibfnamefont{G.~S.} \bibnamefont{Bali}},
  \bibinfo{journal}{Phys. Rev.} \textbf{\bibinfo{volume}{D68}},
  \bibinfo{pages}{071501} (\bibinfo{year}{2003}), \eprint{hep-ph/0305209}.

\bibitem[{\citenamefont{Dougall et~al.}(2003)\citenamefont{Dougall, Kenway,
  Maynard, and McNeile}}]{Dougall:2003hv}
\bibinfo{author}{\bibfnamefont{A.}~\bibnamefont{Dougall}},
  \bibinfo{author}{\bibfnamefont{R.~D.} \bibnamefont{Kenway}},
  \bibinfo{author}{\bibfnamefont{C.~M.} \bibnamefont{Maynard}},
  \bibnamefont{and} \bibinfo{author}{\bibfnamefont{C.}~\bibnamefont{McNeile}}
  (\bibinfo{collaboration}{UKQCD}), \bibinfo{journal}{Phys. Lett.}
  \textbf{\bibinfo{volume}{B569}}, \bibinfo{pages}{41} (\bibinfo{year}{2003}),
  \eprint{hep-lat/0307001}.

\bibitem[{\citenamefont{Green et~al.}(2004)\citenamefont{Green, Koponen,
  McNeile, Michael, and Thompson}}]{Green:2003zz}
\bibinfo{author}{\bibfnamefont{A.~M.} \bibnamefont{Green}},
  \bibinfo{author}{\bibfnamefont{J.}~\bibnamefont{Koponen}},
  \bibinfo{author}{\bibfnamefont{C.}~\bibnamefont{McNeile}},
  \bibinfo{author}{\bibfnamefont{C.}~\bibnamefont{Michael}}, \bibnamefont{and}
  \bibinfo{author}{\bibfnamefont{G.}~\bibnamefont{Thompson}}
  (\bibinfo{collaboration}{UKQCD}), \bibinfo{journal}{Phys. Rev.}
  \textbf{\bibinfo{volume}{D69}}, \bibinfo{pages}{094505}
  (\bibinfo{year}{2004}), \eprint{hep-lat/0312007}.

\bibitem[{\citenamefont{Bowler et~al.}(2001)}]{Bowler:2000xw}
\bibinfo{author}{\bibfnamefont{K.~C.} \bibnamefont{Bowler}}
  \bibnamefont{et~al.} (\bibinfo{collaboration}{UKQCD}),
  \bibinfo{journal}{Nucl. Phys.} \textbf{\bibinfo{volume}{B619}},
  \bibinfo{pages}{507} (\bibinfo{year}{2001}), \eprint{hep-lat/0007020}.

\bibitem[{\citenamefont{Yamada et~al.}(2005)\citenamefont{Yamada, Aoki, and
  Kuramashi}}]{Yamada:2004ri}
\bibinfo{author}{\bibfnamefont{N.}~\bibnamefont{Yamada}},
  \bibinfo{author}{\bibfnamefont{S.}~\bibnamefont{Aoki}}, \bibnamefont{and}
  \bibinfo{author}{\bibfnamefont{Y.}~\bibnamefont{Kuramashi}},
  \bibinfo{journal}{Nucl. Phys.} \textbf{\bibinfo{volume}{B713}},
  \bibinfo{pages}{407} (\bibinfo{year}{2005}), \eprint{hep-lat/0407031}.

\bibitem[{\citenamefont{Choe et~al.}(2003)}]{Choe:2003wx}
\bibinfo{author}{\bibfnamefont{S.}~\bibnamefont{Choe}} \bibnamefont{et~al.}
  (\bibinfo{collaboration}{QCD-TARO}), \bibinfo{journal}{JHEP}
  \textbf{\bibinfo{volume}{08}}, \bibinfo{pages}{022} (\bibinfo{year}{2003}),
  \eprint{hep-lat/0307004}.

\bibitem[{\citenamefont{Gottlieb et~al.}(2006)}]{Gottlieb:2005me}
\bibinfo{author}{\bibfnamefont{S.}~\bibnamefont{Gottlieb}}
  \bibnamefont{et~al.}, \bibinfo{journal}{PoS}
  \textbf{\bibinfo{volume}{LAT2005}}, \bibinfo{pages}{203}
  (\bibinfo{year}{2006}), \eprint{hep-lat/0510072}.

\bibitem[{\citenamefont{Brambilla et~al.}(2004)}]{Brambilla:2004wf}
\bibinfo{author}{\bibfnamefont{N.}~\bibnamefont{Brambilla}}
  \bibnamefont{et~al.} (\bibinfo{year}{2004}), \eprint{hep-ph/0412158}.

\bibitem[{\citenamefont{Rosner et~al.}(2005)}]{Rosner:2005ry}
\bibinfo{author}{\bibfnamefont{J.~L.} \bibnamefont{Rosner}}
  \bibnamefont{et~al.} (\bibinfo{collaboration}{CLEO}), \bibinfo{journal}{Phys.
  Rev. Lett.} \textbf{\bibinfo{volume}{95}}, \bibinfo{pages}{102003}
  (\bibinfo{year}{2005}), \eprint{hep-ex/0505073}.

\bibitem[{\citenamefont{Rubin et~al.}(2005)}]{Rubin:2005px}
\bibinfo{author}{\bibfnamefont{P.}~\bibnamefont{Rubin}} \bibnamefont{et~al.}
  (\bibinfo{collaboration}{CLEO}), \bibinfo{journal}{Phys. Rev.}
  \textbf{\bibinfo{volume}{D72}}, \bibinfo{pages}{092004}
  (\bibinfo{year}{2005}), \eprint{hep-ex/0508037}.

\bibitem[{\citenamefont{Andreotti et~al.}(2005)}]{Andreotti:2005vu}
\bibinfo{author}{\bibfnamefont{M.}~\bibnamefont{Andreotti}}
  \bibnamefont{et~al.}, \bibinfo{journal}{Phys. Rev.}
  \textbf{\bibinfo{volume}{D72}}, \bibinfo{pages}{032001}
  (\bibinfo{year}{2005}).

\bibitem[{\citenamefont{Fang}(2006)}]{Fang:2006bz}
\bibinfo{author}{\bibfnamefont{F.}~\bibnamefont{Fang}} (\bibinfo{year}{2006}),
  \eprint{hep-ex/0605007}.

\bibitem[{\citenamefont{Eidelman et~al.}(2004)}]{Eidelman:2004wy}
\bibinfo{author}{\bibfnamefont{S.}~\bibnamefont{Eidelman}} \bibnamefont{et~al.}
  (\bibinfo{collaboration}{Particle Data Group}), \bibinfo{journal}{Phys.
  Lett.} \textbf{\bibinfo{volume}{B592}}, \bibinfo{pages}{1}
  (\bibinfo{year}{2004}).

\bibitem[{\citenamefont{Martinelli et~al.}(1995)\citenamefont{Martinelli,
  Pittori, Sachrajda, Testa, and Vladikas}}]{Martinelli:1994ty}
\bibinfo{author}{\bibfnamefont{G.}~\bibnamefont{Martinelli}},
  \bibinfo{author}{\bibfnamefont{C.}~\bibnamefont{Pittori}},
  \bibinfo{author}{\bibfnamefont{C.~T.} \bibnamefont{Sachrajda}},
  \bibinfo{author}{\bibfnamefont{M.}~\bibnamefont{Testa}}, \bibnamefont{and}
  \bibinfo{author}{\bibfnamefont{A.}~\bibnamefont{Vladikas}},
  \bibinfo{journal}{Nucl. Phys.} \textbf{\bibinfo{volume}{B445}},
  \bibinfo{pages}{81} (\bibinfo{year}{1995}), \eprint{hep-lat/9411010}.

\bibitem[{\citenamefont{Dawson et~al.}(1998)}]{Dawson:1997ic}
\bibinfo{author}{\bibfnamefont{C.}~\bibnamefont{Dawson}} \bibnamefont{et~al.},
  \bibinfo{journal}{Nucl. Phys.} \textbf{\bibinfo{volume}{B514}},
  \bibinfo{pages}{313} (\bibinfo{year}{1998}), \eprint{hep-lat/9707009}.

\bibitem[{\citenamefont{Blum et~al.}(2002)}]{Blum:2001sr}
\bibinfo{author}{\bibfnamefont{T.}~\bibnamefont{Blum}} \bibnamefont{et~al.},
  \bibinfo{journal}{Phys. Rev.} \textbf{\bibinfo{volume}{D66}},
  \bibinfo{pages}{014504} (\bibinfo{year}{2002}), \eprint{hep-lat/0102005}.

\bibitem[{\citenamefont{Chetyrkin and Retey}(2000)}]{Chetyrkin:1999pq}
\bibinfo{author}{\bibfnamefont{K.~G.} \bibnamefont{Chetyrkin}}
  \bibnamefont{and} \bibinfo{author}{\bibfnamefont{A.}~\bibnamefont{Retey}},
  \bibinfo{journal}{Nucl. Phys.} \textbf{\bibinfo{volume}{B583}},
  \bibinfo{pages}{3} (\bibinfo{year}{2000}), \eprint{hep-ph/9910332}.

\bibitem[{\citenamefont{Aoki et~al.}(2003{\natexlab{a}})\citenamefont{Aoki,
  Izubuchi, Kuramashi, and Taniguchi}}]{Aoki:2002iq}
\bibinfo{author}{\bibfnamefont{S.}~\bibnamefont{Aoki}},
  \bibinfo{author}{\bibfnamefont{T.}~\bibnamefont{Izubuchi}},
  \bibinfo{author}{\bibfnamefont{Y.}~\bibnamefont{Kuramashi}},
  \bibnamefont{and}
  \bibinfo{author}{\bibfnamefont{Y.}~\bibnamefont{Taniguchi}},
  \bibinfo{journal}{Phys. Rev.} \textbf{\bibinfo{volume}{D67}},
  \bibinfo{pages}{094502} (\bibinfo{year}{2003}{\natexlab{a}}),
  \eprint{hep-lat/0206013}.

\bibitem[{\citenamefont{Chetyrkin}(1997)}]{Chetyrkin:1997dh}
\bibinfo{author}{\bibfnamefont{K.~G.} \bibnamefont{Chetyrkin}},
  \bibinfo{journal}{Phys. Lett.} \textbf{\bibinfo{volume}{B404}},
  \bibinfo{pages}{161} (\bibinfo{year}{1997}), \eprint{hep-ph/9703278}.

\bibitem[{\citenamefont{Chetyrkin et~al.}(2000)\citenamefont{Chetyrkin, Kuhn,
  and Steinhauser}}]{Chetyrkin:2000yt}
\bibinfo{author}{\bibfnamefont{K.~G.} \bibnamefont{Chetyrkin}},
  \bibinfo{author}{\bibfnamefont{J.~H.} \bibnamefont{Kuhn}}, \bibnamefont{and}
  \bibinfo{author}{\bibfnamefont{M.}~\bibnamefont{Steinhauser}},
  \bibinfo{journal}{Comput. Phys. Commun.} \textbf{\bibinfo{volume}{133}},
  \bibinfo{pages}{43} (\bibinfo{year}{2000}), \eprint{hep-ph/0004189}.

\bibitem[{\citenamefont{Capitani et~al.}(1999)\citenamefont{Capitani, Luscher,
  Sommer, and Wittig}}]{Capitani:1998mq}
\bibinfo{author}{\bibfnamefont{S.}~\bibnamefont{Capitani}},
  \bibinfo{author}{\bibfnamefont{M.}~\bibnamefont{Luscher}},
  \bibinfo{author}{\bibfnamefont{R.}~\bibnamefont{Sommer}}, \bibnamefont{and}
  \bibinfo{author}{\bibfnamefont{H.}~\bibnamefont{Wittig}}
  (\bibinfo{collaboration}{ALPHA}), \bibinfo{journal}{Nucl. Phys.}
  \textbf{\bibinfo{volume}{B544}}, \bibinfo{pages}{669} (\bibinfo{year}{1999}),
  \eprint{hep-lat/9810063}.

\bibitem[{\citenamefont{Vermaseren et~al.}(1997)\citenamefont{Vermaseren,
  Larin, and van Ritbergen}}]{Vermaseren:1997fq}
\bibinfo{author}{\bibfnamefont{J.~A.~M.} \bibnamefont{Vermaseren}},
  \bibinfo{author}{\bibfnamefont{S.~A.} \bibnamefont{Larin}}, \bibnamefont{and}
  \bibinfo{author}{\bibfnamefont{T.}~\bibnamefont{van Ritbergen}},
  \bibinfo{journal}{Phys. Lett.} \textbf{\bibinfo{volume}{B405}},
  \bibinfo{pages}{327} (\bibinfo{year}{1997}), \eprint{hep-ph/9703284}.

\bibitem[{\citenamefont{Dougall et~al.}(2006)\citenamefont{Dougall, Maynard,
  and McNeile}}]{Dougall:2005ev}
\bibinfo{author}{\bibfnamefont{A.}~\bibnamefont{Dougall}},
  \bibinfo{author}{\bibfnamefont{C.~M.} \bibnamefont{Maynard}},
  \bibnamefont{and} \bibinfo{author}{\bibfnamefont{C.}~\bibnamefont{McNeile}},
  \bibinfo{journal}{JHEP} \textbf{\bibinfo{volume}{01}}, \bibinfo{pages}{171}
  (\bibinfo{year}{2006}), \eprint{hep-lat/0508033}.

\bibitem[{\citenamefont{Booth}(1995)}]{Booth:1994hx}
\bibinfo{author}{\bibfnamefont{M.~J.} \bibnamefont{Booth}},
  \bibinfo{journal}{Phys. Rev.} \textbf{\bibinfo{volume}{D51}},
  \bibinfo{pages}{2338} (\bibinfo{year}{1995}), \eprint{hep-ph/9411433}.

\bibitem[{\citenamefont{Sharpe and Zhang}(1996)}]{Sharpe:1995qp}
\bibinfo{author}{\bibfnamefont{S.~R.} \bibnamefont{Sharpe}} \bibnamefont{and}
  \bibinfo{author}{\bibfnamefont{Y.}~\bibnamefont{Zhang}},
  \bibinfo{journal}{Phys. Rev.} \textbf{\bibinfo{volume}{D53}},
  \bibinfo{pages}{5125} (\bibinfo{year}{1996}), \eprint{hep-lat/9510037}.

\bibitem[{\citenamefont{McNeile and Michael}(2005)}]{McNeile:2004wn}
\bibinfo{author}{\bibfnamefont{C.}~\bibnamefont{McNeile}} \bibnamefont{and}
  \bibinfo{author}{\bibfnamefont{C.}~\bibnamefont{Michael}}
  (\bibinfo{collaboration}{UKQCD}), \bibinfo{journal}{JHEP}
  \textbf{\bibinfo{volume}{01}}, \bibinfo{pages}{011} (\bibinfo{year}{2005}),
  \eprint{hep-lat/0411014}.

\bibitem[{\citenamefont{Artuso et~al.}(2005)}]{Artuso:2005ym}
\bibinfo{author}{\bibfnamefont{M.}~\bibnamefont{Artuso}} \bibnamefont{et~al.}
  (\bibinfo{collaboration}{CLEO}), \bibinfo{journal}{Phys. Rev. Lett.}
  \textbf{\bibinfo{volume}{95}}, \bibinfo{pages}{251801}
  (\bibinfo{year}{2005}), \eprint{hep-ex/0508057}.

\bibitem[{\citenamefont{Aubin et~al.}(2005)}]{Aubin:2005ar}
\bibinfo{author}{\bibfnamefont{C.}~\bibnamefont{Aubin}} \bibnamefont{et~al.},
  \bibinfo{journal}{Phys. Rev. Lett.} \textbf{\bibinfo{volume}{95}},
  \bibinfo{pages}{122002} (\bibinfo{year}{2005}), \eprint{hep-lat/0506030}.

\bibitem[{\citenamefont{Burdman and Shipsey}(2003)}]{Burdman:2003rs}
\bibinfo{author}{\bibfnamefont{G.}~\bibnamefont{Burdman}} \bibnamefont{and}
  \bibinfo{author}{\bibfnamefont{I.}~\bibnamefont{Shipsey}},
  \bibinfo{journal}{Ann. Rev. Nucl. Part. Sci.} \textbf{\bibinfo{volume}{53}},
  \bibinfo{pages}{431} (\bibinfo{year}{2003}), \eprint{hep-ph/0310076}.

\bibitem[{\citenamefont{Stewart}(1998)}]{Stewart:1998ke}
\bibinfo{author}{\bibfnamefont{I.~W.} \bibnamefont{Stewart}},
  \bibinfo{journal}{Nucl. Phys.} \textbf{\bibinfo{volume}{B529}},
  \bibinfo{pages}{62} (\bibinfo{year}{1998}), \eprint{hep-ph/9803227}.

\bibitem[{\citenamefont{Ciuchini et~al.}(1995)\citenamefont{Ciuchini, Franco,
  Martinelli, Reina, and Silvestrini}}]{Ciuchini:1995cd}
\bibinfo{author}{\bibfnamefont{M.}~\bibnamefont{Ciuchini}},
  \bibinfo{author}{\bibfnamefont{E.}~\bibnamefont{Franco}},
  \bibinfo{author}{\bibfnamefont{G.}~\bibnamefont{Martinelli}},
  \bibinfo{author}{\bibfnamefont{L.}~\bibnamefont{Reina}}, \bibnamefont{and}
  \bibinfo{author}{\bibfnamefont{L.}~\bibnamefont{Silvestrini}},
  \bibinfo{journal}{Z. Phys.} \textbf{\bibinfo{volume}{C68}},
  \bibinfo{pages}{239} (\bibinfo{year}{1995}), \eprint{hep-ph/9501265}.

\bibitem[{\citenamefont{Ciuchini et~al.}(1998)}]{Ciuchini:1997bw}
\bibinfo{author}{\bibfnamefont{M.}~\bibnamefont{Ciuchini}}
  \bibnamefont{et~al.}, \bibinfo{journal}{Nucl. Phys.}
  \textbf{\bibinfo{volume}{B523}}, \bibinfo{pages}{501} (\bibinfo{year}{1998}),
  \eprint{hep-ph/9711402}.

\bibitem[{\citenamefont{Bernard et~al.}(1998)\citenamefont{Bernard, Blum, and
  Soni}}]{Bernard:1998dg}
\bibinfo{author}{\bibfnamefont{C.~W.} \bibnamefont{Bernard}},
  \bibinfo{author}{\bibfnamefont{T.}~\bibnamefont{Blum}}, \bibnamefont{and}
  \bibinfo{author}{\bibfnamefont{A.}~\bibnamefont{Soni}},
  \bibinfo{journal}{Phys. Rev.} \textbf{\bibinfo{volume}{D58}},
  \bibinfo{pages}{014501} (\bibinfo{year}{1998}), \eprint{hep-lat/9801039}.

\bibitem[{\citenamefont{Yamada}(2003)}]{Yamada:2002wh}
\bibinfo{author}{\bibfnamefont{N.}~\bibnamefont{Yamada}},
  \bibinfo{journal}{Nucl. Phys. Proc. Suppl.} \textbf{\bibinfo{volume}{119}},
  \bibinfo{pages}{93} (\bibinfo{year}{2003}), \eprint{hep-lat/0210035}.

\bibitem[{\citenamefont{Hashimoto et~al.}(1999)}]{Hashimoto:1999ck}
\bibinfo{author}{\bibfnamefont{S.}~\bibnamefont{Hashimoto}}
  \bibnamefont{et~al.}, \bibinfo{journal}{Phys. Rev.}
  \textbf{\bibinfo{volume}{D60}}, \bibinfo{pages}{094503}
  (\bibinfo{year}{1999}), \eprint{hep-lat/9903002}.

\bibitem[{\citenamefont{Heitger et~al.}(2003)\citenamefont{Heitger, Kurth, and
  Sommer}}]{Heitger:2003xg}
\bibinfo{author}{\bibfnamefont{J.}~\bibnamefont{Heitger}},
  \bibinfo{author}{\bibfnamefont{M.}~\bibnamefont{Kurth}}, \bibnamefont{and}
  \bibinfo{author}{\bibfnamefont{R.}~\bibnamefont{Sommer}}
  (\bibinfo{collaboration}{ALPHA}), \bibinfo{journal}{Nucl. Phys.}
  \textbf{\bibinfo{volume}{B669}}, \bibinfo{pages}{173} (\bibinfo{year}{2003}),
  \eprint{hep-lat/0302019}.

\bibitem[{\citenamefont{Gadiyak and Loktik}(2005)}]{Gadiyak:2005ea}
\bibinfo{author}{\bibfnamefont{V.}~\bibnamefont{Gadiyak}} \bibnamefont{and}
  \bibinfo{author}{\bibfnamefont{O.}~\bibnamefont{Loktik}},
  \bibinfo{journal}{Phys. Rev.} \textbf{\bibinfo{volume}{D72}},
  \bibinfo{pages}{114504} (\bibinfo{year}{2005}), \eprint{hep-lat/0509075}.

\bibitem[{\citenamefont{Gray et~al.}(2005)}]{Gray:2005ad}
\bibinfo{author}{\bibfnamefont{A.}~\bibnamefont{Gray}} \bibnamefont{et~al.}
  (\bibinfo{collaboration}{HPQCD}), \bibinfo{journal}{Phys. Rev. Lett.}
  \textbf{\bibinfo{volume}{95}}, \bibinfo{pages}{212001}
  (\bibinfo{year}{2005}), \eprint{hep-lat/0507015}.

\bibitem[{\citenamefont{Christensen et~al.}(1997)\citenamefont{Christensen,
  Draper, and McNeile}}]{Christensen:1996sj}
\bibinfo{author}{\bibfnamefont{J.~C.} \bibnamefont{Christensen}},
  \bibinfo{author}{\bibfnamefont{T.}~\bibnamefont{Draper}}, \bibnamefont{and}
  \bibinfo{author}{\bibfnamefont{C.}~\bibnamefont{McNeile}},
  \bibinfo{journal}{Phys. Rev.} \textbf{\bibinfo{volume}{D56}},
  \bibinfo{pages}{6993} (\bibinfo{year}{1997}), \eprint{hep-lat/9610026}.

\bibitem[{\citenamefont{Becirevic et~al.}(2006)}]{Becirevic:2005sx}
\bibinfo{author}{\bibfnamefont{D.}~\bibnamefont{Becirevic}}
  \bibnamefont{et~al.}, \bibinfo{journal}{PoS}
  \textbf{\bibinfo{volume}{LAT2005}}, \bibinfo{pages}{218}
  (\bibinfo{year}{2006}), \eprint{hep-lat/0509165}.

\bibitem[{\citenamefont{Aoki et~al.}(2003{\natexlab{b}})}]{Aoki:2002bh}
\bibinfo{author}{\bibfnamefont{S.}~\bibnamefont{Aoki}} \bibnamefont{et~al.}
  (\bibinfo{collaboration}{JLQCD}), \bibinfo{journal}{Phys. Rev.}
  \textbf{\bibinfo{volume}{D67}}, \bibinfo{pages}{014506}
  (\bibinfo{year}{2003}{\natexlab{b}}), \eprint{hep-lat/0208038}.

\bibitem[{\citenamefont{Aoki et~al.}(2003{\natexlab{c}})}]{Aoki:2003xb}
\bibinfo{author}{\bibfnamefont{S.}~\bibnamefont{Aoki}} \bibnamefont{et~al.}
  (\bibinfo{collaboration}{JLQCD}), \bibinfo{journal}{Phys. Rev. Lett.}
  \textbf{\bibinfo{volume}{91}}, \bibinfo{pages}{212001}
  (\bibinfo{year}{2003}{\natexlab{c}}), \eprint{hep-ph/0307039}.

\bibitem[{\citenamefont{Okamoto}(2006)}]{Okamoto:2005zg}
\bibinfo{author}{\bibfnamefont{M.}~\bibnamefont{Okamoto}},
  \bibinfo{journal}{PoS} \textbf{\bibinfo{volume}{LAT2005}},
  \bibinfo{pages}{013} (\bibinfo{year}{2006}), \eprint{hep-lat/0510113}.

\bibitem[{\citenamefont{Hashimoto}(2005)}]{Hashimoto:2004hn}
\bibinfo{author}{\bibfnamefont{S.}~\bibnamefont{Hashimoto}},
  \bibinfo{journal}{Int. J. Mod. Phys.} \textbf{\bibinfo{volume}{A20}},
  \bibinfo{pages}{5133} (\bibinfo{year}{2005}), \eprint{hep-ph/0411126}.

\bibitem[{\citenamefont{Kronfeld}(1998)}]{Kronfeld:1997zc}
\bibinfo{author}{\bibfnamefont{A.~S.} \bibnamefont{Kronfeld}},
  \bibinfo{journal}{Nucl. Phys. Proc. Suppl.} \textbf{\bibinfo{volume}{63}},
  \bibinfo{pages}{311} (\bibinfo{year}{1998}), \eprint{hep-lat/9710007}.

\bibitem[{\citenamefont{Hornbostel}(1999)}]{Hornbostel:1998ki}
\bibinfo{author}{\bibfnamefont{K.}~\bibnamefont{Hornbostel}}
  (\bibinfo{collaboration}{NRQCD}), \bibinfo{journal}{Nucl. Phys. Proc. Suppl.}
  \textbf{\bibinfo{volume}{73}}, \bibinfo{pages}{339} (\bibinfo{year}{1999}),
  \eprint{hep-lat/9809177}.

\bibitem[{\citenamefont{Becirevic et~al.}(2002)\citenamefont{Becirevic, Lubicz,
  and Martinelli}}]{Becirevic:2001yh}
\bibinfo{author}{\bibfnamefont{D.}~\bibnamefont{Becirevic}},
  \bibinfo{author}{\bibfnamefont{V.}~\bibnamefont{Lubicz}}, \bibnamefont{and}
  \bibinfo{author}{\bibfnamefont{G.}~\bibnamefont{Martinelli}},
  \bibinfo{journal}{Phys. Lett.} \textbf{\bibinfo{volume}{B524}},
  \bibinfo{pages}{115} (\bibinfo{year}{2002}), \eprint{hep-ph/0107124}.

\bibitem[{\citenamefont{Juge}(2002)}]{Juge:2001dj}
\bibinfo{author}{\bibfnamefont{K.~J.} \bibnamefont{Juge}},
  \bibinfo{journal}{Nucl. Phys. Proc. Suppl.} \textbf{\bibinfo{volume}{106}},
  \bibinfo{pages}{847} (\bibinfo{year}{2002}), \eprint{hep-lat/0110131}.

\bibitem[{\citenamefont{Rolf and Sint}(2002)}]{Rolf:2002gu}
\bibinfo{author}{\bibfnamefont{J.}~\bibnamefont{Rolf}} \bibnamefont{and}
  \bibinfo{author}{\bibfnamefont{S.}~\bibnamefont{Sint}}
  (\bibinfo{collaboration}{ALPHA}), \bibinfo{journal}{JHEP}
  \textbf{\bibinfo{volume}{12}}, \bibinfo{pages}{007} (\bibinfo{year}{2002}),
  \eprint{hep-ph/0209255}.

\bibitem[{\citenamefont{Nobes and Trottier}(2006)}]{Nobes:2005dz}
\bibinfo{author}{\bibfnamefont{M.}~\bibnamefont{Nobes}} \bibnamefont{and}
  \bibinfo{author}{\bibfnamefont{H.}~\bibnamefont{Trottier}},
  \bibinfo{journal}{PoS} \textbf{\bibinfo{volume}{LAT2005}},
  \bibinfo{pages}{209} (\bibinfo{year}{2006}), \eprint{hep-lat/0509128}.

\end{thebibliography}
\begin{table}
\caption{Ensemble parameters}
\label{tab:params}
\begin{tabular}{c| c}
Quantities &  numbers\\
 \hline\hline
\# of conf. & 103 \\
$a^{-1}$ & 2.914(54)~GeV \\
$L_s$ & 10 \\
$M_5$ & 1.65 \\
$am_{\rm res}$ & $0.9722(27)\times10^{-4}$ \\
$am_{\rm light}$ & 0.008, 0.016, 0.024, 0.032, 0.040 \\
$am_{\rm heavy}$ & 0.1, 0.2, 0.3, 0.4, 0.5\\
\hline
\end{tabular}
\end{table}
\begin{table}
\caption{Meson states in this study are created by local operators
of the form $\bar{\psi}\Gamma\psi$ in this table.  Their spin and
parity are also listed, as well as charge conjugation for
quarkonium cases.} \label{tab:meson_op}
\begin{center}
\begin{tabular}{ccc}
\hline\hline
          $\Gamma$              & $^{2S+1}L_J$   & $J^{PC}$  \\
          \hline
          $\gamma_5$            & $^1S_0$        & $0^{-+}$  \\
          $\gamma_i$            & $^3S_1$        & $1^{--}$  \\
          $1$                   & $^3P_0$        & $0^{++}$  \\
          $\gamma_5\gamma_i$    & $^3P_1$        & $1^{++}$  \\
          $\gamma_i\gamma_j$    &                & $1^{+-}$  \\
\hline\hline
\end{tabular}
\end{center}
\end{table}
\begin{table}
\caption{Heavy-light \(J^{P}=0^{-}\) meson mass obtained from fitting to
 the hyperbolic cosine form in the range of \(15\le t \le 23\).}
\label{tab:HLPSmass}
\begin{tabular}{llll}
\multicolumn{1}{c}{$m_{\rm heavy}$}&
\multicolumn{1}{c}{$m_{\rm light}$}&
\multicolumn{1}{c}{$m_{\rm PS}$}&
\multicolumn{1}{c}{$\chi^{2}$ per d.o.f.}\\
\hline\hline
     & 0.008& 0.3254(15)& 0.22\\
     & 0.016& 0.3373(13)& 0.15\\
0.1& 0.024& 0.3495(12)& 0.09\\
     & 0.032& 0.3617(11)& 0.08\\
     & 0.040& 0.3739(11)& 0.07\\
     \hline
     & 0.008& 0.4715(17)& 0.36\\
     & 0.016& 0.4812(14)& 0.30\\
0.2& 0.024& 0.4913(13)& 0.23\\
     & 0.032& 0.5016(12)& 0.19\\
     & 0.040& 0.5120(11)& 0.18\\
     \hline
     & 0.008& 0.5916(19)& 0.51\\
     & 0.016& 0.6003(15)& 0.46\\
0.3& 0.024& 0.6095(13)& 0.37\\
     & 0.032& 0.6189(12)& 0.32\\
     & 0.040& 0.6285(11)& 0.31\\
     \hline
     & 0.008& 0.6917(21)& 0.71\\
     & 0.016& 0.7000(16)& 0.70\\
0.4& 0.024& 0.7086(14)& 0.59\\
     & 0.032& 0.7175(13)& 0.52\\
     & 0.040& 0.7266(12)& 0.49\\
     \hline
     & 0.008& 0.7734(22)& 0.84\\
     & 0.016& 0.7815(17)& 0.92\\
0.5& 0.024& 0.7898(15)& 0.84\\
     & 0.032& 0.7984(14)& 0.78\\
     & 0.040& 0.8072(13)& 0.75\\
     \hline
\end{tabular}
\end{table}
\begin{table}
\caption{Heavy-light \(J^{P}=1^{-}\) meson mass obtained from fitting to
 the hyperbolic cosine form in the range of \(15\le t \le 23\).}
\label{tab:HLVmass}
\begin{tabular}{llll}
\multicolumn{1}{c}{$m_{\rm heavy}$}&
\multicolumn{1}{c}{$m_{\rm light}$}&
\multicolumn{1}{c}{$m_{\rm PS}$}&
\multicolumn{1}{c}{$\chi^{2}$ per d.o.f.}\\
\hline\hline
     & 0.008& 0.4083(45)& 0.55\\
     & 0.016& 0.4153(35)& 0.40\\
0.1& 0.024& 0.4223(30)& 0.27\\
     & 0.032& 0.4310(26)& 0.20\\
     & 0.040& 0.4401(24)& 0.17\\
     \hline
     & 0.008& 0.5259(37)& 0.65\\
     & 0.016& 0.5324(28)& 0.52\\
0.2& 0.024& 0.5400(24)& 0.36\\
     & 0.032& 0.5484(21)& 0.27\\
     & 0.040& 0.5572(19)& 0.23\\
     \hline
     & 0.008& 0.6304(34)& 0.63\\
     & 0.016& 0.6373(25)& 0.56\\
0.3& 0.024& 0.6450(21)& 0.41\\
     & 0.032& 0.6532(19)& 0.33\\
     & 0.040& 0.6618(17)& 0.29\\
     \hline
     & 0.008& 0.7203(32)& 0.59\\
     & 0.016& 0.7275(24)& 0.57\\
0.4& 0.024& 0.7351(20)& 0.44\\
     & 0.032& 0.7433(18)& 0.36\\
     & 0.040& 0.7518(16)& 0.33\\
     \hline
     & 0.008& 0.7943(31)& 0.64\\
     & 0.016& 0.8017(23)& 0.68\\
0.5& 0.024& 0.8095(20)& 0.57\\
     & 0.032& 0.8176(18)& 0.50\\
     & 0.040& 0.8260(16)& 0.47\\
     \hline
\end{tabular}
\end{table}
\begin{table}
\caption{Heavy-light \(J^{P}=0^{+}\) meson mass obtained from fitting to
 the hyperbolic cosine form in the range of \(10\le t \le 16\).}
\label{tab:HLSmass}
\begin{tabular}{llll}
\multicolumn{1}{c}{$m_{\rm heavy}$}&
\multicolumn{1}{c}{$m_{\rm light}$}&
\multicolumn{1}{c}{$m_{\rm PS}$}&
\multicolumn{1}{c}{$\chi^{2}$ per d.o.f.}\\
\hline\hline
     & 0.024& 0.656(16)& 0.01\\
0.2& 0.032& 0.654(17)& 0.01\\
     & 0.040& 0.658(10)& 0.01\\
     \hline
     & 0.024& 0.762(13)& 0.02\\
0.3& 0.032& 0.760(10)& 0.02\\
     & 0.040& 0.764(8)& 0.02\\
     \hline
     & 0.024& 0.854(14)& 0.01\\
0.4& 0.032& 0.853(11)& 0.01\\
     & 0.040& 0.857(9)& 0.01\\
     \hline
     & 0.024& 0.934(14)& 0.01\\
0.5& 0.032& 0.932(10)& 0.01\\
     & 0.040& 0.936(8)& 0.01\\
     \hline
\end{tabular}
\end{table}
\begin{table}
\caption{Heavy-light \(J^{P}=1^{+}\) meson mass obtained from fitting to
 the hyperbolic cosine form in the range of \(10\le t \le 16\).}
\label{tab:HLAmass}
\begin{tabular}{llll}
\multicolumn{1}{c}{$m_{\rm heavy}$}&
\multicolumn{1}{c}{$m_{\rm light}$}&
\multicolumn{1}{c}{$m_{\rm PS}$}&
\multicolumn{1}{c}{$\chi^{2}$ per d.o.f.}\\
\hline\hline
     & 0.024& 0.689(17)& 0.004\\
0.2& 0.032& 0.688(12)& 0.002\\
     & 0.040& 0.692(10)& 0.002\\
     \hline
     & 0.024& 0.792(16)& 0.011\\
0.3& 0.032& 0.790(12)& 0.006\\
     & 0.040& 0.793(10)& 0.005\\
     \hline
     & 0.024& 0.881(16)& 0.017\\
0.4& 0.032& 0.879(12)& 0.010\\
     & 0.040& 0.881(9)& 0.008\\
     \hline
     & 0.024& 0.955(16)& 0.02\\
0.5& 0.032& 0.952(12)& 0.01\\
     & 0.040& 0.955(9)& 0.01\\
     \hline
\end{tabular}
\end{table}
\begin{table}[hbt]
\caption{Summary of mass splitting results (in MeV).}
\label{tab:splitting}
\begin{tabular}{llr@{}l}
 & \multicolumn{1}{c}{Experiment} & \multicolumn{2}{c}{This work} \\
\hline\hline%
 $\Delta_{0}$ & 444(36) & 533&(90)\\
 $\Delta_{1}$ & 420(36)& 452&(78)\\
$1^{-}-0^{-}$& 140.64(10)& 93&(4) \\
\hline %
 $\Delta_{S0}$ & 348.4(9)& 411&(40)\\
 $\Delta_{S1}$ & 345.9(1.2)& 380&(37)\\
$1^{-}-0^{-}$& 143.8(4) & 82&(2)\\
\hline %
\end{tabular}
\end{table}
\begin{table}[hbt]
\caption{Summary of the heavy-light spectrum (in MeV).}
\label{tab:spectrum}
\begin{tabular}{lll}
\multicolumn{1}{c}{Meson ($J^{P}$)} &
\multicolumn{1}{c}{Experiment} &
\multicolumn{1}{c}{This work} \\
\hline\hline
$D^{\pm}(0^{-})$&1869.4(5)& 1867(4)\\
$D^{*\pm}(1^{-})$&2010.0(5)& 1961(4) \\
$D^{*}_{0}(0^{+})$&2352(50)& 2401(89)\\
$D'_{1}(1^{+})$&2427(26)(25)& 2413(76)\\
\hline %
$D^{\pm}_{s}(0^{-})$&1968.3(5)& $-$\\
$D^{*\pm}_{s}(1^{-})$&2112.1(7)&2051(2)\\
$D^{*\pm}_{s0}(0^{+})$&2317.4(9)&2379(40)\\
$D^{*\pm}_{s1}(1^{+})$&2459.3(1.3)&2431(37)\\
\hline %
\end{tabular}
\end{table}
\begin{table}
\caption{A list of the past charm quark mass estimations from quenched lattice QCD
with continuum extrapolation.}
\label{tab:charmmass}
\begin{center}
\begin{tabular}{clc}
Group &  \multicolumn{1}{c}{$m_{c}^{\overline{\rm {MS}}}(m_{ c})$~GeV} & action \\
\hline\hline%
Kronfeld~\cite{Kronfeld:1997zc}            &  1.33(8)     & Clover\\
Hornbostel et al.~\cite{Hornbostel:1998ki} & 1.20(4)(11)(2)& NRQCD   \\
Becirevic  et al. ~\cite{Becirevic:2001yh} &  1.26(4)(12) & Clover \\
Juge~\cite{Juge:2001dj}                    &  1.27(5)     & Clover\\
Rolf and Sint~\cite{Rolf:2002gu}           &  1.301(34)   & Clover \\
de Divitiis et al.~\cite{deDivitiis:2003iy} & 1.319(28)    & NRQCD \\
Nobes et al.~\cite{Nobes:2005dz}           & 1.22(9)   & Fermilab\\
This work                                  & 1.24(1)(18)   & DWF\\
\hline
\end{tabular}
\end{center}
\end{table}
\begin{table}[hbt]
\caption{\label{tab:Row HHspectrum}Summary of quorkonium spectrum in
 lattice unit.}
\begin{tabular}{llllll}
\multicolumn{1}{c}{$am_{\rm heavy}$} &
\multicolumn{1}{c}{$0^{-+}$} &
\multicolumn{1}{c}{$1^{--}$} &
\multicolumn{1}{c}{$0^{++}$} &
\multicolumn{1}{c}{$1^{++}$} &
\multicolumn{1}{c}{$1^{+-}$} \\
\hline\hline
0.1 &0.4614(12) &0.5113(24) &0.604(11) & 0.632(9) &0.656(17) \\
0.2 &0.7118(10) &0.7392(14) &0.824(7) &0.848(6) &0.858(8) \\
0.3 &0.9226(8) &0.9406(11) &1.026(6) &1.046(5) &1.053(7) \\
0.4 &1.1008(8) &1.1129(9) &1.200(6) &1.218(6) &1.227(6) \\
0.5 &1.2467(7) &1.2541(8) &1.350(6) &1.364(7) &1.377(5)\\
\hline %
\end{tabular}
\end{table}
\begin{table}[hbt]
\caption{\label{tab:HHspectrum}Summary of the charmonium spectrum
(in  MeV).  The last row presents a separate estimate on the mass
difference between \(h_c\) and \(\chi_{c1}\) obtained directly
from themesons propagator ratio: it is likely more reliable than
the calculated excited mesons mass themselves which are
underestimated.}
\begin{tabular}{clr}
Charmonium ($J^{PC}$) & Experiment & This work \\
\hline\hline
$\eta_c (0^{-+})$&2980(1)& 2987(12)\\
$J/\psi (1^{--})$&3096.916(11)& 3030(11) \\
$\chi_{c0} (0^{++})$&3415.19(34)& 3282(21)\\
$\chi_{c1} (1^{++})$&3510.59(12)& 3336(21)\\
$h_{c} (1^{+-})$& not established &  3360(21)\\
\hline
$h_c-\chi_{c1}$& & 22(11)\\
\hline %
\end{tabular}
\end{table}
\begin{table}[hbt]
\caption{\label{tab:decayFit} Fit expressions and various
decay constant related values.} {\footnotesize   
\begin{tabular}{ccccccc}
 fit & $a^{3/2}\sqrt{m_{D}}f_{\rm D}$
     & $a^{3/2}\sqrt{m_{D_{\rm  s}}}f_{D_{\rm  s}}$
     & $\frac{\sqrt{m_{D_{\rm  s}}}f_{D_{\rm  s}}}{\sqrt{m_{D}}f_{D}}$
     & $(a^{3/2}\sqrt{m_{B}}f_{B})^{\rm stat}$
     & $(a^{3/2}\sqrt{m_{B_{\rm  s}}}f_{B_{\rm  s}})^{\rm stat}$
     & $(\frac{\sqrt{m_{B_{\rm  s}}}f_{B_{\rm  s}}}{\sqrt{m_{B}}f_{B}})^{\rm stat}$ \\
\hline\hline%
$a + b m_q  $            & 0.0621(18) & 0.0677(13) & 1.090(19) & 0.0718(31) & 0.0773(25) & 1.073(56)\\
$a + b m_q  + c m_q ^2$     & 0.0621(18) & 0.0688(13) & 1.107(19) & 0.0718(31) & 0.0783(21) & 1.085(37)\\
$a + b m_q  + c m_q  \ln m_q $ & 0.0630(21) & 0.0688(13) & 1.092(26) & 0.0725(39) & 0.0783(21) & 1.076(53) \\
\hline %
\end{tabular}
}
\end{table}
\begin{table}[hbt]
\caption{\label{tab:Bps} List of the $B_{\rm PS}^{\rm lat}$
values.}
\begin{tabular}{ccr@{}lc}
 $am_{\rm heavy}$& $am_{\rm light}$ & \multicolumn{2}{c}{$B_{\rm PS}^{\rm lat}$} & $am_{\rm  PS}$ \\
\hline\hline%
0.2 & 0.016 & 0.792&(10) & 0.4839(14)\\
    & 0.024 & 0.792&(7)  & 0.4940(12)\\
    & 0.032 & 0.795&(6)  & 0.5042(11)\\
    & 0.040 & 0.799&(5)  & 0.5146(11)\\
\hline %
0.3 & 0.016 & 0.830&(11) & 0.6033(15)\\
    & 0.024 & 0.828&(8)  & 0.6124(13)\\
    & 0.032 & 0.830&(6)  & 0.6218(11)\\
    & 0.040 & 0.834&(6)  & 0.6314(11)\\
\hline %
0.4 & 0.016 & 0.855&(12) & 0.7027(15)\\
    & 0.024 & 0.853&(9)  & 0.7113(13)\\
    & 0.032 & 0.855&(7)  & 0.7202(12)\\
    & 0.040 & 0.858&(6)  & 0.7293(11)\\
\hline %
0.5 & 0.016 & 0.874&(13) & 0.7865(23)\\
    & 0.024 & 0.871&(9)  & 0.7943(20)\\
    & 0.032 & 0.873&(7)  & 0.8024(18)\\
    & 0.040 & 0.876&(6)  & 0.8110(17)\\
\hline %
\end{tabular}
\end{table}
\begin{table}[hbt]
\caption{\label{tab:BpsFit} Fit expressions and and various bag
parameter values.}
\begin{tabular}{ccccccc}
 fit & $B_D$ & $B_{D_{\rm  s}}$ & $B_{D_{\rm  s}}/B_D$ & $(B_B)^{\rm static}$ & $(B_{B_{\rm  s}})^{\rm static}$ & $(B_{B_{\rm  s}}/B_B)^{\rm static}$ \\
\hline\hline%
$a + b m_q       $     & 0.845(16) & 0.849(8) & 1.001(12)& 0.923(22) & 0.921(10) & 0.998(14) \\
$a + b m_q  + c m_q ^2$     & 0.859(24) & 0.848(7) & 0.987(22) & 0.940(33) & 0.919(9)  & 0.977(28)\\
$a + b m_q  + c m_q  \ln m_q $ & 0.874(33) & 0.848(7) & 0.970(32) & 0.958(45) & 0.919(9)  & 0.959(39)\\
\hline %
\end{tabular}
\end{table}
\begin{table}[hbt]
\caption{\label{tab:SU3ratio} Fit expressions and various bag
parameter values.}
\begin{tabular}{ccccc}
 fit & $\frac{f_{D_{\rm  s}}}{f_D}\sqrt{\frac{B_{D_{\rm  s}}}{B_D}}$ & $(\frac{f_{D_{\rm  s}}m_{D_{\rm  s}}}{m_D
 f_D})^2{\frac{B_{D_{\rm  s}}}{B_D}}$&
$\frac{f_{B_{\rm  s}}}{f_B}\sqrt{\frac{B_{B_{\rm  s}}}{B_B}}$ &
$(\frac{f_{B_{\rm  s}}m_{B_{\rm  s}}}{m_B f_B})^2{\frac{B_{B_{\rm  s}}}{B_B}}$\\
\hline\hline%
$a + b m_q       $             & 1.064(29) & 1.257(73) & 1.019(53) & 1.27(13)\\
$a + b m_q  + c m_q ^2$        & 1.071(23) & 1.273(65) & 1.019(37) & 1.274(94)\\
$a + b m_q  + c m_q  \ln m_q $ & 1.048(31) & 1.219(80) & 1.001(53) & 1.23(13)\\
\hline %
\end{tabular}
\end{table}

\clearpage

\begin{figure}
\includegraphics[width=0.95\columnwidth,angle = 0]{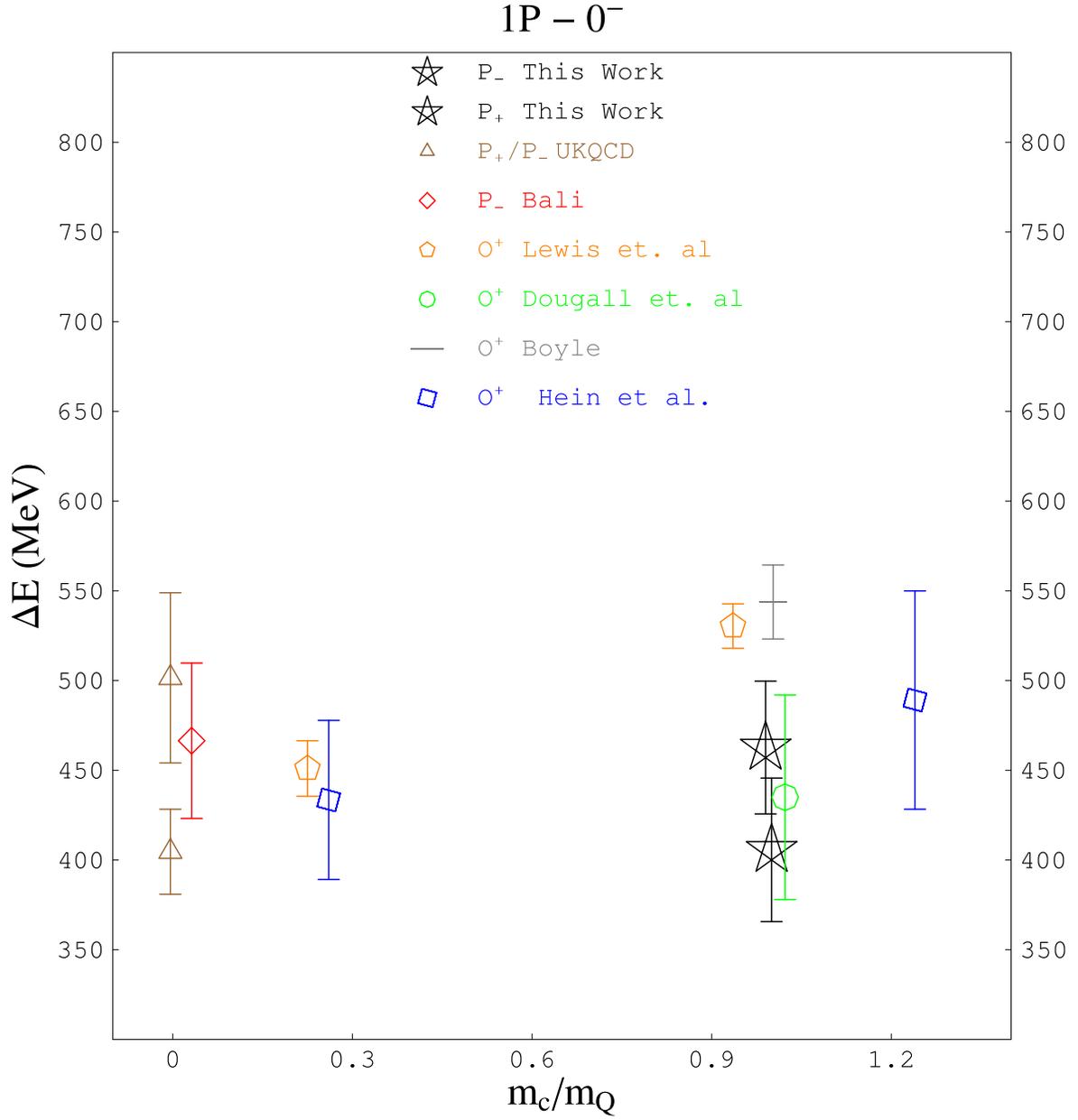} \caption{\small\small
The summary plots of the previous lattice estimate for
$\Delta_{sJ}$ including ours. } \label{summary of lattice
calc}\vskip -3mm
\end{figure}

\clearpage

%

\end{document}